\documentstyle[12pt]{article}
     \def\lsim{\raise0.3ex\hbox{$<$\kern-0.75em\raise-1.1ex\hbox{$\sim$}}}
\def\gsim{\raise0.3ex\hbox{$>$\kern-0.75em\raise-1.1ex\hbox{$\sim$}}}
\def\noi{\noindent}
\def\nn{\nonumber}
\def\bea{\begin{eqnarray}}  \def\eea{\end{eqnarray}}
\def\beq{\begin{equation}}   \def\eeq{\end{equation}}

\def\beeq{\begin{eqnarray}} \def\eeeq{\end{eqnarray}}
\def\R{ {\rm R \kern -.31cm I \kern .15cm}}
\def\C{ {\rm C \kern -.15cm \vrule width.5pt \kern .12cm}}
\def\Z{ {\rm Z \kern -.27cm \angle \kern .02cm}}
\def\N{ {\rm N \kern -.26cm \vrule width.4pt \kern .10cm}}
\def\1{{\rm 1\mskip-4.5mu l} }

\begin{document}
\begin{center}
{\large \bf Sum rules in the heavy quark limit of QCD} \\
\vspace{1 truecm}
{\bf A. Le Yaouanc, L. Oliver and J.-C. Raynal}\\
{\it Laboratoire de Physique Th\'eorique}\footnote{Unit\'e Mixte de Recherche
UMR 8627 - CNRS }\\    {\it Universit\'e de Paris XI, B\^atiment 210, 91405
Orsay Cedex, France}
\end{center}

\begin{abstract}
In the leading order of the heavy quark expansion, we propose a method
within the OPE and the trace formalism, that allows to obtain, in a
systematic way, Bjorken-like sum rules for the derivatives of the
elastic Isgur-Wise function $\xi(w)$ in terms of corresponding
Isgur-Wise functions of transitions to excited states. A key element is
the consideration of the non-forward amplitude, as introduced by
Uraltsev. A simplifying feature of our method is to consider
currents aligned along the initial and final four-velocities. As an
illustration, we give a very simple derivation of Bjorken and Uraltsev
sum rules. On the other hand, we obtain a new class of sum rules that
involve the products of IW functions at zero recoil and IW functions at
any $w$. Special care is given to the needed derivation of the projector
on the polarization tensors of particles of arbitrary integer spin.
The new sum rules give further information on the slope $\rho^2 = - \xi
'(1)$ and also on the curvature $\sigma^2 = \xi '' (1)$, and imply,
modulo a very natural assumption, the inequality $\sigma^2 \geq {5\over
4} \rho^2$, and therefore the absolute bound $\sigma^2 \geq {15 \over
16}$. \end{abstract}

\vskip 1 truecm

\noi LPT Orsay 02-29 \par
\noi July 2002\par
\vskip 1 truecm

\noindent e-mails : leyaouan@th.u-psud.fr, oliver@th.u-psud.fr
\newpage
\pagestyle{plain}
\section{Introduction.}
\hspace*{\parindent}
Since the formulation of Bjorken sum rule \cite{1r}, other sum rules 
(SR) have been derived involving leading and subleading quantities in
the heavy quark expansion \cite{2r,2rnew,3r,4r,5r}. The recent Uraltsev SR 
\cite{5r,6rnew} at leading order came as a big surprise, leading to the
rigorous lower bound for the elastic Isgur-Wise function $\rho^2 \geq 
3/4$ \footnote{This bound was obtained in a class of relativistic 
quark models
(\cite{6r}, \cite{7newr}), that were afterwards shown to satisfy 
Uraltsev sum rule \cite{8newr}.}. As with earlier SR, one gets the 
impression that these results
come out like a fishing in a lake, swarming with sum rules, the 
success of the catch depending on the genius or skill of the 
particular authors. Hence the
necessity of having a systematic way of formulating these SR. This is 
the subject of the present paper, although only in the particular 
case of IW functions in the
heavy quark limit of QCD. The method can be easily applied to 
subleading form factors \cite{6rnewref}. \par

In the derivation of the sum rules we will make use of the Operator 
Product Expansion (OPE) \cite{7r} in heavy quark transitions
\cite{2r,4r,5r,8r}, in a manifestly covariant approach. \par

To be completely general, let us consider the direct graphs $B_i(v_i) 
\displaystyle{\mathrel{\mathop \rightarrow^{\Gamma_1}}} D^{(n)}(v')
\displaystyle{\mathrel{\mathop \rightarrow^{\Gamma_2}}} B_f(v_f)$, 
where $B_i$ and $B_f$ are ground state $B$ or $B^{*}$ mesons and 
$D^{(n)}$ are all possible
ground state or excited $D$ mesons coupled to $B_i$ and $B_f$ through 
the currents $\overline{h}_c(v')\Gamma_1h_b(v_i)$ and
$\overline{h}_b(v_f)\Gamma_2h_c(v')$. The Dirac matrices $\Gamma_i$ 
($i=1,2$) are arbitrary and can be chosen to derive relations
involving definite current matrix elements. \par

Let us summarize the general argument. We consider two arbitrary currents~:
\beq
\label{1e}
J_1 (x) = \overline{c} (x) \Gamma_1 b(x) \qquad J_2 (y) = 
\overline{b} (y) \Gamma_2 c(y)
		 \eeq

\noi and their $T$ product
\beq
\label{2e}
T_{fi}(q) = i \int d^4x\ e^{-iq\cdot x} <B_f |T\left [ J_2(x) J_1(0) 
\right ] |B_i> \ .
		 \eeq

As explained in detail for example in ref. \cite{4r}, inserting in 
this expression intermediate states, $x < 0$ receives contributions 
from the direct channel
with a single heavy quark $c$, while $x > 0$ receives contributions 
from intermediate states with $b\overline{c}b$ quarks, the $Z$ 
diagrams.
The energy denominators are $M_B - q^0 - E_{X_c}$ for the direct 
graphs and $M_B + q^0 - (E_{X'_c} + 2M_B)$ for the $Z$ diagrams. 
Taking the typical
virtuality of the direct channels ${V} = M_B - q^0 - E_{X_c}$ such 
that $\Lambda_{QCD} \ll {V} \ll M_B$, one sees that the direct 
channels
contribute at the order $1/{V}$ and the $Z$ diagrams at the order 
$1/(-{V}-2M_D)$. In both cases the absolute value of the denominator
is $\gg \Lambda_{QCD}$. This allows to approximate (\ref{2e}) with 
the leading contribution to the OPE \cite{8r}~:
\beq
\label{3e}
T_{fi}(q) = i \int d^4x\ e^{-iq\cdot x} <B_f |\overline{b}(x) 
\Gamma_2 S_c(x,0) \Gamma_1 b(0) |B_i> + \ O(1/m_c^2)
		 \eeq

\noi where $S_c(x,0)$ is the free charm quark propagator if 
$O(\alpha_s)$ corrections are neglected. The $c$ quark propagator has 
two terms, a positive
energy denominator $\sim {V}$ and a negative energy denominator $\sim 
(- {V} - 2m_c)$. Varying ${V}$ independently of $m_c$ one can equate
the direct channel contribution to (\ref{2e}) to the one of the positive energy pole 
of the $c$ quark propagator in (\ref{3e}), the so-called OPE side, 
giving the following result
that involves only the direct channel~:
	\bea
\label{4e}
&&{1 \over 2v'^{0} \sqrt{4v_i^0v_f^0}} \nn \\
&&\Big \{ \sum_{D=P,V} \sum_n Tr \left [ \overline{\cal B}_f (v_f) 
\overline{\Gamma}_2 {\cal D}^{(n)}(v')  \right ] Tr \left [ 
\overline{\cal D}^{(n)} (v') \Gamma_1
{\cal B}_i(v_i) \right ] \xi^{(n)}(w_i)\xi^{(n)*} (w_f) \nn \\ 
&&\hbox{+ Contribution from other excited states + $O(1/m_Q)$} \Big 
\}\nn \\
&&= - {1 \over \sqrt{4v_i^0v_f^0}} \ \xi(w_{if})\  Tr \left [ 
\overline{\cal B}_f (v_f ) {\Gamma}_2 {{/\hskip - 2 truemm v'}_c
+ 1 \over 2v_c^{'0} } \Gamma_1 {\cal B}_i(v_i) \right ] + O(1/m_Q) \ .\eea

\noi In this equation, ${{/\hskip - 1.5 truemm v'}_c + 1 \over 
2v_c^{'0} }$ is the positive energy residue of the $c$ quark 
propagator and the
l.h.s. is the sum over all possible ground state or excited $D$ 
mesons. We have adopted the trace formalism for the current matrix
elements \cite{3r,9r} and made explicit in (\ref{4e}) the sum over pseudoscalar and 
vector $D(D^*)$ mesons and their radial quantum number.  In relation 
(\ref{4e})
\beq
\label{5e}
w_i = v_i \cdot v' \quad , \quad w_f = v_f \cdot v' \quad ,  \quad 
w_{if} = v_i \cdot v_f \ .
\eeq

\noi In the l.h.s. there are also leading order contributions of 
excited states and subleading terms coming from the ground state or 
from transitions between the
ground state and excited states, denoted by $O(1/m_Q)$, where $m_Q$ 
can be $m_c$ or $m_b$. \par

One main point we want to emphasize is that in the OPE side the 
ground state IW function $\xi(w_{if})$ appears since, following Uraltsev \cite{5r} we assume in 
general $v_i \not=
v_f$ and take $B_i$ and $B_f$ to be ground state $B$ mesons. Of 
course, for $w_{if} = 1$ one gets $\xi (1) = 1$, $w_i = w_f =
w$ and the general formula (\ref{4e}) takes the more familiar form \cite{4r}~:
 
\bea
\label{6e}
&&{1 \over 4v^{0} v'^{0}} \Big \{ \sum_{D=P,V} \sum_n Tr \left [ 
\overline{\cal B}_f(v) \overline{\Gamma}_2 {\cal D}^{(n)}(v') \right 
] Tr \left [ \overline{\cal
D}^{(n)} (v') \Gamma_1 {\cal B}_i(v) \right ]  |\xi^{(n)}(w)|^2 \nn 
\\ &&\hbox{+ Contribution from other excited states + $O(1/m_Q)$} 
\Big \}\nn \\
&&= - {1 \over 2v^0} Tr \left [ \overline{\cal B}_f (v ) {\Gamma}_2 
{{/\hskip - 2 truemm v'}_c
+ 1 \over 2v_c^{'0} } \Gamma_1 {\cal B}_i(v) \right ] + O(1/m_Q^2) \ .\eea
\vskip 3 truemm

  But let us keep to the general case $v_i \not= v_f$. By choosing in a convenient  way the initial and final mesons $B_i$ and $B_f$ and
  the Dirac matrices $\Gamma_1$ and $\Gamma_2$, one can derive sum rules at 
the leading order (Bjorken SR \cite{1r} and Uraltsev SR
\cite{5r}) and also SR involving subleading Isgur-Wise functions, as 
we have obtained in ref. \cite{4r}. To illustrate the method,
we will limit ourselves in this paper to the heavy quark limit. \par

In the heavy quark limit, since we can make the four-velocity of the 
intermediate quark equal to the intermediate hadron velocity, $v'_c
= v'$, relation (\ref{4e}) writes, multiplying by $2v'_0\sqrt{4v_i^0v_f^0}$
\beq
\label{7e}
L\left ( w_{if}, w_i, w_{f} \right ) = R \left ( w_{if} , w_i, w_{f} 
\right ) \ ,
			\eeq

\noi where $L(w_{if}, w_i, w_f)$ stands for the l.h.s. (the sum over 
intermediate states ${\cal D}^{(n)}(v')$) and $R(w_{if}, w_i, w_f)$ 
stands for the r.h.s. (the
OPE side, proportional to $\xi(w_{if})$). \par

The variables $w_{if}$, $w_i$ and $w_f$ are independent within a 
certain domain. Indeed, without loss of
generality one can take \beq
\label{8e}
v_i = (1,0,0,0) \qquad v_f = \left ( \sqrt{1 + a^2}, 0, 0, a \right ) 
\qquad v' = \left ( \sqrt{1 + b^2 + c^2}, 0, b, c \right )
\eeq

\noi giving
\beq
\label{9e}
w_{if} = \sqrt{1 + a^2} \qquad w_i = \sqrt{1 + b^2 + c^2} \qquad w_f 
= \sqrt{1 + b^2 + c^2} \sqrt{1 + a^2} - ac \ .
\eeq

\noi One has three independent parameters $a$, $b$ and $c$ or 
equivalently $w_i$, $w_f$ and $w_{if}$ that lie within a limited 
domain. The domain of
($w_{if}$, $w_i$, $w_f$) is
\beq
\label{10e}
w_{if} \geq 1 \quad , \quad 2w_{if} w_i w_f - w_{if}^2 - w_i^2 - 
w_f^2 + 1 \geq 0
\eeq

\noi that implies
\beq
\label{11e}
w_i \geq 1 \quad , \quad w_f \geq 1
\eeq

\noi and is equivalent to
$$w_i \geq 1 \quad , \qquad w_f \geq 1$$

\beq
\label{12e}
  w_i w_f - \sqrt{(w_i^2 - 1) (w_f^2-1)} \leq w_{if} \leq w_i  w_f + 
\sqrt{(w_i^2 - 1)(w_f^2-1)} \ .
\eeq

\noi There is a subdomain for $w_i = w_f = w$, namely~:
\beq
\label{13e}
w \geq 1 \quad , \qquad 1 \leq w_{if} \leq 2w^2-1 \ .
\eeq

\noi Within this domain one can differentiate relatively to any of these variables,
\beq
\label{14e}
{\partial^{p+q+r} L \over \partial w_{if}^p \ \partial w_i^q \ 
\partial w_f^r} = {\partial^{p+q+r} R \over \partial w_{if}^p \ 
\partial w_i^q \ \partial w_f^r}
			\eeq

\noi and obtain different sum rules taking different limits to the 
frontier of the domain, e.g.,
\bea
\label{15e}
&\qquad &w_{if} \to 1 \ , \  w_i = w_f = w\nn \\
&\hbox{or} \qquad &w_{i} \to 1 \ , \  w_{if} = w_f = w\nn \\
&\hbox{or} \qquad  &w_{f} \to 1 \ , \  w_{if} = w_i = w \ .
\eea
  \vskip 3 truemm

A last general remark. In the SR we will consider the sum over discrete
intermediate ground state or excited D mesons. However, our results have
a wider generality, as they can include a possible continuum. Such a
continuum would only be a slight technical complication, as it can also
be expanded into $j^P$ states, and the sum over discrete states would
become an integral, without any conceptual change in the final results.\par

The paper is organized as follows. In Section 2 we write down the
general form of the SR in the heavy quark limit for a general pair of
currents $\bar{h}_{v'}^{(c)}\Gamma_1 h_{v_i}^{(b)}$,
$\bar{h}_{v_f}^{(b)}\Gamma_2 h_{v'}^{(c)}$, making explicit the
intermediate states ${1\over 2}^-$, ${1 \over 2}^+$, ${3 \over 2}^+$,
and ${3 \over 2}^-$ as well, in order to have a control on high powers
of the recoil $(w-1)$. In Section 3 we derive the sum rules (in
particular Bjorken and Uraltsev SR) for the axial currents $\{\Gamma_1,
\Gamma_2\} = \{ {/\hskip-2 truemm v}_i\gamma_5, {/\hskip-2 truemm
v}_f\gamma_5\}$ and in Section 4 similarly for the vector currents
$\{\Gamma_1, \Gamma_2\} = \{ {/\hskip-2 truemm v}_i, {/\hskip-2 truemm
v}_f\}$. In Section 5 we underline a new class of sum rules with
implications, in particular, for the slope and curvature of $\xi (w)$.
Moreover, we demonstrate that higher excited states give a vanishing
contribution to these SR. In Section 6 we write down a lower bound on
the curvature of $\xi (w)$ and in Section 7 we point out some
phenomenological remarks in connection with the Bakamjian-Thomas class
of relativistic quark models. In Section 8 we conclude. In Appendix A we
construct the general formula for the projector on the polarization
tensors of particles of arbitrary spin. With it, we deduce a formula
that is needed in the calculation  of the contributions to the sum rules
of higher excited states. Using this general result, we have recently
obtained rigorous bounds on all derivatives of the IW function $\xi (w)$
\cite{6rbis}. For the curvature $\sigma^2 = \xi '' (1)$ we find in the
present paper the same bound using a different method and making a sensible
phenomenological hypothesis. Finally, in Appendix B we give a derivation
of Bjorken and Uraltsev SR with the currents $\{\Gamma_1, \Gamma_2\} =
\{ {/\hskip-2 truemm v}_i, {/\hskip-2 truemm v}_i\}$ and initial and
final states $B^{*(\lambda_i)}(v_i)$, $B^{*(\lambda_f)}(v_f)$, a
manifestly covariant version of those states and currents used by
Uraltsev, $\{\Gamma_1, \Gamma_2\} = \{\gamma^0, \gamma^0\}$ in the rest
frame of the initial $B^{*(\lambda_i)}(1, {\bf 0})$. Of course, this
choice of the vector current would make simpler the calculation of
radiative corrections to the sum rules than in the case, say, of the
axial current. But radiative corrections are outside the scope of the
present paper, that adopts the strict heavy quark limit.

\section{General form of the sum rules in the heavy quark limit.}
\hspace*{\parindent} The r.h.s. writes, in the heavy quark limit, since
then $v'_c = v'$~: \beq
\label{16e}
R(w_{if},w_i,w_f) = - \ 2 \xi(w_{if}) Tr \left [ \overline{\cal 
B}_f(v_f) {\Gamma}_2 P'_+ \Gamma_1 {\cal B}_i(v_i)\right ]  \ .
  \eeq

\noi Let us decompose the l.h.s. into contributions of the different 
intermediate states~: as intermediate states, we will consider the 
$0^-_{1/2}$, $1^-_{1/2}$,
and the orbitally excited states $2^+_{3/2}$, $1^+_{3/2}$, 
$0^+_{1/2}$, $1^+_{1/2}$ (with the tower of their radial 
excitations). Moreover, to have some control of
the SR near zero recoil, it is important to have an idea of the 
contributions of higher orbital excitations. To this
purpose, we will take into account the ${3\over 2}^-$ intermediate 
states, namely the states $2^-_{3/2}$ and $1^-_{3/2}$. \par

Let us now write down the $4 \times 4$ matrices of the lower $j^P$
states \cite{3r,9r}. The matrices for the ${1 \over 2}^-$ mesons read~: 
\bea \label{17e} &0_{1/2}^- : &\qquad {\cal M}(v) = P_+ (-\gamma_5) \nn \\
&1_{1/2}^- : &\qquad {\cal M}(v) = P_+ \varepsilon_v^{\mu} \gamma_{\mu}
\eea

\noi where $P_+$ is the projector~:
\beq
\label{18e}
P_+ = {1 + {/\hskip - 2 truemm v} \over 2} \ .
\eeq

\noi The $4 \times 4$ matrices of the ${3 \over 2}^+$ states are 
given by the four-vectors~:
\bea
\label{19e}
&2_{3/2}^+ : &\qquad \qquad {\cal M}^{\mu} (v) = P_+ 
\varepsilon_v^{\mu\nu} \gamma_{\mu} \nn \\
&1_{3/2}^+ : &\qquad \qquad {\cal M}^{\mu} (v) = - \sqrt{{3 \over 2}} 
P_+ \varepsilon_v^{\nu} \gamma_5 \left [ g_{\nu}^{\mu} - {1 \over 3} 
\gamma_{\nu} (
\gamma^{\mu} - v^{\mu}) \right ] \eea

\noi and those of the ${1 \over 2}^+$ states are given by \cite{3r}~:
\bea
\label{20e}
&0_{1/2}^+ : &\qquad {\cal M}(v) = P_+ \nn \\
&1_{1/2}^+ : &\qquad {\cal M}(v) = P_+ \varepsilon_v^{\mu} \gamma_5 
\gamma_{\mu} \ .
\eea

\noi Finally, those of the ${3 \over 2}^-$ states will be obtained 
from (\ref{19e}) by multiplying on the right by $(- \gamma_5)$~:
\bea
\label{21e}
&2_{3/2}^- : &\qquad \qquad {\cal M}^{\mu} (v) = P_+ 
\varepsilon_v^{\mu\nu} \gamma_{\nu} (- \gamma_5)\nn \\
&1_{3/2}^- : &\qquad \qquad {\cal M}^{\mu} (v) = \sqrt{{3 \over 2}} 
P_+ \varepsilon_v^{\nu} \left [ g_{\nu}^{\mu} - {1 \over 3} 
\gamma_{\nu} (
\gamma^{\mu} + v^{\mu}) \right ] \ .
\eea

The corresponding matrix elements, for a current given by the 
Dirac matrix $\Gamma$, read \cite{3r}~:
\beq
\label{22e}
<D^{(n)} \left ( \scriptstyle{{1 \over 2}}^- \right ) (v') 
|\overline{h}_{v'}^{(c)} \Gamma h_v^{(b)}|B \left ( \scriptstyle{{1 
\over 2}}^-\right )(v) > \
=  \xi^{(n)}(w) Tr \left [ \overline{\cal D}(v') \Gamma {\cal B} (v) 
\right ]\eeq
\beq
\label{23e}
<D^{(n)} \left ( \scriptstyle{{3 \over 2}}^+ \right ) (v') 
|\overline{h}_{v'}^{(c)} \Gamma h_v^{(b)}|B \left ( \scriptstyle{{1 
\over 2}}^-\right )(v) > \ = \sqrt{3}
\tau_{3/2}^{(n)}(w) Tr \left [ v_{\mu} \overline{\cal D}^{\mu}(v') 
\Gamma {\cal B}(v) \right ]\eeq
\beq
\label{24e}
<D^{(n)} \left ( \scriptstyle{{1 \over 2}}^+ \right ) (v') 
|\overline{h}_{v'}^{(c)} \Gamma h_v^{(b)}|B \left ( \scriptstyle{{1 
\over 2}}^-\right )(v) > \ =
2 \tau_{1/2}^{(n)}(w) Tr \left [ \overline{\cal D}(v') \Gamma {\cal 
B}(v) \right ]  \eeq
\beq
\label{25e}
<D^{(n)} \left ( \scriptstyle{{3 \over 2}}^- \right ) (v') 
|\overline{h}_{v'}^{(c)} \Gamma h_v^{(b)}|B \left ( \scriptstyle{{1 
\over 2}}^-\right )(v) > \ =
\sqrt{3} \sigma_{3/2}^{(n)}(w) Tr \left [ v_{\mu} \overline{\cal 
D}^{\mu}(v') \Gamma {\cal B}(v) \right ]
\eeq

\vskip 3 truemm

\noi where $w = v \cdot v'$, $n$ is a radial quantum number and, in 
analogy with $\tau_{3/2}(w)$, we have called $\sigma_{3/2}(w)$ the IW 
function between the ground
state and the ${3 \over 2}^-$ states. As pointed out in \cite{3r}, 
$\sigma_{3/2}(w)$ need not to vanish at $w = 1$, since the current 
matrix elements vanish in the
heavy quark limit. The notation $\xi^{(n)}(w)$, $\tau_{1/2}^{(n)}(w)$ 
and $\tau_{3/2}^{(n)}(w)$ is the one of Isgur and Wise \cite{1r}.\par

In what follows, we set the different IW functions
to be real. \par

The contributions of the $0^-_{1/2}$, $1^-_{1/2}$ states write~:
	\beq
\label{26e}
L(0^-_{1/2}) =  Tr \left [ \overline{\cal B}_f(v_f)\overline{\Gamma}_2
P'_+(-\gamma_5)  \right ] Tr \left [ (-\gamma_5)P'_+\Gamma_1 {\cal 
B}_i(v_i)\right ] \sum_n \xi^{(n)}(w_i)\xi^{(n)}(w_f)
			\eeq
\beq
\label{27e}
L(1^-_{1/2}) =
\sum_{\lambda} \varepsilon '^{(\lambda )\mu}\varepsilon '^{(\lambda 
)\nu} Tr \left [ \overline{\cal B}_f (v_f) \overline{\Gamma}_2 P'_+ 
\gamma_{\nu} \right ] Tr
\left [ \gamma_{\mu} P'_+ \Gamma_1 {\cal B}_i (v_i) \right ]  \sum_n 
\xi^{(n)}(w_i)\xi^{(n)}(w_f) \ .
			\eeq

The contribution of the parity + excited states $2^+_{3/2}$, 
$1^+_{3/2}$, $0^+_{1/2}$, $1^+_{1/2}$ is given by the
following expressions~:
\beq
\label{28e}
L(2^+_{3/2}) =  \sum_{\lambda} \varepsilon '^{(\lambda 
)\mu\nu}\varepsilon '^{(\lambda )*\rho\sigma} Tr \left [ v_{f\rho} 
\overline{\cal B}_f\overline{\Gamma}_2
P'_+  \gamma_{\sigma} \right ]  Tr \left [ v_{i\mu} \gamma_{\nu} P'_+ 
\Gamma_1 {\cal B}_i\right ] 3 \sum_n \tau_{3/2}^{(n)}(w_i) 
\tau_{3/2}^{(n)} (w_f)
			\eeq
\bea
\label{29e}
&&L(1^+_{3/2}) = \sum_{\lambda} \varepsilon '^{(\lambda 
)\nu}\varepsilon '^{(\lambda )*\sigma} {3 \over 2} Tr \left \{ 
\overline{\cal B}_f \overline{\Gamma}_2 P'_+
\gamma_5 \left [ v_{f\sigma} - {1 \over 3} \gamma_{\sigma} ({/\hskip 
- 2 truemm v}_f - w_f ) \right ] \right \}  \nn \\
&&Tr \left \{ \left [ v_{i\nu} - {1 \over 3} ({/\hskip - 2 truemm 
v}_i - w_i) \gamma_{\nu} \right ] (-\gamma_5)P'_+ \Gamma_1 {\cal 
B}_i\right \} 3 \sum_n \tau_{3/2}^{(n)}(w_i)
\tau_{3/2}^{(n)} (w_f)  \eea

\beq
\label{30e}
L(0^+_{1/2}) =  Tr \left [ \overline{\cal B}_f \overline{\Gamma}_2 
P'_+ \right ] Tr \left [ P'_+ \Gamma_1 {\cal B}_i \right ] 4 \sum_n 
\tau_{1/2}^{(n)} (w_i)
\tau_{1/2}^{(n)}(w_f)
			\eeq

\beq
\label{31e}
L(1^+_{1/2}) = \sum_{\lambda} \varepsilon '^{(\lambda 
)\mu}\varepsilon '^{(\lambda )*\nu} Tr \left [  \overline{\cal 
B}_f\overline{\Gamma}_2 P'_+  \gamma_5
\gamma_{\mu} \right ] Tr \left [ \gamma_{\nu}(-\gamma_5) P'_+ 
\Gamma_1 {\cal B}_i\right ] 4 \sum_n \tau_{1/2}^{(n)}(w_i) 
\tau_{1/2}^{(n)} (w_f)   \ .
			\eeq

\bea
\label{32e}
&&L(2^-_{3/2}) = \sum_{\lambda} \varepsilon '^{(\lambda 
)\mu\nu}\varepsilon '^{(\lambda )*\rho\sigma} Tr \left [ v_{f\rho} 
\overline{\cal B}_f\overline{\Gamma}_2
P'_+  \gamma_{\sigma} (-\gamma_5) \right ] \nn \\
&&Tr \left [ v_{i\mu} \gamma_5 \gamma_{\nu} P'_+ \Gamma_1 {\cal 
B}_i\right ] 3 \sum_n \sigma_{3/2}^{(n)}(w_i)
\sigma_{3/2}^{(n)} (w_f) \eea
\bea
\label{33e}
&&L(1^-_{3/2}) = \sum_{\lambda} \varepsilon '^{(\lambda 
)\nu}\varepsilon '^{(\lambda )*\sigma} {3 \over 2} Tr \left \{ 
\overline{\cal B}_f \overline{\Gamma}_2 P'_+
\left [ v_{f\sigma} - {1 \over 3} \gamma_{\sigma} ({/\hskip - 2 
truemm v}_f + w_f ) \right ] \right \} \nn \\
&&  Tr \left \{ \left [ v_{i\nu} - {1
\over 3} ({/\hskip - 2 truemm v}_i + w_i) \gamma_{\nu} \right ] P'_+ 
\Gamma_1 {\cal B}_i\right \} 3 \sum_n \sigma_{3/2}^{(n)}(w_i) 
\sigma_{3/2}^{(n)} (w_f) \ .
\eea

\noi It is convenient to introduce the tensors
\bea
\label{34e}
&&T^{\mu\nu} = \sum_{\lambda} \varepsilon '^{(\lambda )\mu} 
\varepsilon '^{(\lambda )*\nu} \\
&&T^{\mu\nu , \rho \sigma} = \sum_{\lambda} \varepsilon '^{(\lambda 
)\mu\nu} \varepsilon '^{(\lambda )*\rho\sigma} \ .
\label{35e}
			\eea

The polarizations of the vector and tensor intermediate 
states of velocity $v'$ satisfy $\varepsilon 
'^{(\lambda
)} \cdot v' = \varepsilon '^{(\lambda )\mu\nu}v'_{\nu} = 0$. 
Moreover, the polarization tensor $\varepsilon '^{(\lambda)\mu\nu}$ 
is symmetric in $(\mu\nu)$ and
traceless, $\varepsilon '^{(\lambda)\mu}_{\ \ \ \ \ \mu} = 0$. One 
can show that these tensors write~:  \bea
\label{36e}
&&T^{\mu\nu} = - g^{\mu\nu} + v'^{\mu}v'^{\nu} \\
&& \nn \\
&&T^{\mu\nu , \rho \sigma} = {1 \over 6} \Big \{ 
-2g^{\rho\sigma}g^{\mu\nu} + 3 \left [ g^{\rho\mu} g^{\sigma \nu} + 
g^{\rho\nu}
g^{\sigma \mu} \right ] + 2 \left [ g^{\rho\sigma} v'^{\mu} v'^{\nu} 
+ v'^{\rho} v'^{\sigma}g^{\mu\nu} \right ] \nn \\
&&- 3 \left [ g^{\rho\mu} v'^{\sigma} v'^{\nu} + g^{\sigma\nu} 
v'^{\rho} v'^{\mu} + g^{\rho\nu} v'^{\sigma} v'^{\mu} + g^{\sigma \mu}
v'^{\rho} v'^{\nu}\right ] + 4 v'^{\mu} v'^{\nu} v'^{\rho} 
v'^{\sigma} \Big \} 
\label{37e}
\eea

\noi and have the following properties~: $T^{\mu\nu}$ is 
symmetric and $T_{\ \mu}^{\mu} = - 3$ while $T^{\mu\nu , \rho \sigma}$
is symmetric in the exchanges $(\mu \nu \leftrightarrow \rho \sigma 
)$, $(\mu \leftrightarrow \nu)$ and $(\rho \leftrightarrow \sigma)$
and satisfies $T^{\mu\nu ,}_{\ \ \ \mu\nu} = + 5$ (the $+$ sign comes 
from the fact that the polarization of a spin 2 particle can be seen 
as
a symmetric combination of the polarizations of two spin 1 
particles). With these expressions for the polarization tensors one 
can
make more explicit the contributions of the intermediate states in 
the l.h.s. of the SR (\ref{4e}). After some algebra one gets, from 
(\ref{26e})-(\ref{33e}), for
arbitrary Dirac matrices $\Gamma_1$ and $\Gamma_2$, the following SR
\bea
\label{38e}
&&\left \{ \left [ Tr \left [ \overline{\cal B}_f\overline{\Gamma}_2 
P'_+\gamma_5 \right ] Tr  \left [ (-\gamma_5) P'_+ \Gamma_1 {\cal 
B}_i\right ] \right
] \right .\nn \\
&&+ \Big [ - Tr \left [ \overline{\cal B}_f \overline{\Gamma}_2 
P'_+\gamma_{\mu} \right ] Tr \left [ \gamma^{\mu} P'_+ \Gamma_1 {\cal 
B}_i \right ] \nn
\\ &&+ \ Tr \left [ \overline{\cal B}_f \overline{\Gamma}_2 
P'_+\right ] Tr \left [ P'_+ \Gamma_1 {\cal B}_i \right ] \Big ] \Big 
\} \sum_n \xi^{(n)} (w_i)
\xi^{(n)}(w_f) \nn \\
&&+ {1 \over 2} \left \{ \Big [ 3 \left ( w_{if} - w_f w_i \right ) Tr \left
[ \overline{\cal B}_f \overline{\Gamma}_2 P'_+\gamma_{\sigma} \right 
] Tr \left [ \gamma^{\sigma} P'_+ \Gamma_1 {\cal B}_i \right ] \right 
.\nn \\
&&+ \left ( - 2 - 2w_i-2w_f - 3 w_{if} + 4 w_iw_f \right ) Tr \left [ 
\overline{\cal B}_f
\overline{\Gamma}_2 P'_+\right ] Tr \left [ P'_+ \Gamma_1 {\cal B}_i 
\right ]  \nn \\
&&+ \ 3 \ Tr \left [ \overline{\cal B}_f \overline{\Gamma}_2 P'_+ {/
\hskip - 2 truemm v}_i \right ] Tr \left [ {/ \hskip - 2 truemm v}_f 
P'_+ \Gamma_1 {\cal B}_i \right ]  - 3w_i \ Tr \left [ \overline{\cal 
B}_f \overline{\Gamma}_2 P'_+ \right ]
Tr \left [ {/ \hskip - 2 truemm v}_f P'_+ \Gamma_1 {\cal B}_i \right ]  \nn \\
&&- \ 3w_f \ Tr \left [ \overline{\cal B}_f \overline{\Gamma}_2 P'_+ 
{/ \hskip - 2 truemm v}_i \right ] Tr \left [ P'_+ \Gamma_1 {\cal 
B}_i \right ]   \Big ] \nn \\
&&+ \Big [ - (1 + w_i) (1 + w_f) Tr \left [ \overline{\cal B}_f 
\overline{\Gamma}_2 P'_+\gamma_5 \gamma_{\sigma} \right ] Tr \left [ 
\gamma^{\sigma} (-\gamma_5) P'_+
\Gamma_1 {\cal B}_i \right ] \nn \\
&&+ \left ( 1 - 9w_{if} + 4w_i w_f - 2 w_i - 2w_f \right ) Tr \left [ 
\overline{\cal B}_f \overline{\Gamma}_2 P'_+\gamma_5 \right ] Tr 
\left [ (-\gamma_5) P'_+
\Gamma_1 {\cal B}_i \right ] \nn \\
&&- \ 3 (1 + w_f) \ Tr \left [ \overline{\cal B}_f \overline{\Gamma}_2
P'_+ \gamma_5 {/ \hskip - 2 truemm v}_i \right ] Tr \left [ 
(-\gamma_5) P'_+ \Gamma_1 {\cal B}_i \right ] \nn \\
&&- \ 3(1 + w_i) \ Tr \left [
\overline{\cal B}_f \overline{\Gamma}_2 P'_+ \gamma_5 \right ] Tr 
\left [ {/ \hskip - 2 truemm v}_f (-\gamma_5) P'_+ \Gamma_1 {\cal 
B}_i \right ] \Big ]  \Big \}
\sum_n \tau_{3/2}^{(n)}(w_i) \tau_{3/2}^{(n)} (w_f) \nn \\
&& + \ 4 \Big \{ \Big [ Tr \left [ \overline{\cal B}_f \overline{\Gamma}_2
P'_+\right ]Tr \left [ P'_+ \Gamma_1 {\cal B}_i \right ] \Big ] \nn \\
&&+ \Big [ - Tr \left [ \overline{\cal B}_f \overline{\Gamma}_2 P'_+ 
\gamma_5 \gamma_{\sigma} \right ] Tr \left [ \gamma^{\sigma} 
(-\gamma_5) P'_+ \Gamma_1 {\cal
B}_i \right ]  \nn \\
&&+ \ Tr \left [ \overline{\cal B}_f \overline{\Gamma}_2 P'_+ 
\gamma_5  \right ] Tr \left [ (-\gamma_5) P'_+ \Gamma_1 {\cal B}_i 
\right ]  \Big ] \Big \} \sum_n
\tau_{1/2}^{(n)}(w_i) \tau_{1/2}^{(n)} (w_f) \nn \\
&&+ {1 \over 2} \Big \{ \Big [ 3 \left ( w_{if} - w_f w_i \right ) Tr \left
[ \overline{\cal B}_f \overline{\Gamma}_2 P'_+\gamma_{\sigma} 
(-\gamma_5) \right ] Tr \left [ \gamma_5\gamma^{\sigma} P'_+ \Gamma_1 
{\cal B}_i \right ] \nn \\
&&+ \left ( - 2 + 2w_i+2w_f - 3 w_{if} + 4 w_iw_f \right ) Tr \left [ 
\overline{\cal B}_f
\overline{\Gamma}_2 P'_+(-\gamma_5) \right ] Tr \left [ \gamma_5 P'_+ 
\Gamma_1 {\cal B}_i \right ]  \nn \\
&&+ \ 3 \ Tr \left [ \overline{\cal B}_f \overline{\Gamma}_2 P'_+ {/
\hskip - 2 truemm v}_i (-\gamma_5) \right ] Tr \left [ \gamma_5 {/ 
\hskip - 2 truemm v}_f P'_+ \Gamma_1 {\cal B}_i \right ] \nn \\
&&- \ 3w_i \ Tr \left [ \overline{\cal B}_f
\overline{\Gamma}_2 P'_+ (-\gamma_5) \right ] Tr \left [ \gamma_5 {/ 
\hskip - 2 truemm v}_f P'_+ \Gamma_1 {\cal B}_i \right ]  \nn \\
&&- \ 3w_f \ Tr \left [ \overline{\cal B}_f \overline{\Gamma}_2 P'_+ 
{/ \hskip - 2 truemm v}_i (-\gamma_5) \right ] Tr \left [ \gamma_5 
P'_+ \Gamma_1 {\cal B}_i
\right ]   \Big ] \nn \\
&&+ \Big [ - (w_i-1) (w_f-1) Tr \left [ \overline{\cal B}_f 
\overline{\Gamma}_2 P'_+\gamma_{\sigma} \right ] Tr \left [
\gamma^{\sigma} P'_+ \Gamma_1 {\cal B}_i \right ] \nn \\
  &&+ \left ( 1 - 9w_{if} + 4w_i w_f + 2 w_i + 2w_f \right ) Tr \left 
[ \overline{\cal B}_f \overline{\Gamma}_2 P'_+ \right ] Tr \left [ 
P'_+
\Gamma_1 {\cal B}_i \right ] \nn \\
&&+ \ 3 (w_f-1) \ Tr \left [ \overline{\cal B}_f \overline{\Gamma}_2
P'_+ {/ \hskip - 2 truemm v}_i \right ] Tr \left [ P'_+ \Gamma_1 
{\cal B}_i \right ] \nn \\
&&+ \ 3(w_i-1) \ Tr \left [
\overline{\cal B}_f \overline{\Gamma}_2 P'_+ \right ] Tr \left [ {/ 
\hskip - 2 truemm v}_f P'_+ \Gamma_1 {\cal B}_i \right ] \Big ]  \Big 
\}
\sum_n \sigma_{3/2}^{(n)}(w_i) \sigma_{3/2}^{(n)} (w_f) \nn \\
&&+ \ \hbox{Contribution from other excited states} \nn \\
&& = - 2 \xi(w_{if}) \ Tr \left [ \overline{\cal B}_f (v_f) 
\overline{\Gamma}_2 P'_+  \Gamma_1 {\cal B}_i(v_i) \right ] \ .
\eea

\noi In the r.h.s., the function $\xi(w_{if})$ must match the 
corresponding function of $w_{if}$ that one would get summing over all
possible intermediate states. In this formula, the coefficient of 
$\sum\limits_n \xi^{(n)} (w_{i}) \xi^{(n)} (w_{f})$ is the 
contribution of the $0^-_{1/2}$ (first
bracket) and the $1^-_{1/2}$ (second bracket) states. Likewise, the 
coefficient of $\sum\limits_n \tau_{3/2}^{(n)}(w_i) \tau_{3/2}^{(n)} 
(w_f)$ is the contribution
of the $2_{3/2}^+$ (first bracket) and the $1_{3/2}^+$ states (second 
bracket). The coefficient of $\sum\limits_n \tau_{1/2}^{(n)}(w_i) 
\tau_{1/2}^{(n)}
(w_f)$ is the contribution of the $0_{1/2}^+$ (first bracket) and the 
$1_{1/2}^+$ states (second bracket). Finally, the coefficient of 
$\sum\limits_n
\sigma_{3/2}^{(n)}(w_i) \sigma_{3/2}^{(n)}(w_f)$ is the contribution 
of the $2^-_{3/2}$ (first bracket) and the $1^-_{3/2}$ (second 
bracket).\par

What we did call $L(w_{if}, w_i, w_f)$ and $R(w_{if},w_i,w_f)$ in 
Section 1 are given now explicitely by (\ref{38e}). We will now 
consider the sum rules given by
(\ref{14e}). However, since we have included only a limited number of 
intermediate states, it would be dangerous to draw conclusions from 
sum rules for $p, q, r \geq
2$, because missing intermediate states could contribute to the 
desired order. Therefore, we will limit ourselves to $(p, q, r) = (0, 
0, 0), (1, 0, 0), (0, 1, 0),
(0, 0, 1)$. The consideration of the states ${3 \over 2}^-$ will give 
us some control over higher powers of $(w-1)$. In the main text, we 
will limit ourselves to
currents that give functions $L(w_{if}, w_i, w_{f})$ and $R(w_{if}, 
w_i, w_{f})$ symmetric in $w_i$, $w_f$. We are then limited to the 
following relations
from the different derivatives and boundary conditions~:  \bea \label{39e}
&&\left . L(w_{if}, w_i, w_f)\right |_{w_{if}=1,w_i=w_f=w} = \left . 
R(w_{if}, w_i, w_f)\right |_{w_{if}=1,w_i=w_f=w}\\
&&\left . L(w_{if}, w_i, w_f)\right |_{w_i=1, w_{if}=w_f=w} = \left . 
R(w_{if}, w_i, w_f)\right |_{w_i=1, w_{if}=w_f=w}
\label{40e}
\eea

\bea
\label{41e}
&&\left . {\partial L \over \partial w_{if}} \right |_{w_{if}=1,w_i=w_f=w} = \left . {\partial R \over \partial w_{if}} \right |_{ 
w_{if}=1,w_i=w_f=w}\\
&&\nn \\
\label{42e}
&&\left . {\partial L \over \partial w_{if}} \right |_{w_i=1, 
w_{if}=w_f=w} = \left . {\partial R \over \partial w_{if}} \right 
|_{w_i=1, w_{if}=w_f=w}\\
&&\nn \\
\label{43e}
&&\left . {\partial L \over \partial w_{i}} \right |_{w_i=1, 
w_{if}=w_f=w} = \left . {\partial R \over \partial w_{i}} \right 
|_{w_i=1, w_{if}=w_f=w}\\
&&\nn \\
\label{44e}
&&\left . {\partial L \over \partial w_{i}} \right |_{w_f=1, 
w_{i}=w_{if}=w} = \left . {\partial R \over \partial w_{i}} \right 
|_{w_f=1, w_{i}=w_{if}=w}\\
&&\nn \\
&&\left . {\partial L \over \partial w_{i}} \right 
|_{w_{if}=1,w_i=w_f=w} = \left . {\partial R \over \partial w_{i}} 
\right |_{w_{if}=1,w_i=w_f=w} \ .
\label{45e}
\eea

\vskip 3 truemm

In Appendix B, we will consider a manifestly covariant version of Uraltsev 
case, where the functions $L(w_{if}, w_i, w_{f})$, $R(w_{if}, w_i, 
w_{f})$ are not symmetric in
$w_i$, $w_f$. In our conclusion we discuss the perspectives and outlook of
these non-symmetric cases.

\section{The axial current : a simple covariant derivation of Bjorken 
of Uraltsev sum rules.}
\hspace*{\parindent}
  To illustrate the method, let us now particularize to the simple case~:
\bea
\label{46e}
&{\cal B}_i = P_{i+} (- \gamma_5) &\qquad {\cal B}_f = P_{f+} 
(-\gamma_5) \nn \\
&\Gamma_1 = {/ \hskip - 2 truemm v}_i \gamma_5\qquad  &\qquad \Gamma_2 = {/ \hskip - 2 truemm v}_f \gamma_5 \ .
\eea

\noi where the currents are projected along the initial and final velocities. \par

In this symmetric situation between currents and initial and final 
states, a number of intermediate states do not contribute, and the 
calculation simplifies
considerably. One has, namely~: \beq
\label{47e}
L(0^-_{1/2}) = L (1^+_{3/2}) = L(1^+_{1/2}) = L(2^-_{3/2}) = 0
\eeq

\noi  and the SR (\ref{38e}) writes
\bea
\label{48e}
&&\left ( w_i w_f - w_{if} \right ) \sum_n \xi^{(n)}(w_i) \xi^{(n)} (w_f)\nn \\
&&+ \left [ 3 (w_i w_f -  w_{if})^2 - (w_i^2 -1) (w_f^2 - 1)\right ] 
\sum_n \tau_{3/2}^{(n)}(w_i) \tau_{3/2}^{(n)} (w_f)\nn \\
&&+ \ 4 (w_i - 1) (w_f - 1) \sum_n \tau_{1/2}^{(n)}(w_i) 
\tau_{1/2}^{(n)} (w_f) \nn \\
&&+ \ 2  (w_i - 1) (w_f - 1) (w_i w_f - w_{if}) \sum_n
\sigma_{3/2}^{(n)}(w_i) \sigma_{3/2}^{(n)} (w_f) \nn \\
  &&+ \ \hbox{Contribution from other excited states} \nn \\
&&= - (1 - w_i - w_f + w_{if})\xi(w_{if})\ .
\eea

\noi The symmetry of (\ref{48e}) in $(w_i, w_f)$ follows from the 
symmetric choice (\ref{46e}) of currents and states. \par

We assume now that the higher states contributions are, 
at most, of the same order in $(w-1)$ as the ${3\over2}^-$ states, 
that are included in the
calculation. This conjecture will be demonstrated in Section 5. The 
equations (\ref{40e}), (\ref{42e}) and (\ref{44e}) are trivial 
(giving $0 = 0$, or $\xi
(w)= \xi(w)$), while (\ref{39e}), (\ref{41e}), (\ref{43e}) and 
(\ref{45e}) give, respectively (the contribution of higher excited 
states is denoted by $+
\cdots$)~:
  \bea \label{49e}
&&(w^2 -1) \sum_n \left [\xi^{(n)}(w)\right ]^2 + 2 (w^2-1)^2 \sum_n 
\left [\tau_{3/2}^{(n)}(w)\right ]^2\nn \\
&&+ \ 4(w - 1)^2 \sum_n \left [\tau_{3/2}^{(n)}(w)\right ]^2 + 2  (w 
+ 1) (w-1)^3 \sum_n \left [\sigma_{3/2}^{(n)}(w)\right ]^2 + 
\cdots\nn \\
&&= 2(w-1)
\eea
\bea
\label{51e}
&&- \sum_n \left [ \xi^{(n)}(w)\right ]^2 - 6(w^2 - 1) \sum_n \left 
[\tau_{3/2}^{(n)}(w)\right ]^2 - 2 (w-1)^2 \sum_n \left 
[\sigma_{3/2}^{(n)}(w)\right ]^2 +
\cdots \nn \\ &&= - 1 - 2 \rho^2 (w-1)
			\eea

\beq
\label{53e}
2(w+1) \sum_n \tau_{3/2}^{(n)}(1) \tau_{3/2}^{(n)}(w) - 4 \sum_n 
\tau_{1/2}^{(n)}(1) \tau_{1/2}^{(n)}(w) + \cdots =  \xi(w)
\eeq

\bea
\label{55e}
&&w \sum_n \left [\xi^{(n)}(w) \right ]^2 + (w^2-1) \sum_n 
\xi^{(n)}(w) \xi^{(n)'}(w)\nn \\
&&+ \ 2(w^2-1) \left \{ 2w \sum_n \left [ \tau_{3/2}^{(n)}(w)\right 
]^2 + (w^2 - 1) \sum_n \tau_{3/2}^{(n)}(w) \tau_{3/2}^{(n)'}(w)\right 
\} \nn \\
&&+ \ 4(w-1) \left \{ \sum_n \left [ \tau_{1/2}^{(n)}(w)\right ]^2 + 
(w - 1) \sum_n \tau_{1/2}^{(n)}(w) \tau_{1/2}^{(n)'}(w)\right \} \nn 
\\
&&+ \ 2  (w-1)^2 \left \{ (2 w+1) \sum_n \left 
[\sigma_{3/2}^{(n)}(w)\right ]^2 + (w^2 - 1) \sum_n 
\sigma_{3/2}^{(n)}(w) \sigma_{3/2}^{(n)'}(w)\right \} + \cdots\nn
\\ &&= 1 \ .
\eea

Dividing (\ref{49e}) by $2(w-1)$ one gets
Bjorken SR \cite{1r}~:
\bea
\label{56e}
&&{w+1 \over 2} \sum_n \left [ \xi^{(n)}(w)\right ]^2 + (w-1) \left 
\{ 2 \sum_n \left [\tau_{1/2}^{(n)}(w)\right ]^2 + (w+1)^2 \sum_n 
\left
[\tau_{3/2}^{(n)}(w)\right ]^2 \right \} \nn \\ &&+ \ (w+1) (w-1)^2 
\sum_n \left [\sigma_{3/2}^{(n)}(w)\right ]^2 + \cdots = 1\ .
			\eea

\noi where the ${3\over 2}^-$ states have been included explicitly. \par

Equation (\ref{51e}) gives,  at order $(w-1)$~:
\beq
\label{57e}
1 - 2 \rho^2(w-1) + 12 \sum_n \left [ \tau_{3/2}^{(n)}(1)\right ]^2 
(w-1) = 1 + 2 \rho^2 (w-1)
			\eeq

\noi implying~:
\beq
\label{58e}
\rho^2 = 3 \sum_n \left [\tau_{3/2}^{(n)}(1)\right ]^2
			\eeq
\noi that, combined with the first order in $(w-1)$ of Bjorken SR (\ref{56e})
\beq
\label{59e}
\rho^2  = {1 \over 4} + \sum_n \left [\tau_{1/2}^{(n)}(1)\right ]^2 + 
2 \sum_n \left [\tau_{3/2}^{(n)}(1)\right ]^2
			\eeq

\noi gives Uraltsev SR \cite{5r}~:
\beq
\label{60e}
\sum_n \left [\tau_{3/2}^{(n)}(1)\right ]^2 - \sum_n \left 
[\tau_{1/2}^{(n)}(1)\right ]^2 = {1 \over 4} \ .
			\eeq

  Equation (\ref{53e}) yields also Uraltsev SR for $w = 1$. Notice the 
important point that in this equation the contribution of the IW 
functions
$\sigma_{3/2}^{(n)}(w)$ {\it vanishes identically}. \par

Finally, equation (\ref{55e}) at $O[(w-1)]$ gives again Bjorken SR 
under the form (\ref{59e}).

\section{The case of the vector current.}
\hspace*{\parindent} Let us now consider the vector
current, i.e.
\bea
\label{65e}
&{\cal B}_i = P_{i+} (-\gamma_5) &\qquad {\cal B}_f = P_{f+} (-\gamma_5) \nn \\
&\Gamma_1 = {/ \hskip - 2 truemm v}_i\qquad \qquad  &\qquad \Gamma_2 
= {/ \hskip - 2 truemm v}_f \ .
\eea

\noi In this particular case, a number of different intermediate 
states do not contribute, namely~:
\beq
\label{66e}
L(1_{1/2}^-) = L(2_{3/2}^+) = L(0_{1/2}^+) = L(1_{3/2}^-) = 0
\eeq

\noi and the SR (\ref{38e}) writes~:
\bea
\label{67e}
&&(w_i + 1) (w_f + 1) \sum_n \xi^{(n)}(w_i) \xi^{(n)}(w_f) \nn \\
&&+ \ 2 (w_i + 1) (w_f + 1) (w_i w_f-w_{if}) \sum_n \tau_{3/2}^{(n)} 
(w_i) \tau_{3/2}^{(n)} (w_f) \nn \\
&&+ \ 4 (w_i w_f - w_{if}) \sum_n \tau_{1/2}^{(n)} 
(w_i)\tau_{1/2}^{(n)} {(w_f)} \nn \\
&&+ \ [3 (w_i w_f - w_{if})^2 - (w_i^2 -1) (w_f^2 - 1)] \sum_n 
\sigma_{3/2}^{(n)} (w_i) \sigma_{3/2}^{(n)} (w_f) \nn \\
&&+ \ \hbox{Contribution from other excited states}\nn \\
&&= (w_{if} + 1 + w_f + w_i) \xi(w_{if})
\eea

\noi where the first, second, third and fourth term in the r.h.s. 
comes from the states $0_{1/2}^-$, $1_{3/2}^+$, $1_{1/2}^+$ and 
$2_{3/2}^-$, respectively. \par

Equation (\ref{40e}) is trivial ($\xi (w) = \xi (w)$), 
while equations (\ref{39e}), (\ref{41e})-(\ref{45e}) give now, 
respectively~:
\bea
\label{68e}
&&(w+1)^2 \sum_n \left [ \xi^{(n)}(w) \right ]^2 +2 (w^2 - 1) \left 
\{ (w+1)^2  \sum_n \left [ \tau_{3/2}^{(n)}(w)\right ]^2 + 2\sum_n 
\left [
\tau_{1/2}^{(n)}(w)\right ]^2 \right \}\nn \\
&&+ \ 2(w^2-1)^2 \sum_n \left [ \sigma_{3/2}^{(n)}(w)\right ]^2 + 
\cdots = 2(w+1)
\eea
\bea
\label{70e}
&&- \ 2(w+1)^2 \sum_n \left [ \tau_{3/2}^{(n)}(w) \right ]^2 - 4 
\sum_n \left [ \tau_{1/2}^{(n)}(w)\right ]^2 \nn \\
&&- \ 6(w^2-1) \sum_n \left [ \sigma_{3/2}^{(n)}(w)\right ]^2 + 
\cdots = 1 - 2(w+1) \rho^2
\eea

\bea
\label{71e}
&&-\ 4(w+1) \sum_n \tau_{3/2}^{(n)}(1) \tau_{3/2}^{(n)}(w) - 4 \sum_n 
\tau_{1/2}^{(n)}(1) \tau_{1/2}^{(n)}(w) + \cdots \nn \\
&&= \xi(w) + 2(w+1) \xi '(w)
\eea

\bea
\label{72e}
&&w \xi (w) + 2(w+1) \sum_n \xi^{(n)'}(1) \xi^{(n)}(w) \nn \\
&&+\ 4w(w+1) \sum_n \tau_{3/2}^{(n)}(1) \tau_{3/2}^{(n)}(w) + 4 w 
\sum_n \tau_{1/2}^{(n)}(1) \tau_{1/2}^{(n)}(w) \nn \\
&&- \ 2(w^2-1) \sum_n \sigma_{3/2}^{(n)}(1) \sigma_{3/2}^{(n)}(w) + \cdots = 0
\eea

\bea
\label{73e}
&&\xi(w) + 2(w+1) \xi '(w) \nn \\
&&+\ 4(w+1)\sum_n \tau_{3/2}^{(n)}(1) \tau_{3/2}^{(n)}(w) + 4  \sum_n 
\tau_{1/2}^{(n)}(1) \tau_{1/2}^{(n)}(w) + \cdots = 0
\eea

\bea
\label{74e}
&&(w+1) \sum_n \left [ \xi^{(n)}(w) \right ]^2 + (w + 1)^2 \sum_n 
\xi^{(n)}(w) \xi^{(n)'}(w) \nn \\
&&+ \ 2 (w+1)^2 \left \{ (2w-1) \sum_n \left [ 
\tau_{3/2}^{(n)}(w)\right ]^2 + (w^2-1) \sum_n
\tau_{3/2}^{(n)}(w) \tau_{3/2}^{(n)'}(w) \right \}\nn \\
&&+\  4 \left \{ w\sum_n \left [ \tau_{1/2}^{(n)}(w)\right ]^2 + 
(w^2- 1) \sum_n \tau_{1/2}^{(n)}(w) \tau_{1/2}^{(n)'}(w) \right \}\nn 
\\
&&+\  (w^2 - 1) \left \{ 4w \sum_n \left [ 
\sigma_{3/2}^{(n)}(w)\right ]^2 + 2(w^2 - 1) \sum_n 
\sigma_{3/2}^{(n)}(w) \sigma_{3/2}^{(n)'}(w)\right \} + \cdots = 1\
.\nn \\ \eea

Notice an important point, namely that in equation (\ref{71e}), 
identical to equation (\ref{73e}), the
contribution of the IW functions $\sigma_{3/2}^{(n)}(w)$ {\it 
vanishes identically}. \par

Dividing (\ref{68e}) by $2 (w+1)$ one gets Bjorken SR for all $w$ 
(\ref{56e}). Equations (\ref{70e})-(\ref{74e}) imply, for $w = 1$, 
Bjorken SR (\ref{59e}) for the
elastic slope $\rho^2$.

\section{A new class of sum rules and the contribution of higher 
excited states.}
\hspace*{\parindent}
Among the SR that we have obtained in Sections 3 and 4, there is a 
new class that involves the IW functions $\xi^{(n)}(w)$, 
$\tau_{3/2}^{(n)}(w)$,
$\tau_{1/2}^{(n)}(w)$, ... for any $w$ and at zero recoil $w = 1$. 
The relation that we got from the axial currents is~:
\beq
\label{75e}
2(w+1) \sum_n \tau_{3/2}^{(n)}(1) \tau_{3/2}^{(n)}(w) - 4 \sum_n 
\tau_{1/2}^{(n)}(1) \tau_{1/2}^{(n)}(w) + \cdots = \xi (w)
\eeq

\noi while we obtained, from the vector current~:
\beq
\label{76e}
- 4 (w+1) \sum_n \tau_{3/2}^{(n)}(1) \tau_{3/2}^{(n)}(w) - 4 \sum_n 
\tau_{1/2}^{(n)}(1) \tau_{1/2}^{(n)}(w) + \cdots
= \xi (w) + 2(w+1) \xi '(w) \ .
\eeq
\bea
\label{77e}
&&w \xi (w) + 2(w+1) \sum_n \xi^{(n)'} (1) \xi^{(n)}(w) \nn \\
&&+ \ 4 w(w+1) \sum_n \tau_{3/2}^{(n)}(1) \tau_{3/2}^{(n)}(w) + 4w 
\sum_n \tau_{1/2}^{(n)}(1) \tau_{1/2}^{(n)}(w) \nn \\
&&-\ 2 (w^2-1) \sum_n \sigma_{3/2}^{(n)}(1) \sigma_{3/2}^{(n)}(w) + 
\cdots = 0 \ .
\eea

\noi The first equation (\ref{75e}) is a generalization of Uraltsev 
SR for $w \not=  1$, that reduces to (\ref{60e}) for $w = 1$, while 
the other two (\ref{76e}) and
(\ref{77e}) give, taking $w= 1$, Bjorken SR (\ref{59e}) for the slope 
$\rho^2$. \par

Let us concentrate on equations (\ref{75e}) and (\ref{76e}). An 
important feature of these relations is that the contribution from 
the ${3 \over 2}^-$ states
vanishes identically. This is not the case however for relation 
(\ref{77e}). \par

We will now give a proof that no other higher intermediate states 
contribute to the sum rules (\ref{75e}) and (\ref{76e}). \par

Following the work of A. Falk \cite{10r}, we write first the $4 
\times 4$ matrices of the whole tower of $j^P$ states, generalizing 
the notation we have given above
(\ref{17e})-(\ref{21e}), where $k = j - {1 \over 2}$, $J$ is the spin 
of the state, and $\ell$ is the orbital angular momentum~:\\

\noi $j = \ell + {1 \over 2}$, $J = j + {1 \over 2}$~:
\beq
\label{77eq}
{\cal M}^{\mu_1 \cdots \mu_k}(v)= P_+ \ \varepsilon_v^{\mu_1 \cdots 
\mu_{k+1}}\ \gamma_{\mu_{k+1}}
\eeq

\noi $j = \ell + {1 \over 2}$, $J = j - {1 \over 2}$~:

$${\cal M}^{\mu_1 \cdots \mu_k}(v) = - \ \sqrt{{2k+1 \over k+1}} \ 
P_+ \ \gamma_5\  \varepsilon_v^{\nu_1 \cdots \nu_{k}}$$
\beq
\label{78eq}
\left [ g_{\nu_1}^{\mu_1} \cdots g_{\nu_k}^{\mu_k} - {1 \over 2k+1} \ 
\gamma_{\nu_1} \left ( \gamma^{\mu_1} - v^{\mu_1} \right ) 
g_{\nu_2}^{\mu_2} \cdots
g_{\nu_k}^{\mu_k} - \cdots - {1 \over 2k+1}\ g_{\nu_1}^{\mu_1} \cdots 
g_{\nu_{k-1}}^{\mu_{k-1}} \ \gamma_{\nu_k} \left ( \gamma^{\mu_{k}} -
v^{\mu_k} \right ) \right ]  \eeq

\noi $j = \ell - {1 \over 2}$, $J = j + {1 \over 2}$~:
\beq
\label{79eq}
{\cal M}^{\mu_1 \cdots \mu_k}(v)= P_+ \ \varepsilon_v^{\mu_1 \cdots 
\mu_{k+1}}\ \gamma_5\ \gamma_{\mu_{k+1}}
\eeq

\noi $j = \ell - {1 \over 2}$, $J = j - {1 \over 2}$~:

$${\cal M}^{\mu_1 \cdots \mu_k}(v) =  \sqrt{{2k+1 \over k+1}} \ P_+ \ 
\varepsilon^{\nu_1 \cdots \nu_{k}}$$
\beq
\label{80eq}
\left [ g_{\nu_1}^{\mu_1} \cdots g_{\nu_k}^{\mu_k} - {1 \over 2k+1} \ 
\gamma_{\nu_1} \left ( \gamma^{\mu_1} + v^{\mu_1} \right ) 
g_{\nu_2}^{\mu_2} \cdots
g_{\nu_k}^{\mu_k} - \cdots - {1 \over 2k+1}\ g_{\nu_1}^{\mu_1} \cdots 
g_{\nu_{k-1}}^{\mu_{k-1}} \ \gamma_{\nu_k} \left ( \gamma^{\mu_{k}} + 
v^{\mu_k} \right )
\right ] \eeq

For a transition of the type ${\cal B}^{\mu_1 \cdots \mu_k}(v) \to 
{\cal D}^{\nu_1 \cdots \nu_{k'}}(v')$, the preceding expressions have 
to be contracted with the
tensor containing all possible independent IW functions $(k' \geq k)$~:
\bea
\label{81eq}
&&\xi_{\nu_1 \cdots \nu_{k'}, \mu_1 \cdots \mu_k} = (-1)^k 
(v-v')_{\nu_{k+1}} \cdots (v-v')_{\nu_{k'}} \Big [ \xi_0^{(k',k)}(w) 
\ g_{\nu_1\mu_1} \cdots
g_{\nu_k\mu_k} \nn \\
&&+ \ \xi_1^{(k',k)}(w) (v-v')_{\nu_1} (v-v')_{\mu_1} \ 
g_{\nu_2\mu_2} \cdots g_{\nu_k\mu_k} + \cdots \nn \\
&&+ \ \xi_k^{(k',k)}(w) (v-v')_{\nu_1} (v-v')_{\mu_1} \cdots 
(v-v')_{\nu_k} (v-v')_{\mu_k}\Big ]\ . \eea

\noi However, we are here interested in the transitions between the 
ground state and the excited states ${1 \over 2}^- \to j^P$, i.e. $k 
= 0$, and the tensor
(\ref{81eq}) becomes, in this case,
\beq
\label{82eq}
\xi_{\mu_1 \cdots \mu_{k}} = (v-v')_{\mu_{1}} \cdots (v-v')_{\mu_{k}} 
\ \xi_0^{(k)}(w)\ .
\eeq

\noi Then, the matrix elements will write, for the different cases~:

\bea
\label{83eq}
&&<D^{(n)} \left ( \scriptstyle{ j = \ell +
{1\over 2} , J = j +  {1 \over 2}} \right ) (v') 
|\overline{h}_{v'}^{(c)} \Gamma h_v^{(b)} | B^{(*)}(v) > \nn \\
&&=  \tau_{\ell + {1 \over 2}}^{(\ell )(n)} (w) \ v_{\mu_1} \cdots 
v_{\mu_k} \ \varepsilon '^{*\mu_1 \cdots \mu_{k+1}} \ Tr \left [ 
\gamma_{\mu_{k+1}} \
P'_+ \ \Gamma {\cal B}(v) \right ] \eea

$$<D^{(n)} \left ( \scriptstyle{j = \ell +
{1\over 2} , J = j -  {1 \over 2}} \right ) (v') 
|\overline{h}_{v'}^{(c)} \Gamma h_v^{(b)} | B^{(*)}(v) > \ =$$
$$\sqrt{{2k+1 \over k+1}}\ \tau_{\ell + {1 \over 2}}^{(\ell ) (n)} 
(w) \varepsilon '^{*\nu_1 \cdots \nu_k} \ Tr
\Big \{ \Big [  v_{\nu_1} \cdots v_{\nu_k} - {1 \over 2k+1} ({/\hskip 
- 2 truemm v} - w) \gamma_{\nu_{1}}\  v_{\nu_2} \cdots v_{\nu_k} - 
\cdots$$
\beq
\label{84eq}
- {1 \over 2k+1} \ v_{\nu_1} \cdots v_{\nu_{k-1}} ({/\hskip - 2 
truemm v} - w) \gamma_{\nu_{k}} \Big ] \gamma_5 \ P'_+ \Gamma {\cal 
B}(v) \Big \}\eeq

  \bea
\label{85eq}
&&<D^{(n)} \left ( \scriptstyle{j = \ell - {1\over 2} , J = j +  {1 
\over 2}} \right ) (v') |\overline{h}_{v'}^{(c)} \Gamma h_v^{(b)} | 
B^{(*)}(v) >
\nn \\ &&= \tau_{\ell - {1 \over 2}}^{(\ell ) (n)}(w) \ v_{\mu_1} 
\cdots v_{\mu_k} \ \varepsilon '^{*\mu_1 \cdots \mu_{k+1}} \ Tr \left 
[
\gamma_{\mu_{k+1}}(-\gamma_5) \ P'_+ \ \Gamma {\cal B}(v) \right ] \eea

$$<D^{(n)} \left ( \scriptstyle{j = \ell - {1\over 2} , J = j -  {1 
\over 2}} \right ) (v') |\overline{h}_{v'}^{(c)} \Gamma h_v^{(b)} | 
B^{(*)}(v)
> \ =$$ $$\sqrt{{2k+1 \over k+1}}\ \tau_{\ell - {1 \over 2}}^{(\ell )(n)} (w) \varepsilon '^{*\nu_1 \cdots \nu_k} \ Tr
\Big \{ \Big [  v_{\nu_1} \cdots v_{\nu_k} - {1 \over 2k+1} ({/\hskip 
- 2 truemm v} + w) \gamma_{\nu_{1}}\  v_{\nu_2} \cdots v_{\nu_k} - 
\cdots$$
\beq
\label{86eq}
- {1 \over 2k+1} \ v_{\nu_1} \cdots v_{\nu_{k-1}} ({/\hskip - 2 
truemm v} + w) \gamma_{\nu_{k}} \Big ] P'_+ \Gamma  {\cal B}(v)\Big 
\}\ .\eeq
\vskip 3 truemm

\noi In all these relations we have made use of the orthogonality condition
\beq
\label{87eq}
v'_{\nu_i} \ \varepsilon '^{*\nu_1 \cdots \nu_k} = 0 \qquad (i = 1, 
\cdots , k) \ .
\eeq

\noi ${\cal B}(v)$ denotes the $4 \times 4$ matrix of the ground 
state, $B$ or $B^*$ (\ref{17e}). The functions $\tau_{j= \ell \pm {1 
\over 2}}^{(\ell ) (n)}(w)$ are
the generalizations to arbitrary $j$ of the IW functions introduced 
above, namely

\beq
\label{88}
\tau_{1/2}^{(0)}(w) \equiv \xi (w) \ , \ \tau_{3/2}^{(1)}(w) \equiv 
\sqrt{3}\tau_{3/2}(w) \ ,  \ \tau_{1/2}^{(1)}(w) \equiv 2 
\tau_{1/2}(w) \ ,
\ \tau_{3/2}^{(2)}(w) \equiv \sqrt{3} \sigma_{3/2}(w) \ ,  \eeq

\noi with an implicit radial quantum number $n$. Therefore, 
$\tau_{3/2}^{(1)}(w)$ and $\tau_{1/2}^{(1)}(w)$ are respectively 
identical to the functions $\tau
(w)$ and $\zeta (w)$ defined by Leibovich {\it et al.} \cite{3r}. The
superindex $\ell$ in $\tau_{\ell \pm 1/2}^{(\ell)(n)}(w)$ is necessary
as indicates the parity, since for a given $j = \ell \pm {1 \over 2}
\geq {1 \over 2}$, there are two possible values for $\ell = j \pm {1
\over 2}$, and therefore two possible parities $P = (-1)^{\ell +
1}$.\par

Considering now the $B$ meson, as in the preceding Sections,
\beq
\label{88eq}
{\cal B}(v) = P_+ (-\gamma_5)
\eeq

\noi we compute the different matrix elements. Remembering that $k = 
j - {1 \over 2}$ one obtains the following results.\\

\noi Vector current~:
\bea
\label{89eq}
&&<D^{(n)} \left ( \scriptstyle{j = \ell + {1\over 2} , J = j +  {1 
\over 2}} \right ) (v') |\overline{h}_{v'}^{(c)} {/ \hskip - 2
truemm v} h_v^{(b)} | B(v) > \nn \\
&&= \ <D^{(n)} \left ( \scriptstyle{j = \ell - {1\over 2} , J = j - 
{1 \over 2}} \right ) (v') |\overline{h}_{v'}^{(c)} {/ \hskip - 2 
truemm v}
h_v^{(b)} | B(v) > \ = 0
\eea

\bea
\label{90eq}
&&<D^{(n)} \left ( \scriptstyle{j = \ell + {1\over 2} , J = j -  {1 
\over 2}} \right ) (v') |\overline{h}_{v'}^{(c)} {/\hskip - 2 truemm 
v}
h_v^{(b)} | B(v) > \nn \\
&&= - \ \sqrt{{\ell +1 \over 2\ell +1}}\ (w+1) \ \tau_{\ell + {1 
\over 2}}^{(\ell ) (n)} (w) v_{\mu_1} \cdots v_{\mu_{\ell}}\ 
\varepsilon '^{*\mu_1 \cdots
\mu_{\ell}}  \eea

\bea
\label{91eq}
&&<D^{(n)} \left ( \scriptstyle{j = \ell - {1\over 2} , J = j +  {1 
\over 2}} \right ) (v') |\overline{h}_{v'}^{(c)} {/\hskip - 2 truemm 
v}
h_v^{(b)} | B(v) > \nn \\
&&= - \ \tau_{\ell - {1 \over 2}}^{(\ell ) (n)} (w) \ v_{\mu_1} 
\cdots v_{\mu_{\ell} }\ \varepsilon '^{*\mu_1 \cdots \mu_{\ell}}
\eea

\noi Axial current~:

\bea
\label{92eq}
&&<D^{(n)} \left ( \scriptstyle{j = \ell + {1\over 2} , J = j -  {1 
\over 2}} \right ) (v') |\overline{h}_{v'}^{(c)} {/\hskip - 2 truemm 
v} \gamma_5
h_v^{(b)} | B(v) > \nn \\
&&= \ <D^{(n)} \left ( \scriptstyle{j = \ell - {1\over 2} , J = j + 
{1 \over 2}} \right ) (v') |\overline{h}_{v'}^{(c)} {/\hskip - 2 
truemm v}
\gamma_5 h_v^{(b)} | B^{(*)}(v) >\ = 0
\eea

\bea
\label{93eq}
&&<D^{(n)} \left ( \scriptstyle{j = \ell + {1\over 2} , J = j +  {1 
\over 2}} \right ) (v') |\overline{h}_{v'}^{(c)} {/\hskip - 2 truemm 
v} \gamma_5
h_v^{(b)} | B(v) > \nn \\
&&= -  \ \tau_{\ell + {1 \over 2}}^{(\ell ) (n)} (w) \ v_{\mu_1} 
\cdots v_{\mu_{\ell+1} }\ \varepsilon '^{*\mu_1 \cdots \mu_{\ell+1}}
\eea

\bea
\label{94eq}
&&<D^{(n)} \left ( \scriptstyle{j = \ell - {1\over 2} , J = j -  {1 
\over 2}} \right ) (v') |\overline{h}_{v'}^{(c)} {/\hskip - 2 truemm 
v} \gamma_5
h_v^{(b)} | B(v) > \nn \\
&&= - \ \sqrt{{\ell \over 2\ell - 1}}\ (w-1) \ \tau_{\ell - {1 \over 
2}}^{(\ell ) (n)} (w) \ v_{\mu_1} \cdots v_{\mu_{\ell - 1}}\ 
\varepsilon '^{*\mu_1 \cdots
\mu_{\ell - 1}} \ . \eea

\noi We can now write down the contributions to the l.h.s. of the SR. 
We proceed as in Sections 3 and 4 adopting the symmetric cases 
(\ref{46e}) and
(\ref{65e}). In an obvious notation, one finds the following results. \\

\noi Vector current~:
\beq
\label{95eq}
L \left ( \scriptstyle{j = \ell + {1 \over 2}, J = j + {1 \over 2}} 
\right ) = L \left ( \scriptstyle{j = \ell - {1 \over 2}, J = j - {1 
\over 2}} \right ) = 0
\eeq

\beq
\label{96eq}
L \left ( \scriptstyle{j = \ell + {1 \over 2}, J = j - {1 \over 2}} 
\right ) = {\ell +1\over 2 \ell + 1} (w_i+1)(w_f+1) 
S_{\ell}(w_i,w_f,w_{if}) \sum_n \tau_{\ell +
{1 \over 2}}^{(\ell )(n)}(w_i) \ \tau_{\ell + {1 \over 2}}^{(\ell 
)(n)}(w_f)  \eeq

\beq
\label{97eq}
L \left ( \scriptstyle{j = \ell - {1 \over 2}, J = j + {1 \over 2}} 
\right ) = S_{\ell}(w_i,w_f,w_{if}) \sum_n \tau_{\ell - {1 \over 
2}}^{(\ell
) (n)}(w_i) \ \tau_{\ell - {1 \over 2}}^{(\ell ) (n)}(w_f)  \eeq
\vskip 3 truemm

\noi Axial current~:
\beq
\label{98eq}
L \left ( \scriptstyle{j = \ell + {1 \over 2}, J = j - {1 \over 2}} 
\right ) = L \left ( \scriptstyle{j = \ell - {1 \over 2}, J = j + {1 
\over 2}} \right ) = 0
\eeq

\beq
\label{99eq}
L \left ( \scriptstyle{j = \ell + {1 \over 2}, J = j + {1 \over 2}} 
\right ) = S_{\ell +1}(w_i,w_f,w_{if}) \sum_n \tau_{\ell + {1 \over
2}}^{(\ell ) (n)}(w_i) \ \tau_{\ell + {1 \over 2}}^{(\ell )(n)}(w_f)  \eeq

\beq
\label{100eq}
L \left ( \scriptstyle{j = \ell - {1 \over 2}, J = j - {1 \over 2}} 
\right ) = {\ell \over 2 \ell - 1} (w_i-1)(w_f-1) S_{\ell - 
1}(w_i,w_f,w_{if}) \sum_n \tau_{\ell
- {1 \over 2}}^{(\ell )(n)}(w_i) \ \tau_{\ell - {1 \over 2}}^{(\ell ) 
(n)}(w_f) \ . \eeq
\vskip 3 truemm

\noi In all these relations, the quantity $S_n$ defined by
$$S_n = v_{i\nu_1} \cdots  v_{i\nu_n}\  v_{f\mu_1} \cdots v_{f\mu_n} \ T^{\nu_1 \cdots \nu_n, \mu_1 \cdots \mu_n}$$
\beq
\label{101eq} 
T^{\nu_1 \cdots \nu_n, \mu_1 \cdots \mu_n} = \sum_{\lambda} \varepsilon '^{(\lambda )*\nu_1 \cdots 
\nu_{n}}\ \varepsilon '^{(\lambda )\mu_1 \cdots \mu_{n}}
\eeq

\noi depends only on the four-velocity $v'$ and $\varepsilon '^{(\lambda
)\mu_1 \cdots \mu_n}$ is a symmetric tensor with vanishing contractions
and transverse to $v'$ (see Appendix A).\par

It can be shown, as demonstrated below in Appendix A, that the scalar quantity
\beq
\label{102eq}
S_n = v_{i\nu_1} \cdots  v_{i\nu_n}\  v_{f\mu_1} \cdots v_{f\mu_n} \ 
T^{\nu_1 \cdots \nu_n, \mu_1 \cdots \mu_n}
\eeq

\noi can be computed and is given by the following expression
\beq
\label{103eq}
S_n = \sum_{0 \leq k \leq {n \over 2}} C_{n,k} \ (w_i^2 - 1)^k \ 
(w_f^2 - 1)^k \ (w_iw_f - w_{if})^{n-2k} \ .
\eeq

\noi where
\beq
\label{neweq}
C_{n,k} = (-1)^k \  {(n!)^2 \over (2n)!}\  {(2n-2k)! \over k! 
(n-k)!(n-2k)!} \ .
\eeq

\vskip 3 truemm
We did find that in the SR (\ref{53e}) and (\ref{71e}) or (\ref{73e})
the contribution of the states ${3 \over 2}^-$ is identically zero.
Using now the preceding general formulas, let us prove that not only the
contribution of the states $j^P = {3 \over 2}^-$ vanishes, but that this
is also the case for any $j \geq {5 \over 2}$. This result will imply
that the SR (\ref{75e}) and (\ref{76e}) are exact equations, i.e. we can
drop out the $+ \cdots$\par

Let us begin with equation (\ref{53e}), that was found with the axial 
current by differentiating with respect to $w_i$, and taking the limit 
$w_i = 1$, $w_f = w_{if} = w$.
Notice first that
\beq
\label{104eq}
\left . S_n(w_i, w_f, w_{if}) \right |_{w_i=1,w_f = w_{if}= w} = 0 
\qquad (n \geq 1)
\eeq

\noi because of the orthogonality condition (\ref{87eq}). \par

 From eqs. (\ref{103eq}) and (\ref{neweq}) we need to prove that
\beq
\label{105eq}
\left . {\partial \over \partial w_i} \ S_{n+1}(w_i, w_f, w_{if}) 
\right |_{w_i=1,w_f = w_{if}= w} = 0 \qquad (n \geq 2)
\eeq

\beq
\label{106eq}
\left . {\partial \over \partial w_i} \ (w_i-1) (w_f-1) S_n(w_i, w_f, 
w_{if})\right |_{w_i=1,w_f = w_{if}= w} = 0 \qquad (n \geq 1) \ .
\eeq

\noi The second condition (\ref{106eq}) is obviously held because of 
the factor $(w_i-1)$ and the orthogonality condition (\ref{87eq}). 
\par

The first condition (\ref{105eq}) holds also, as can be seen from the 
explicit formula (\ref{103eq})~:

$${\partial \over \partial w_i} \ S_{n+1}(w_i, w_f, w_{if}) = 
\sum_{0\leq k \leq {n+1 \over 2}} C_{n+1,k} (w_f^2 - 1)^k \Big [ 2k 
w_i (w_i^2 - 1)^{k-1} (w_iw_f
- w_{if})^{n+1-2k}$$
\beq
\label{107eq}
+ (n+1 -2k ) w_f (w_i^2 - 1)^k (w_i w_f - w_{if})^{n-2k} \Big ]
\eeq

\noi that vanishes for $w_i = 1$, $w_f = w_{if} = w$ when $n \geq 2$. 
Notice that this expression does not vanish for $n = 1$, that 
corresponds to the contribution
of the ${3 \over 2}^+$ states to the SR. \par

Let us now consider equation (\ref{71e}), that was found with the 
vector current by derivation with respect to $w_{if}$, and taking the 
limit $w_i = 1$, $w_f = w_{if}
= w$, or (\ref{73e}) by derivation with respect to $w_i$, and taking 
the limit $w_f = 1$, $w_i = w_{if} =w$. From eq. (\ref{103eq}) we 
need to prove
\beq
\label{108eq}
\left . {\partial \over \partial w_{if}} \ (w_i+1) (w_f+1) S_n(w_i, 
w_f, w_{if})\right |_{w_i=1,w_f = w_{if}= w} = 0 \qquad (n \geq 2)
\eeq

\beq
\label{109eq}
\left . {\partial \over \partial w_{if}} \ S_{n+1}(w_i, w_f, w_{if}) 
\right |_{w_i=1,w_f = w_{if}= w} = 0 \qquad (n \geq 1) \ .
\eeq

\vskip 3 truemm
\noi This is indeed the case, since

\beq
\label{111}
{\partial \over \partial w_{if}} \ S_{n}(w_i, w_f, w_{if}) = - 
\sum_{0\leq k \leq {n \over 2}} C_{n,k} (w_i^2 - 1)^k (w_f^2 - 1)^k \ 
(n-2k)
(w_iw_f - w_{if})^{n-2k-1}
\eeq

\noi vanishes for $w_i = 1$, $w_f = w_{if} = w$ when $n \geq 2$. 
Notice that this quantity does not vanish for $n = 1$, corresponding 
again to the contribution of
the ${3 \over 2}^+$ states to the SR. The proof can be done also by 
derivation with respect to $w_i$, and taking the limit $w_f = 1$, 
$w_i = w_{if} = w$. \par

In conclusion, we have demonstrated that in the SR (\ref{75e}) and 
(\ref{76e}) there are no contributions from higher excitations.

We must make an important distinction between the different SR that we
have obtained. On the one hand, there are the SR to which contribute the whole series of
$j^P$ excitations. On the other hand, we have obtained two special SR
(\ref{75e}) and (\ref{76e}) where only a limited number of intermediate states
contribute.\par

One can understand the truncation of the series in this latter case
because the SR correspond to the boundary condition $w_i = 1$, $w_{if} =
w_f = w$. Therefore, the matrix element $<D^{(n}(v')|\bar{h}_{v'}^{(c)}
\Gamma_1 h_{v_i}^{(b)}|B(v_i)>$ is computed at zero recoil, hence the
finite number of terms in the expansion. As we have seen, SR (\ref{75e})
obtained with the axial currents implies at zero recoil Uraltsev SR
(\ref{60e}). The corresponding SR from the vector current (\ref{76e}) is
the Bjorken-type counterpart and indeed implies, at zero recoil, Bjorken
SR for the slope (\ref{59e}). \par

On the other hand, since all the SR that we have obtained are exact
relations, we can derive them relatively to $w$ and, for a given
derivative, taking the zero recoil limit $w = 1$, the series will be
truncated due to the higher powers of the recoil $(w-1)^{\ell}$ as
$\ell$ increases. Therefore, one can expect to obtain information on
higher derivatives of the elastic IW function $\xi (w)$, as we have done
in ref. \cite{6rbis}.

\section{A bound on the curvature from the new sum rules.}
\hspace*{\parindent}
In the preceding Section we have demonstrated that the SR (\ref{75e}) 
and (\ref{76e}) do not have contributions from higher excited states, 
i.e. we can omit $+
\cdots$ in these equations. This is an important result that means 
that {\it these SR, involving only $\xi (w)$, 
$\tau_{1/2}^{(n)}(w)$ and
$\tau_{3/2}^{(n)}(w)$, are exact relations for all $w$}, namely~:

\beq
\label{75ebis}
2(w+1) \sum_n \tau_{3/2}^{(n)}(1) \tau_{3/2}^{(n)}(w) - 4 \sum_n 
\tau_{1/2}^{(n)}(1) \tau_{1/2}^{(n)}(w) = \xi (w)
\eeq

\beq
\label{76ebis}
- 4 (w+1) \sum_n \tau_{3/2}^{(n)}(1) \tau_{3/2}^{(n)}(w) - 4 \sum_n 
\tau_{1/2}^{(n)}(1) \tau_{1/2}^{(n)}(w)
= \xi (w) + 2(w+1) \xi '(w) \ .
\eeq

\vskip 3 truemm
\noi These relations are the main result of this paper. \par

Therefore, we can still differentiate relation (\ref{76ebis}) relatively to $w$~: \bea
\label{78e}
&&-\ 4 \sum_n \tau_{3/2}^{(n)}(1) \tau_{3/2}^{(n)}(w) - 4(w+1) \sum_n 
\tau_{3/2}^{(n)}(1) \tau_{3/2}^{(n)'}(w) \nn \\
&&- \ 4 \sum_n \tau_{1/2}^{(n)}(1) \tau_{1/2}^{(n)'}(w) = \xi '(w) + 
2 \xi ' (w) + 2(w+1) \xi ''(w)
\eea

\noi expanding the elastic IW function $\xi (w)$ in powers of $(w-1)$,
\beq
\label{79e}
\xi (w) = 1 - \rho^2 (w-1) + {\sigma^2 \over 2} (w-1)^2 + \cdots
\eeq

\noi one obtains, at zero recoil
\beq
\label{80e}
-4 \sum_n \left [ \tau_{3/2}^{(n)}(1)\right ]^2 - 8 \sum_n 
\tau_{3/2}^{(n)}(1) \tau_{3/2}^{(n)'}(1)
- 4 \sum_n \tau_{1/2}^{(n)}(1) \tau_{1/2}^{(n)'}(1) = - 3 \rho^2 + 4 \sigma^2
\eeq

\noi and from relation (\ref{58e}) for $\rho^2$ one obtains
\beq
\label{81e}
\sigma^2 = {5 \over 12} \rho^2  - 2 \sum_n \tau_{3/2}^{(n)}(1) 
\tau_{3/2}^{(n)'}(1) -  \sum_n \tau_{1/2}^{(n)}(1) 
\tau_{1/2}^{(n)'}(1) \ .
\eeq

\noi We can also differentiate relation (\ref{75ebis}) relatively to $w$ and 
take the zero recoil limit~: \beq
\label{82e}
2 \sum_n \left [ \tau_{3/2}^{(n)}(1)\right ]^2  + 4 \sum_n 
\tau_{3/2}^{(n)}(1) \tau_{3/2}^{(n)'}(1) - 4 \sum_n 
\tau_{1/2}^{(n)}(1) \tau_{1/2}^{(n)'}(1) = - \rho^2
\eeq

\noi and from (\ref{58e}) we obtain~:
\beq
\rho^2 = - {12 \over 5} \left [  \sum_n \tau_{3/2}^{(n)}(1) 
\tau_{3/2}^{(n)'}(1) - \sum_n \tau_{1/2}^{(n)}(1) 
\tau_{1/2}^{(n)'}(1) \right ]\ .
\label{83e}
\eeq

  Combining relations (\ref{81e}) and (\ref{83e}) one obtains~:
\beq
\label{84e}
\sigma^2  = - 3 \sum_n \tau_{3/2}^{(n)}(1) \tau_{3/2}^{(n)'}(1) \ .
\eeq

Equations (\ref{83e}) and (\ref{84e}) are important results of the 
present paper. We must insist on the fact that they are exact 
relations, as no other higher
excited states contribute to the sums in the r.h.s. \par

Let us now 
discuss these formulas. If we make the plausible assumption~:
\beq
\label{88e}
- \sum_n \tau_{1/2}^{(n)}(1)\ \tau_{1/2}^{(n)'}(1) > 0 \ .
\eeq

\noi the following inequality follows from (\ref{83e}) and (\ref{84e})~:
\beq
\label{89e}
  \sigma^2 \geq {5 \over 4} \rho^2 \ .
\eeq

\noi This inequality, from the lower bound $\rho^2 \geq {3 \over 4}$ 
(\cite{5r}, \cite{7newr}), implies the absolute bound

\beq
\label{116}
\sigma^2 \geq {15 \over 16}
\eeq

The assumption (\ref{88e}) would be valid if the $n = 0$ state 
dominates the sum, and if $\tau_{1/2}^{(0)'}(1) < 0$. This latter 
condition is very natural, since
it concerns transitions between states of radial quantum number $n = 
0$, and therefore with no nodes in the wave function.

\section{Phenomenological remarks.}
\hspace*{\parindent}
In the Bakamjian-Thomas type of relativistic quark models, we have 
shown that Bjorken and Uraltsev SR are satisfied \cite{14r}. 
Moreover, these SR are approximately
saturated by the $n = 0$ states. We can add that the slopes of all 
three IW functions $\xi(w)$, $\tau_{3/2}^{(0)}(w)$ and 
$\tau_{1/2}^{(0)}(w)$ are negative
\cite{15r}. Namely, a good approximate parametrization of these 
functions is given by
\beq
\label{85e}
\xi (w) = \left ( {2 \over w+1} \right )^{2\rho^2} \quad , \quad 
\tau_{3/2}^{(0)}(w) = \left ( {2 \over w+1} \right 
)^{2\sigma_{3/2}^2} \quad , \quad
\tau_{1/2}^{(0)}(w) = \left ( {2 \over w+1} \right )^{2\sigma_{1/2}^2} \ .\eeq

\noi In the spectroscopic model of Godfrey and Isgur, one finds the results
\bea
\label{86e}
&\xi (1) = 1 &\qquad \qquad \rho^2 = 1.02 \nn \\
&\tau_{1/2}^{(0)}(1) = 0.22  &\qquad \qquad \sigma_{1/2}^2 = 0.83 \nn \\
&\tau_{3/2}^{(0)}(1) = 0.54  &\qquad \qquad \sigma_{3/2}^2 = 1.50 \ .
\eea

\noi We observe that approximating the r.h.s. of (\ref{83e}) with the 
$n = 0$ states this SR writes~:
\beq
\label{87e}
\rho^2 = 1.02 = 0.95 + \ \hbox{Contributions from $n \not= 0$ excitations} \ .
\eeq

  \noi The inequality (\ref{89e}) is satisfied also in the BT scheme, 
since, for example in the GI spectroscopic model~:
\beq
\label{90e}
\rho^2 \cong 1 \qquad \qquad \qquad \sigma^2 \cong {3 \over 2}
\eeq

\noi and the inequality (\ref{89e}) writes $3/2 > 5/4$. Therefore the
conjecture (\ref{88e}) is satisfied in the model. Notice that BT quark
models satisfy Bjorken and Uraltsev SR \cite{14r}. \par

Although it remains to be proved, it is highly plausible that these
models satisfy the whole set of SR of QCD in the heavy quark limit, and
therefore the new class (\ref{83e}) and (\ref{84e}).\par

Finally, from relation (\ref{84e}) we get the following result for 
the curvature, compared with the direct result (\ref{90e}) from the 
elastic IW function
$\xi (w)$ (\ref{85e}),
\beq
\label{91e}
\sigma^2 \cong 1.5 = 1.31 + \ \hbox{Contributions from $n \not= 0$ 
excitations} \ .
\eeq

We can conclude that there is an excellent qualitative agreement 
between the slope and the curvature of the elastic IW function as 
given directly from its
calculation and as estimated from the SR (\ref{83e}) and (\ref{84e}), 
if one assumes that the $n = 0$ states dominate, as already has been 
checked from the Bjorken
and Uraltsev sum rules.

\section{Conclusions and outlook.}
\hspace*{\parindent}
In conclusion, within the OPE, we have presented a covariant method, 
using the trace formalism, to obtain sum rules in the heavy quark 
limit that relate the elastic
Isgur-Wise $\xi (w)$ to IW functions of transitions to excited states.\par

A main ingredient has been the introduction of the domain of the 
three variables $(w_i, w_f, w_{if})$, that allows a systematic way of 
exploring all possible SR.
In particular, we have given a simple and direct deduction of Bjorken 
and Uraltsev SR, with generalizations of the latter for $w \not= 1$. 
The simplicity of the
proof relies on the choice of the pseudoscalar $B$ meson $B(v_i) \to 
D^{(n)} (v') \to B(v_f)$ and of symmetric currents projected on the 
initial and final
velocities $v_i$ and $v_f$, like $(\Gamma_1, \Gamma_2) = ({/ \hskip - 
2 truemm v}_i, {/ \hskip - 2 truemm v}_f)$ or $({/ \hskip - 2 truemm 
v}_i\gamma_5, {/ \hskip -
2 truemm v}_f\gamma_5)$. This simplifies enormously the calculation, 
since it gives vanishing contributions for half of the possible 
intermediate states. Notice
that we obtain the same SR (\ref{48e}) and (\ref{67e}), if we use 
$(\Gamma_1, \Gamma_2 ) = ({/ \hskip - 2 truemm v}', {/\hskip - 2 
truemm v}')$ or $(\1, \1)$
and $(\Gamma_1, \Gamma_2 ) = ({/ \hskip - 2 truemm v}'\gamma_5 , 
{/\hskip - 2 truemm v}'\gamma_5)$ or $(i\gamma_5, i\gamma_5)$.\par

Moreover, a new class of SR, involving on the one hand IW functions 
at zero recoil and, on the other hand, IW functions for any $w$ have 
been obtained. These SR
reduce to known results for $w = 1$. \par

Among these new SR, we have found two new relations that involve only 
the elastic IW function $\xi (w)$, and the excited 
$\tau_{1/2}^{(n)}(w)$ and
$\tau_{3/2}^{(n)}(w)$, with {\it vanishing} contributions for all 
other IW functions between the ground state and higher excited 
states. The vanishing of the states
${3 \over 2}^-$ has been shown explicitely, using the corresponding 
wave function. We have generalized this result, demonstrating that 
all contributions of higher
states with $j^{\pm}$, $j \geq {5 \over 2}$ vanish identically. An 
important ingredient in the proof has been a compact formula for the 
polarization tensor
saturated with initial and final four-velocities. \par

These new SR are therefore very strong and provide new results that 
relate the slope $\rho^2$ and the curvature $\sigma^2$ of $\xi (w)$ to $\tau_{1/2}^{(n)}(1)$,
$\tau_{3/2}^{(n)}(1)$ and $\tau_{1/2}^{(n)'}( 
1)$, $\tau_{3/2}^{(n)'}(1)$. Modulo a very natural assumption, these 
SR imply the bound
$\sigma^2 \geq {5 \over 4} \rho^2$. \par

On the other hand, as a phenomenological remark, we have shown that 
these new SR for $\rho^2$ and $\sigma^2$ are in good agreement with 
the numerical results
obtained within the Bakamjian-Thomas relativistic quark models, that 
satisfy Isgur-Wise scaling. In this framework, the SR are saturated 
to a great accuracy by the
$n = 0$ intermediate states. \par

Which are the prospects of this work~? The main aim would be to 
obtain all possible usable SR. By usable we mean SR that involve only 
$\xi (w)$ and
$\tau_{1/2}^{(n)}(w)$, $\tau_{3/2}^{(n)}(w)$. \par

For the moment, we have concentrated mainly to the case, that appears 
to be simple, $B(v_i) \to B(v_f)$ with symmetric currents, projected 
along $v_i$ and $v_f$.
One should also study, on the one hand, the case of the transitions 
$B(v_i) \to B^*(v_f)$ and $B^*(v_i) \to B^*(v_f)$ and non-symmetric 
currents like $(\Gamma_1,
\Gamma_2) = ({/ \hskip - 2 truemm v}_i, {/ \hskip - 2 truemm v}_i)$, 
$({/ \hskip - 2 truemm v}_i, {/ \hskip - 2 truemm v}')$, etc, or 
equivalently  $(\Gamma_1,
\Gamma_2) = (\gamma_{\mu}, \gamma_{\nu})$, 
$(\gamma_{\mu}\gamma_5,\gamma_{\nu}\gamma_5)$ ... for which in 
general all intermediate states contribute. We have
explored a number of these non-symmetric situations for the 
pseudoscalar $B$-meson and found confirmation of the results 
presented here. \par

The case of the $B^*$ is rather involved because of the polarization, 
mainly in the case of non-symmetric currents, as used by Uraltsev in 
the finding of his SR. We
have given in Appendix B our covariant version of his calculation. \par

A systematic complete study remains to be done and may be worth. In 
particular, it would be interesting to check if the conjecture 
(\ref{88e}) on
$\tau_{1/2}^{(n)}(w)$, satisfied by BT quark models, that leads from 
the SR obtained here to $\sigma^2 \geq {5 \over 4} \rho^2$, is or is 
not a result of heavy
quark symmetry \cite{19r}. \par

\vskip 2 truecm

\noi {\large \bf Appendix A. Projector on the polarization tensors} \\

The polarization state of a relativistic boson is commonly described by
a polarization tensor, generalizing the polarization vector of  a spin 1
particle. The polarization tensors of a particle of integer spin $J$ are
the tensors $\varepsilon_{\mu_1 \cdots \mu_J}$ of tank $J$ which satisfy
the following conditions~: \par

1) Symmetry : $\varepsilon_{\mu_1 \cdots \mu_J} =
\varepsilon_{\mu_{\sigma (1)}  \cdots \mu_{\sigma (J)}}$ for any
permutation $\sigma$ of $1, \cdots , J$. \par

2) Vanishing contractions (or tracelessness)~: $g^{\mu\mu '}
\varepsilon_{\mu\mu ' \mu_3 \cdots \mu_J} = 0$ (when $J \geq 2$). \par

3) Transversity~: $v^{\mu} \varepsilon_{\mu\mu_2 \cdots \mu_J} = 0$,
where $v$ is the 4-velocity of the particle. \par

An orthornormal set of $2J + 1$ polarization states will be described by
a set $\varepsilon_{\mu_1 \cdots \mu_J}^{(\lambda)}$ of polarization
tensors satisfying the following normalisation conditions~:

$$g^{\mu_1\nu_1} \cdots g^{\mu_J\nu_J} \ \varepsilon_{\mu_1 \cdots
\mu_J}^{(\lambda)} \ \left ( \varepsilon_{\nu_1 \cdots \nu_J}^{(\lambda
')} \right )^* = (-1)^J \ \delta_{\lambda \lambda '} \ . \eqno({\rm
A}.1)$$

\noi Then, when summing over the intermediate states of a particle of
integer spin $J$, one has to deal with the {\it projector on
polarization tensors} $\prod^{(v)}$ defined by~:

$$\prod\nolimits_{\mu_1 \cdots \mu_J ; \nu_1 \cdots \nu_J}^{(v)} =
\sum_{\lambda = - J}^J \varepsilon_{\mu_1 \cdots \mu_J}^{(\lambda )}  \
\left ( \varepsilon_{\nu_1 \cdots \nu_J}^{(\lambda )} \right )^* \ .
\eqno({\rm A}.2)$$

In this appendix, we intend to deduce an explicit expression for this
tensor. The basic result is~:

$$v_f^{\mu_1} \cdots v_f^{\mu_J} \prod\nolimits_{\mu_1 \cdots
\mu_J;\nu_1 \cdots \nu_J}^{(v)} v_i^{\nu_1} \cdots v_i^{\nu_J}$$ $$ =
2^J {(J!)^2 \over (2J)!}  ( w_i^2 - 1)^{J/2} ( w_f^2 - 1 )^{J/2} \ P_J
\left ( {w_iw_f - w_{if} \over \sqrt{(w_i^2-1) (w_f^2 - 1)}} \right ) 
\eqno({\rm A}.3)$$

\noi where $v_i$ and $v_f$ are arbitrary velocity 4-vectors, $w_i$,
$w_f$, $w_{if}$ are the following scalar products~:

$$w_i = v \cdot v_i \ , \quad w_f = v \cdot v_f \ , \quad w_{if} = v_i
\cdot v_f\eqno({\rm A}.4)$$

\noi and $P_n$ is the usual Legendre polynomial.\par

The matrix element (A.3) is a polynomial in $w_i$, $w_f$, $w_{if}$.
Using explicit expressions of $P_n$, one has the two following useful
expressions of this polynomial~:

$$v_f^{\mu_1} \cdots v_f^{\mu_J} \prod\nolimits_{\mu_1 \cdots
\mu_J;\nu_1 \cdots \nu_J}^{(v)} v_i^{\nu_1} \cdots v_i^{\nu_J}$$ $$=
\sum_{0 \leq k \leq J/2} C_{J,k}  ( w_i^2 - 1)^k  ( w_i w_f - w_{if}
)^{J-2k}  ( w_f^2 - 1)^k \eqno({\rm A}.5)$$

$$v_f^{\mu_1} \cdots v_f^{\mu_J} \prod\nolimits_{\mu_1 \cdots
\mu_J;\nu_1 \cdots \nu_J}^{(v)} v_i^{\nu_1} \cdots v_i^{\nu_J}$$ $$=
\sum_{0 \leq k \leq J/2} C'_{J,k} ( w_i w_f - w_{if} )^{J-2k} \left [ (
w_i^2 - 1)    ( w_f^2 - 1) - ( w_i w_f - w_{if} )^2 \right ]^k
\eqno({\rm A}.6)$$

\vskip 8 truemm \noi where the $C_{j,k}$ and $C'_{J,k}$ are the
numerical coefficients given by~: 
$$C_{J,k} = (-1)^k \  {(J!)^2 \over 
(2J)!}\  {(2J-2k)! \over k! (J-k)!(J-2k)!} \eqno({\rm A}.7)$$ 
$$C'_{J,k} = (-1)^k \ 2^{J-2k}\ {(J!)^2 \over (2J)!}\  {J! \over (k!)^2
(J-2k)!} \ . \eqno({\rm A}.8)$$

\noi The expression (A.5) is useful when considering the limit $v_i \to
v$ in which $w_i \to 1$ and $w_{if} \to w_f$ (or as well $v_f \to v$ in
which $w_f \to 1$ and $w_{if} \to w_i$). The expression (A.6) is useful
when considering the limit $v_f \to v_i$ in which $w_{if} \to 1$ and
$w_f \to w_i$. Indeed the $k^{\rm th}$ term in (A.5) or (A.6) vanishes
at order $k$, and only the $k = 0$ term survives in the considered
limit. \par

In this paper the matrix elements (A.3) are all we need. However, as we
shall see the full expression of $\prod^{(v)}$ can be deduced from these
particular matrix elements. For brevity, we write this full expression
for a particle at rest, in which case only the purely spatial components
are non-vanishing. The tensor $\prod_{\mu_1 , \cdots , \mu_n ; \nu_1 ,
\cdots , \nu_n}^{(v)}$ for an arbitrary 4-velocity $v$ is readily
obtained from the formula below by the substitutions~:

$$\begin{array}{l} \delta_{i_ri_{r'}} \to - g_{\mu_r\mu_{r'}} +
v_{\mu_r} v_{\mu_{r'}} \\ \delta_{i_sj_{s'}} \to - g_{\mu_s\nu_{s'}} +
v_{\mu_s} v_{\nu_{s'}} \\ \delta_{j_tj_{t'}} \to - g_{\nu_t\nu_{t'}} +
v_{\nu_t} v_{\nu_{t'}}  \ . \\ \end{array} \eqno({\rm A}.9)$$

\noi The formula is~:

$$\prod\nolimits_{i_1 \cdots i_n;j_1\cdots j_n} = \sum_{0 \leq k \leq
n/2}\ f_{n,k}\ \sum_{I,J\subset \{1 \cdots n\} \atop{|I| = |J| = 2k}}$$
$$\left ( \sum_{{\cal J} \in {\cal P}_2(I)} \ \prod_{\{r,r'\} \in {\cal
J}} \ \delta_{i_r i_{r'}}\right ) \left ( \sum_{\sigma \in {\cal B}({\bf
C}I,{\bf C}J)} \prod_{s \notin I} \ \delta_{i_sj_{\sigma (s)}} \right )
\left (  \sum_{{\cal J}' \in {\cal P}_2(J)} \ \prod_{\{t,t'\} \in {\cal
J}'} \ \delta_{j_tj_{t'}}\right )\eqno({\rm A}.10)$$

\vskip 3 truemm \noi with

$$f_{n,k} = (-1)^k \ 2^{2k}\ {k!(2n-2k)! \over (n-k)!(2n)!} \
.\eqno({\rm A}.11)$$

\noi In this formula (A.10), ${\bf C}I$ and ${\bf C}J$ are the
complementary sets in $\{ 1, 2, \cdots n \}$ of the subsets $I$ and $J$.
For a set $X$, ${\cal P}_2(X)$ is the set of partitions of $X$ by
two-element subsets (precisely unordered partitions or, partitions as
sets of subsets). For two sets $X$ and $Y$, ${\cal B}(X,Y)$ is the set
of bijections $X \to Y$. \par

(There is a logical subtlety in the (important) terms $I = J =
\emptyset$ in (A.10)). Namely, one has

$$\sum_{{\cal J} \in {\cal P}_2(\emptyset )} \prod_{\{r, r'\} \in {\cal
J}} \delta_{i_r,i_{r'}} = 1 \ .$$

\noi The reason is that the set (of sets) ${\cal P}_2(\emptyset )$ is
not the empty set (else the sum would be 0), but is $\{\emptyset \}$. It
contains the only element ${\cal J} = \emptyset$ (the empty set), and
the product $\prod\limits_{\{r,r'\} \in {\cal J}} \delta_{i_r,i_{r'}}$
of an empty family is conventionally 1). \par

In words, a term in (A.10) is obtained as follows. Select an even number
$2k$ of indices among the $i$'s and also among the $j$'s. Match the
remaining $i$'s with the remaining $j$'s and include a factor
$\delta_{ij}$ for each matched pair $(i, j)$. Divide the $2k$ selected
$i$'s into pairs and include a factor $\delta_{ii'}$ for each pair $(i,
i')$. Divide the $2k$ selected $j$'s into pairs and include a factor
$\delta_{jj'}$ for each pair $(j,j')$. Different terms correspond to
different such products of $\delta$'s. \par

\noi The lower rank $(n \leq 3)$ tensors write~:

$$ \prod\nolimits_{i_1;j_1} = \delta_{i_1j_1} \eqno({\rm A}.12)$$

$$\prod\nolimits_{i_1,i_2;j_1,j_2} =  \displaystyle{{1 \over 2}} \left (
\delta_{i_1j_1}\ \delta_{i_2j_2} + \delta_{i_1j_2}\ \delta_{i_2j_1}
\right ) -  \displaystyle{{1 \over 3}} \ \delta_{i_1i_2}\
\delta_{j_1j_2} \eqno({\rm A}.13)$$

$$\prod\nolimits_{i_1,i_2,i_3;j_1,j_2,j_3} =  \displaystyle{{1 \over 6}}
\Big ( \delta_{i_1j_1}\ \delta_{i_2j_2} \ \delta_{i_3j_3}+
\delta_{i_1j_1}\ \delta_{i_2j_3} \ \delta_{i_3j_2} + \ \delta_{i_1j_2} \
\delta_{i_2j_1}\ \delta_{i_3j_3} \eqno({\rm A}.14)$$ $$+ \
\delta_{i_1j_2}\ \delta_{i_2j_3} \ \delta_{i_3j_1} + \delta_{i_1j_3}\
\delta_{i_2j_1} \ \delta_{i_3j_2} + \delta_{i_1j_3}\ \delta_{i_2j_2} \
\delta_{i_3j_1} \Big ) $$ $$  -  \displaystyle{{1 \over 15}} \ \Big (
\delta_{i_1i_2}\ \delta_{i_3j_3} \ \delta_{j_1j_2} + \delta_{i_1i_2}\
\delta_{i_3j_2} \ \delta_{j_1j_3} + \delta_{i_1i_2}\ \delta_{i_3j_1} \
\delta_{j_2j_3}$$ $$+ \ \delta_{i_1i_3}\ \delta_{i_2j_3} \
\delta_{j_1j_2} + \delta_{i_1i_3}\ \delta_{i_2j_2} \ \delta_{j_1j_3} +
\delta_{i_1i_3}\ \delta_{i_2j_1} \ \delta_{j_2j_3}$$ $$  + \
\delta_{i_2i_3}\ \delta_{i_1j_3} \ \delta_{j_1j_2} + \delta_{i_2i_3}\
\delta_{i_1j_2} \ \delta_{j_1j_3} + \delta_{i_2i_3}\ \delta_{i_1j_1} \
\delta_{j_2j_3}\Big ) \ .$$ \vskip 5 truemm

\noi{\bf Reduction to a 3-dimensional problem} \vskip 5 truemm

We now turn to proofs of the above results. As a preliminary step,
observe that the problem reduces itself to a 3-dimensional problem.
Indeed, due to Lorentz covariance, it is enough to consider a particle
at rest, namely $v = (1, \vec{0})$. Then the transversity condition 3)
amounts saying that $\varepsilon_{\mu_1 \cdots \mu_J} = 0$ if {\it one}
of the indices is 0. Therefore, a polarization tensor is completely
determined by its purely spatial components $\varepsilon_{i_1 \cdots
i_J}$, all the other components being zero. On these 3-dimensional
tensors, the conditions of symmetry 1) and of tracelessness 2) write~:
\par

1$'$) Symmetry~: $\varepsilon_{i_1 \cdots i_J} = \varepsilon_{i_{\sigma
(1)}  \cdots i_{\sigma (J)}}$ for any permutation $\sigma$ of $1, \cdots
, J$. \par

2$'$) Tracelessness~: $\sum\limits_i \varepsilon_{iii_3 \cdots i_J} = 0$
(when $J \geq 2$). \par

\noi The orthonormality conditions (A.1) writes

$$ \sum_{i_1 \cdots i_J} \varepsilon_{i_1 \cdots i_J}^{(\lambda)} \left
( \varepsilon_{i_1 \cdots i_J}^{(\lambda ')}\right )^* = \delta_{\lambda
\lambda '} \ . \eqno({\rm A}.15)$$

\noi The tensor $\prod$ is also purely spatial and, according to (A.2),
is given by~:

$$\prod\nolimits_{i_1 \cdots i_J;i'_1 \cdots i'_J} = \sum_{\lambda = -
J}^J \varepsilon_{i_1 \cdots i_J}^{(\lambda)} \left ( \varepsilon_{i'_1
\cdots i'_J}^{(\lambda )}\right )^* \ .  \eqno({\rm A}.16)$$

\noi But the preceding consideration identifies this spatial $\prod$ as
the projection operator, in the space of tensors of rank $J$, on the
subspace of traceless symmetric tensors. Indeed, according to (A.16),
the tensor $\sum_{i'_1 \cdots i'_J} \prod_{i_1 \cdots i_J;i'_1 \cdots
i'_J} T_{i'_1 \cdots i'_J}$ is traceless symmetric for any tensor
$T_{i_1 \cdots i_J}$ and, according to (A.15) and (A.16), one has

$$\sum_{i'_1 \cdots i'_J} \prod\nolimits_{i_1 \cdots i_J;i'_1 \cdots
i'_J} \varepsilon_{i'_1 \cdots i'_J} = \varepsilon_{i_1 \cdots i_J}
\eqno({\rm A}.17)$$

\noi for any traceless symmetric tensor $\varepsilon_{i_1 \cdots i_J}$.
\par

The problem of finding the projector on the polarization tensors is now
reduced to the problem of finding the projector on the {\it spatial}
symmetric traceless tensors. \par

\vskip 5 truemm

\noi{\bf Deduction of the projector's particular matrix elements by
standard methods of angular momentum coupling} \vskip 5 truemm

Let us now turn to a proof of (A.3). The space of rank $J$ tensors is
just the tensor product of a number $J$ of the angular momentum 1
representation of the rotation group. The subspace of traceless
symmetric tensors is just the subspace of angular momentum $J$, as can
be understood since this subspace is used to describe the spin states of
a particle of spin $J$. \par

Our problem is now reduced to the coupling of $J$ copies of the angular
momentum 1 into a total angular momentum $J$. We now on use standard
methods of angular momentum coupling. \par

The Clebsch-Gordan coefficients for the coupling of two angular momenta
$J_1$ and $J_2$ to the maximum value $J_1 + J_2$ has the simple
following factorized form~:

$$<J_1 + J_2, M|J_1,J_2,M_1,M_2> = \delta_{M,M_1 + M_2} \ {c(J_1,M_1) \
c(J_2,M_2) \over c (J_1 + J_2, M)}\eqno({\rm A}.18)$$

\noi with

$$c(J,M) = {\sqrt{(2J)!} \over \sqrt{(J+M)!(J-M)!}} \ . \eqno({\rm
A}.19)$$

\noi The coupling coefficients of three angular momenta $J_1$, $J_2$,
$J_3$ to the maximum value $J_1 + J_2 + J_3$, defined by

$$<J_1 + J_2 + J_3, M|J_1, J_2, J_3, M_1,M_2,M_3> =$$ $$
\sum\nolimits_{M'} <J_1 + J_2 + J_3, M|J_1 + J_2, J_3, M', M_3 > <J_1 +
J_2, M'|J_1, J_2, M_1, M_2 > \eqno({\rm A}.20)$$ \vskip 3 truemm

\noi is easily calculated from (A.17)~:

$$<J_1 + J_2 + J_3,M|J_1,J_2,J_3,M_1,M_2,M_3> = \delta_{M,M_1 + M_2 +
M_3} {c(J_1,M_1)c(J_2,M_2)c(J_3,M_3) \over c(J_1 + J_2 + J_3, M)} \ .
\eqno({\rm A}.21)$$

\noi Moreover, these coefficients do not depend on the particular order
of coupling chosen in (A.19) (first coupling $J_1$ and $J_2$, and then
coupling the result to $J_3$). \par

By a simple recursive argument, one finds from (A.19) that the coupling
coefficients of $n$ angular momenta $J_1, \cdots, J_n$ to the maximum
value $J_1 + \cdots + J_n$ is given by

$$<J_1 + \cdots + J_n,M|J_1,\cdots ,J_n,M_1,\cdots ,M_n> = \delta_{M,M_1
+ \cdots + M_n} {c(J_1,M_1)\cdots c(J_n,M_n)\over c(J_1 + \cdots +
J_n,M_n)}  \eqno({\rm A}.22)$$

\noi and is independent of the order of the couplings. Remind that the
$|J_1 + \cdots + J_n,M>$ are basis states of the $J_1 + \cdots + J_n$
angular momentum subspace in the tensorial product of the $J_1 \cdots
J_n$ representation spaces of $SU(2)$, and that the coefficient $<J_1 +
\cdots + J_n,M|J_1, \cdots , J_n, M_1 , \cdots , M_n>$ is the scalar
product of the state $|J_1 + \cdots + J_n,M>$ with the basis state

$$|J_1 , \cdots , J_n,M_1,\cdots ,M_n> \ = |J_1, M_1> \otimes \cdots
\otimes |J_n, M_n> \eqno({\rm A}.23)$$

\noi in the tensorial product space. \par

Now we may take the case of interest to us, $J_1 = \cdots = J_n = 1$,
with the $J = 1$ representation of $SU(2)$ in the form of the ordinary
rotations in $\C^3$ space (complexified ordinary three-dimensional
space). The tensorial product space is just the space of tensors of
order $n$, and the $J_1 + \cdots + J_n = n$ subspace is just the
subspace of traceless symmetric tensors. The states $|1, \cdots  , 1,
M_1 , \cdots , M_n>$ are the tensorial products of standard basis
vectors $|1, M>$ of $\C^3$, and the states $|n, M>$ constitute a
standard basis of symmetric tensors. We are interested by the scalar
product of the tensors $|n,M>$ with the tensors $\vec{x}^{\otimes n}$
$((\vec{x}^{\otimes n})_{i_1 \cdots i_n} = x_{i_1} \cdots x_{i_n})$ for
any $\vec{x} \in \C^3$. Therefore, we have to expand the tensors
$\vec{x}^{\otimes n}$ in the basis $|1, \cdots , 1, M_1, \cdots , M_n>$.
\par

The qualifier ``standard'' above means in conformity to the standard
definition of the Clebsch-Gordan coefficients. The standard basis if
$\C^3$ is~:

$$\begin{array}{l} |1,  1 > = \vec{f}_{1} = - {1 \over \sqrt{2}} \
(\vec{e}_1 + i\vec{e}_2)\\ \\ |1, 0> = \vec{f}_0 = \vec{e}_3\\ \\ |1, -
1> = \vec{f}_{-1} = {1 \over \sqrt{2}}\ (\vec{e}_1 - i
\vec{e}_2)\end{array} \eqno({\rm A}.24)$$

\noi where $(\vec{e}_1, \vec{e}_2, \vec{e}_3)$ is the Cartesian basis.
Then a vector $\vec{x} = x_1 \vec{e}_1 + x_2 \vec{e}_2 + x_3 \vec{e}_3
\in \C^3$ writes

$$\vec{x} = - {x_1 - ix_2 \over \sqrt{2}} \ \vec{f}_1 + {x_1 + ix_2
\over \sqrt{2}} \ \vec{f}_{-1} + x_3 \ \vec{f}_0 \eqno({\rm A}.25)$$

\noi and the tensor $(\vec{x})^{\otimes n}$ is expanded as~:

$$(\vec{x})^{\otimes n} = \sum_{k,k'} {n! \over k!k'!(n-k-k')!}\
(-1)^{k} \ \left ( {x_1 - ix_2 \over \sqrt{2}} \right )^k \ \left ( {x_1
+ ix_2 \over \sqrt{2}} \right )^{k'} \ (x_3)^{n-k-k'}$$ $${\rm Sym} \
(\vec{f}_1)^{\otimes k} \otimes (\vec{f}_{-1})^{\otimes k'} \otimes
(\vec{f}_0)^{\otimes n - k - k'} \eqno({\rm A}.26)$$

\noi where Sym is the projector on symmetric tensors

$$({\rm Sym}\ T)_{i_1 \cdots i_n} = {1 \over n !} \ \sum_{\sigma} \
T_{i_{\sigma (1)} \cdots i_{\sigma(n)}} \ . $$

\noi Actually, equipped with the symmetrized product, the symmetric
tensors constitute a commutative algebra, so that formula (A.26) is just
obtained by multinomial expansion. \par

Then we have

$$<n,M|(\vec{x})^{\otimes n}> = \sum_{k,k'} {n! \over k! k'! (n-k-k')!}
\ (-1)^{k} \ \left ( {x_1 - ix_2 \over \sqrt{2}}\right )^k \ \left (
{x_1 + ix_2 \over \sqrt{2}} \right )^{k'} \ (x_3)^{n-k-k'}$$
$$<n,M|(\vec{f}_1)^{\otimes k} \otimes (\vec{f}_{-1})^{\otimes k'}
\otimes (\vec{f}_0)^{\otimes n- k-k'}> \eqno({\rm A}.27)$$

\noi where the Sym operator has been dropped because the coupling
coefficients do not depend on the order of the couplings. Formula (A.22)
now readily gives~: 

$$<n,M|(\vec{x})^{\otimes n}> = \sum_{k,k'}
{n! \over k! k'! (n-k-k')!} \ (-1)^{k} \ \left ( {x_1 - ix_2 \over
\sqrt{2}}\right )^k \ \left ( {x_1 + ix_2 \over \sqrt{2}} \right )^{k'}
\ (x_3)^{n-k-k'}$$ $$\delta_{M,k-k'} \ {c(1,1)^k \ c(1,0)^{n-k-k'} \
c(1,-1)^{k'} \over c(n, M)} \ . \eqno({\rm A}.28)$$

\noi An easy calculation (just undoing the multinomial expansion) gives
the following generating function for these $<n,M|(\vec{x})^{\otimes
n}>$~:

$$\sum_{M=-n}^n c(n,M) <n,M|(\vec{x})^{\otimes n}> \ s^M=$$ $$\left [ -
c(1,1) {x_1 - ix_2 \over \sqrt{2}} \ s + c(1,0) x_3 - c(1,-1) \ {x_1 +
ix_2 \over \sqrt{2}} \ s^{-1} \right ]^{n} \ .\eqno({\rm A}.29)$$

\noi According to formula (A.19), we have

$$\begin{array}{l} c(n,M) = \displaystyle{{\sqrt{(2n)!} \over
\sqrt{(n+M)!(n-M)!}}}\\ \\ c(1,1) = 1\ , \quad c(1,0) = \sqrt{2} \ ,
\quad c(1,-1) = 1 \ .\end{array} \eqno({\rm A}.30)$$

\noi Therefore

$$\sum_{M=-n}^n\ {\sqrt{(2n)!} \over \sqrt{(n+M)!(n-M)!}} \
<n,M|(\vec{x})^{\otimes n}> \ s^M =$$ $$2^{n/2} \ \left (- {x_1 - ix_2
\over 2}\ s + x_3 + {x_1 + ix_2 \over 2}\ s^{-1} \right )^{n} \ .
\eqno({\rm A}.31)$$

\noi Comparing this to the generating function for the solid spherical
harmonics ${\cal Y}_L^M(\vec{x}) = |\vec{x}|^L \ Y_L^M(\widehat{x})$,
which is~:

$$\sum_{M=-L}^L\ {(L)! \over \sqrt{(L-M)!(L+M)!}}\ {\cal Y}_L^M(\vec{r})
\ s^M = {\sqrt{2L+1} \over \sqrt{4\pi}} \ \left ( {x_1 - ix_2 \over 2}\
s^{-1} + x_3 - {x_1 + ix_2 \over 2}\ s \right )^{L}  \eqno({\rm A}.32)$$

\noi we arrive at the fundamental result~:

$$<n,M|(\vec{x})^{\otimes n}> \ = 2^{n/2} \ {n! \over \sqrt{(2n+1)!}}\
\sqrt{4\pi}\  {\cal Y}_n^M(\vec{x})^*  \ . \eqno({\rm A}.33)$$

\noi From this we compute the matrix element $<(\vec{y})^{\otimes
n}|\prod_n|(\vec{x})^{\otimes n}>$ of the seeked projector $\prod_n$ (on
the traceless symmetric tensors). One has

$$<(\vec{y})^{\otimes n}|\prod\nolimits_n|(\vec{x})^{\otimes n}> \ =
\sum_{M=-n}^n \ <n,M|(\vec{y})^{\otimes n}>^*\ <n,M|(\vec{x})^{\otimes
n}>$$ $$ = 2^n\ {(n!)^2 \over (2n+1)!} \ 4 \pi \ \sum_{M=-n}^n \ {\cal
Y}_n^M(\vec{y})\ {\cal Y}_n^M(\vec{x})^* \eqno({\rm A}.34)$$

\noi and using

$$\sum_{M=-n}^n \ {\cal Y}_n^M(\vec{x})\ {\cal Y}_n^M(\vec{y})^* = {2n+1
\over 4 \pi} \ |\vec{x}|^n\ |\vec{y}|^n \ P_n \left ( {\vec{x} \cdot
\vec{y} \over |\vec{x}|\ |\vec{y}|} \right ) \eqno({\rm A}.35)$$

\noi one readily obtains~:

$$<(\vec{y})^{\otimes n}|\prod\nolimits_n|(\vec{x})^{\otimes n}> \ = 2^n
\ {(n!)^2 \over (2n)!} \ |\vec{x}|^n\ |\vec{y}|^n \ P_n \left ( {\vec{x}
\cdot \vec{y} \over |\vec{x}|\ |\vec{y}|} \right ) \ . \eqno({\rm
A}.36)$$

\noi One may then introduce explicit expressions for the Legendre
polynomials $P_n$~:

$$P_n(x) = {1 \over 2^n} \ \sum_{0 \leq k \leq n/2}\ (-1)^k\ {(2n-2k)!
\over k! (n-k)! (n-2k)!} \ x^{n-2k}\eqno({\rm A}.37)$$ 

$$P_n(x) = \sum_{0 \leq k \leq n/2}\ (-1)^k\  {1 \over 2^{2k}} \  {n!
\over (k!)^2 (n-2k)!}  \ x^{n-2k}\ (1 - x^2)^k \eqno({\rm A}.38)$$

\noi and we obtain the following explicit expressions for
$<(\vec{y})^{\otimes n}|\prod_n|(\vec{x})^{\otimes n}>$~:

$$\begin{array}{ll}<(\vec{y})^{\otimes
n}|\prod\nolimits_n|(\vec{x})^{\otimes n}> &=  \displaystyle{\sum_{0
\leq k \leq n/2}}\ C_{n,k} \ (\vec{x}^{\, 2})^k \ (\vec{x}\cdot
\vec{y})^{n-2k} (\vec{y}^{\, 2})^k \\ \\ &= \displaystyle{\sum_{0 \leq k
\leq n/2}}\ C'_{n,k} \ (\vec{x}\cdot \vec{y})^{n-2k} \left [ \vec{x}^{\,
2} \vec{y}^{\, 2} - (\vec{x} \cdot \vec{y})^2\right ]^k \\ \end{array}
\eqno({\rm A}.39)$$ \vskip 3 truemm

$$\begin{array}{l} C_{n,k} = (-1)^k \  \displaystyle{{(n!)^2 \over
(2n)!}}\  \displaystyle{{(2n-2k)! \over k! (n-k)!(n-2k)!}} \\ \\
C'_{n,k} = (-1)^k \ 2^{n-2k}\ \displaystyle{{(n!)^2 \over (2n)!}}\ 
\displaystyle{{n! \over (k!)^2 (n-2k)!}} \ . \\ \end{array} \eqno({\rm
A}.40)$$

\noi To go back to an arbitrary velocity $v$ and obtain (A.3), just set
$\vec{x} = \vec{v}_i$, $\vec{y} = \vec{v}_f$ and use the following
formulae~:

$$\begin{array}{l} \vec{v}_i^{\, 2} = (v_i \cdot v)^2 - v_i^2 = w_i^2 -
1 \\ \vec{v}_f^{\, 2} = (v_f \cdot v)^{2} - v_f^2 = w_f^2 - 1 \\
\vec{v}_i \cdot \vec{v}_f = (v_i \cdot v) (v_f \cdot v) - v_i \cdot v_f
= w_i w_f - w_{if} \ . \\ \end{array} \eqno({\rm A}.41)$$

\vskip 5 truemm

\noi {\bf Deduction of the projector itself from its particular matrix
elements} \vskip 5 truemm

We now present a deduction of $\prod_{i_1, \cdots , i_n;j_1, \cdots ,
j_n}$ from the matrix elements\break \noindent $<(\vec{y})^{\otimes
n}|\prod_n|(\vec{x})^{\otimes n}>$. To see how to proceed, let us
consider a multilinear function $F(\vec{x}_1, \cdots , \vec{x}_n)$,
which is {\it symmetric} in the permutations of its $n$ vector variables
$\vec{x}_1, \cdots , \vec{x}_n$. Then it can be recovered from its
diagonal values $F((\vec{x})_n) = F(\vec{x} , \cdots , \vec{x})$ by the
following formula~:

$$F(\vec{x}_1 , \cdots , \vec{x}_n) = {(-1)^n \over n!} \sum_{{s_1,
\cdots , s_n}\atop 0 \leq s_i \leq  1} (-1)^{s_1 + \cdots + s_n} \
F\left ((s_1 \vec{x}_1 + \cdots + s_n \vec{x}_n )_n\right ) \ .
\eqno({\rm A}.42)$$

\noi Indeed, expanding $ F((s_1 \vec{x}_1 + \cdots + s_n \vec{x}_n )_n)$
by multinearity and collecting terms equal by symmetry, one has~:

$$F\left ((s_1 \vec{x}_1 + \cdots + s_n \vec{x}_n )_n\right ) =
\sum_{{q_1, \cdots ,q_n}\atop q_1 + \cdots + q_n = n} {n! \over q_1!
\cdots q_n!} \ s_1^{q_1} \cdots s_n^{q_n} \  F\left ((\vec{x}_1)_{q_1}, 
\cdots , (\vec{x}_n )_{q_n}\right ) \eqno({\rm A}.43)$$

\noi where the notation $(\vec{x}_i)_{q_i}$ (also used in (A.42)) stands
for the $q_i$-uple $(\vec{x}_i, \cdots , \vec{x}_i)$. A term in (A.43)
with {\it some} $q_i$ vanishing gives no contribution to (A.42) because
it does not depend on $s_i$, and the corresponding $s_i = 0$ and $s_i =
1$ terms in (A.42) cancels. Then, since $q_1 + \cdots + q_n = n$, the
only term of (A.43) contributing to (A.42) is $q_1 = \cdots = q_n = 1$.
The right-hand side of (A.42) is therefore equal to~:

$${(-1)^n \over n!} \sum_{{s_1, \cdots, s_n}\atop 0 \leq s_i \leq 1}
(-1)^{s_1 + \cdots + s_n} \ n! s_1 \cdots s_n \ F(\vec{x}_1, \cdots ,
\vec{x}_n) = F(\vec{x}_1 , \cdots , \vec{x}_n) \ .$$

Using (A.42), we can now deduce $\prod_{i_1 , \cdots , i_n;j_1, \cdots ,
j_n}$ from the matrix elements in two steps. As a first step, let us
apply formula (A.42) to the multilinear symmetric function

$$\left ( \vec{y}_1 , \cdots , \vec{y}_n \right ) \to <\vec{y}_1 \otimes
\cdots \otimes \vec{y}_n | \prod\nolimits_n | (\vec{x})^{\otimes n}>
\eqno({\rm A}.44)$$

\noi with $\vec{x}$ fixed. This gives~:

$$<\vec{y}_1 \otimes \cdots \otimes \vec{y}_n|\prod\nolimits_n|
(\vec{x})^{\otimes n}> = {(-1)^n \over n!} \sum_{{s_1, \cdots, s_n}\atop
0 \leq s_i \leq 1} (-1)^{s_1 + \cdots + s_n} \langle \left ( \sum_i s_i
\vec{y}_i\right )^{\otimes n} |\prod\nolimits_n| (\vec{x})^{\otimes
n}\rangle \eqno({\rm A}.45)$$

\noi or, using (A.39)

$$<\vec{y}_1 \otimes \cdots \otimes \vec{y}_n|\prod\nolimits_n|
(\vec{x})^{\otimes n}> = {(-1)^n \over n!} \sum_{0 \leq k \leq n/2}
C_{n,k} (\vec{x}^{\, 2})^k  \eqno({\rm A}.46)$$ $$\sum_{{s_1, \cdots,
s_n}\atop 0 \leq s_i \leq 1} (-1)^{s_1 + \cdots + s_n} \left ( \sum_i
s_i (\vec{y}_i \cdot \vec{x}) \right )^{n-2k} \left ( \left ( \sum_i s_i
\vec{y}_i \right )^2 \right )^k \ .$$

\noi Then we work out multinomial expansions~:

$$\left ( \sum_i s_i (\vec{y}_i \cdot \vec{x}) \right )^{n-2k} =
\sum_{{u_1, \cdots ,u_n \geq 0}\atop u_1 + \cdots + u_n = n-2k} {(n-2k)!
\over u_1! \cdots u_n!} \prod_{i=1}^n (s_i)^{u_i} (\vec{y}_i \cdot
\vec{x})^{u_i} \eqno({\rm A}.47)$$

$$\left ( \left ( \sum_i s_i \vec{y}_i \right )^2 \right )^k = \left (
\sum_{1 \leq i, i'\leq n}  s_is_{i'}  (\vec{y}_i \cdot \vec{y}_{i'})
\right )^{k}\eqno({\rm A}.48)$$ $$ = \sum_{{v_{11}, v_{12}, \cdots ,
v_{n-1,n}, v_{nn \geq 0}}\atop v_{11} + v_{12} + \cdots + v_{nn} = k}
{k! \over v_{11}! v_{12}! \cdots v_{nn} !} \prod_{i,i'=1}^n
(s_is_{i'})^{v_{ii'}} (\vec{y}_i \cdot \vec{y}_{i'})^{v_{ii'}} \ .$$

\noi Using this in (A.46) and collecting the powers of the $s_i$, we
have~:

$$<\vec{y}_1 \otimes \cdots \otimes \vec{y}_n|\prod\nolimits_n|
(\vec{x})^{\otimes n}> = {(-1)^n \over n!} \sum_{0 \leq k \leq n/2}
C_{n,k} (\vec{x}^{\, 2})^k\eqno({\rm A}.49)$$ $$\sum_{{u_1, \cdots ,u_n
\geq 0}\atop u_1 + \cdots + u_n = n-2k} \sum_{{v_{11}, v_{12}, \cdots ,
v_{n-1,n}, v_{nn \geq 0}}\atop v_{11} + v_{12} + \cdots + v_{nn} = k}   
         {(n-2k)! \over u_1! \cdots u_n!} {k! \over v_{11}! v_{12}!
\cdots v_{nn} !}$$ $$\sum_{{s_1, \cdots, s_n}\atop 0 \leq s_i \leq 1}
(-1)^{s_1 + \cdots + s_n} \left [ \prod_{i=1}^n (s_i)^{p_i} \right ]
\left [ \prod_{i=1}^n (\vec{y}_i \cdot \vec{x})^{u_i} \right ] \left [
\prod_{i,i'=1}^n (\vec{y}_i \cdot \vec{y}_{i'})^{v_{ii'}}\right ]$$

\noi where the exponent $p_i$ of $s_i$ is~:

$$p_i = u_i + \sum_{i'=1}^n (v_{i'i} + v_{ii'})\ . \eqno({\rm A}.50)$$

\noi Notice now that, for values of the $u_i$'s and of the $v_{ii'}$'s
such that some exponent $p_{i_0}$ vanishes, the $s_{i_0} = 0$ and the
$s_{i_0} = 1$ terms cancel. Since, according to the constraints on the
$u_i$'s and the $v_{ii'}$'s, one has

$$\sum_{i=1}^n \ p_i = \sum_{i=1}^n \ u_i + 2 \sum_{i,i'=1}^n v_{ii'} =
n \ , \eqno({\rm A}.51)$$

\noi we are left with the values of the $u_i$'s and of the $v_{ii'}$'s
such that $p_1 = \cdots = p_n = 1$. These values can then be only 0 or
1, so that $u_i! = 1$ and $v_{ii'} ! = 1$. Moreover we have then a
factor $s_1 \cdots s_n$, so that only $s_1 = \cdots = s_n = 1$
contributes. So, (A.49) reduces to~:

$$<\vec{y}_1 \otimes \cdots \otimes \vec{y}_n|\prod\nolimits_n|
(\vec{x})^{\otimes n}> = \sum_{0 \leq k \leq n/2} {k! (n-2k)! \over n!}
C_{n,k} (\vec{x}^{\, 2})^k\eqno({\rm A}.52)$$ $$\sum_{{u_1, \cdots ,u_n,
v_{11}, v_{11}, v_{12}, \cdots , v_{n-1,n}, v_{nn} \geq 0\atop u_1 +
\cdots + u_n = n-2k}\atop u_i + v_{1i} + \cdots + v_{ni} + v_{i1} +
\cdots + v_{in} = 1} \left [ \prod_{i=1}^n (\vec{y}_i \cdot
\vec{x})^{u_i} \right ] \left [ \prod_{i,i'=1}^n (\vec{y}_i \cdot
\vec{y}_{i'})^{v_{ii'}}\right ]$$

\noi where we have dropped the constraint $\sum\limits_{i,i'=1}^n
v_{ii'} = k$ since it is implied by the remaining constraints. \par

Since the $u_i$'s take only the values 0 or 1, we can replace $(u_1 ,
\cdots , u_n)$, as summation variable, by subsets $I$ of $\{1, \cdots
,n\}$. It will be convenient to use the subset related to $(u_1 , \cdots
, u_n)$ by $I = \{ i|u_i=0\}$. The constraint $u_1 + \cdots + u_n =
n-2k$ is translated into the constraint $|I| = 2k$, and formula (A.52)
becomes~:

$$<\vec{y}_1 \otimes \cdots \otimes \vec{y}_n|\prod\nolimits_n|
(\vec{x})^{\otimes n}> = \sum_{0 \leq k \leq n/2} {k! (n-2k)! \over n!}
C_{n,k} \eqno({\rm A}.53)$$ $$(\vec{x}^{\, 2})^k \sum_{{I\subset \{1
\cdots n\}}\atop |I| = 2k} \left [ \prod_{i\notin I} (\vec{y}_i \cdot
\vec{x}) \right ] \sum_{{v_{11}, v_{11}, v_{12}, \cdots , v_{n-1,n},
v_{nn} \geq 0 \atop v_{1i} + \cdots v_{ni} + v_{i1} + \cdots + v_{in} =
0 (i \notin I)} \atop v_{1i} + \cdots + v_{ni} + v_{i1} + \cdots +
v_{in} = 1 (i \in I)} \left [ \prod_{i,i'\in I} (\vec{y}_i \cdot
\vec{y}_{i'})^{v_{ii'}}\right ] \ .$$

Consider now the constraints on the $v_{ii'}$'s. If $i \notin I$ or $i'
\notin I$, we have $v_{ii'} = 0$. We are left with the $v_{ii'}$ for
$i$, $i' \in I$, constrained by~:

$$\sum_{i' \in I} (v_{ii'} + v_{i'i}) = 1 \qquad \hbox{(for any $i \in
I$)} \ . \eqno({\rm A}.54)$$

\noi Notice that $v_{ii} = 0$, since $v_{ii}$ occurs twice in this sum.
Let us then replace the $v_{ii'}$'s, as summation variable, by the set
${\cal J}$ of two-element subsets of $I$ related to the $v_{ii'}$'s by

$${\cal J} = \{ \{i,i'\} |v_{ii'} + v_{i'i} = 1 ) \ .\eqno({\rm A}.55)$$

\noi For any $i \in I$, there is, according to (A.54), one and only one
$i' \in I$ such that $\{i,i'\} \in {\cal J}$. In other words, ${\cal J}$
belongs to the set ${\cal P}_2(I)$ of partitions of $I$ by two-element
subsets. Conversely, if ${\cal J} \in {\cal P}_2(I)$, the values of
$v_{ii'} + v_{i'i}$ defined by (A.55), namely

$$v_{ii'} + v_{i'i} = 1 \ {\rm if} \ \{i, i'\} \in {\cal J}\ , \qquad 
v_{ii'} + v_{i'i} = 0 \ {\rm if} \ \{i, i'\} \notin {\cal J} \eqno({\rm
A}.56)$$

\noi do satisfy the constraints (A.54), since for any $i \in I$, there
is one and only one $i' \in I$ such that $\{i, i'\} \in {\cal J}$. Now,
(A.56) does not determine completely the values of the $v_{ii'}$'s. When
$\{i, i'\} \notin {\cal J}$ we must have $v_{ii'} = v_{i'i} = 0$, but
when $\{i, i'\} \in {\cal J}$ we have two solutions~: $v_{ii'} = 1$,
$v_{i'i} = 0$ and $v_{ii'} = 0$, $v_{i'i} = 1$. Since $|{\cal J}| = k$,
we have in all $2^k$ values of the $v_{ii'}$'s corresponding to each
${\cal J} \in {\cal P}_2(I)$. However, values of the $v_{ii'}$'s
corresponding to a same ${\cal J}$ give equal terms in (A.53) and,
lumping these terms together, formula (A.53) becomes~:

$$<\vec{y}_1 \otimes \cdots \otimes \vec{y}_n|\prod\nolimits_n|
(\vec{x})^{\otimes n}> = \sum_{0 \leq k \leq n/2} 2^k \ {k! (n-2k)!
\over n!} C_{n,k} \eqno({\rm A}.57)$$ $$(\vec{x}^{\, 2})^k
\sum_{{I\subset \{1 \cdots n\}}\atop |I| = 2k} \left [ \prod_{i\notin I}
(\vec{y}_i \cdot \vec{x}) \right ] \left [ \sum_{{\cal J} \in {\cal
P}_2(I)} \prod_{\{i,i'\}\in {\cal J}} (\vec{y}_i \cdot
\vec{y}_{i'})\right ] \ .$$

As a second (and last) step, let us apply formula (A.42) to the
multilinear symmetric function

$$\left ( \vec{x}_1, \cdots ,\vec{x}_n \right ) \to <\vec{y}_1 \otimes
\cdots \otimes \vec{y}_n|\prod\nolimits_n| \vec{x}_1 \otimes \cdots
\otimes \vec{x}_n>  \eqno({\rm A}.58)$$

\noi with $(\vec{y}_1, \cdots ,\vec{y}_n )$ fixed. This gives~:

$$<\vec{y}_1 \otimes \cdots \otimes \vec{y}_n|\prod\nolimits_n|
\vec{x}_1 \otimes \cdots \otimes \vec{x}_n> = \eqno({\rm A}.59)$$
$${(-1)^n \over n !} \sum_{{s_1, \cdots , s_n}\atop 0 \leq s_i \leq 1}
(-1)^{s_1 + \cdots + s_n}  <\vec{y}_1 \otimes \cdots \otimes
\vec{y}_n|\prod\nolimits_n| \left ( \sum_j s_j \vec{x}_j \right
)^{\otimes n} >$$

\noi or, using (A.57)

$$<\vec{y}_1 \otimes \cdots \otimes \vec{y}_n|\prod\nolimits_n|
\vec{x}_1 \otimes \cdots \otimes \vec{x}_n> = {(-1)^n \over n !}
\sum_{0\leq k \leq n/2} 2^k \ {k! (n-2k)! \over n!} \ C_{n,k} \eqno({\rm
A}.60)$$ $$\sum_{{I\subset \{1 \cdots n\}}\atop |I| = 2k}  \left [
\sum_{{\cal J} \in {\cal P}_2(I)} \prod_{\{i,i'\}\in {\cal J}}
(\vec{y}_i \cdot \vec{y}_{i'})\right ]$$ $$\sum_{{s_1, \cdots ,
s_n}\atop 0 \leq s_j \leq 1} (-1)^{s_1 + \cdots + s_n} \left ( \left (
\sum_j s_j \vec{x}_j \right )^2\right )^k \left [ \prod_{i\notin I}
\left ( \sum_j s_j (\vec{y}_i \cdot \vec{x}_j ) \right ) \right ] \ .$$

\noi Using the expansions

$$\prod_{i\notin I} \left ( \sum_j s_j (\vec{y}_i \cdot \vec{x}_j )
\right )  = \sum_{{u_{ij}\geq 0 (i \notin I, 1 \leq j \leq n)}\atop
u_{i1} + \cdots + u_{in} = 1} \left [ \prod_{{i\notin I}\atop 1 \leq j
\leq n} (s_j)^{u_{ij}} (\vec{y}_i \cdot \vec{x}_j)^{u_{ij}} \right ]
\eqno({\rm A}.61)$$

$$\left ( \left (  \sum_j s_j \vec{x}_j \right )^2\right )^k = \left (
\sum_{1\leq j, j' \leq n} s_js_{j'} (\vec{x}_j \cdot \vec{x}_{j'})
\right )^k \eqno({\rm A}.62)$$ $$= \sum_{{v_{11}, v_{12}, \cdots ,
v_{n-1,n},v_{nn} \geq 0}\atop v_{11} + v_{12} + \cdots + v_{nn} = k} {k!
\over v_{11}!v_{12}! \cdots v_{nn}!} \left [ \prod_{j,j'=1}^n
(s_js_{j'})^{v_{jj'}} (\vec{x}_j \cdot \vec{x}_{j'})^{v_{jj'}} \right ]
$$

\noi in (A.60) and collecting the powers of the $s_i$, we have~:

$$<\vec{y}_1 \otimes \cdots \otimes \vec{y}_n|\prod\nolimits_n|
\vec{x}_1 \otimes \cdots \otimes \vec{x}_n> = {(-1)^n \over n !}
\sum_{0\leq k \leq n/2} 2^k {k! (n-2k)! \over n!} C_{n,k} \eqno({\rm
A}.63)$$ $$\sum_{{I\subset \{1 \cdots n\}}\atop |I| = 2k} \left [
\sum_{{\cal J} \in {\cal P}_2(I)} \prod_{\{i,i'\}\in {\cal J}}
(\vec{y}_i \cdot \vec{y}_{i'})\right ]$$ $$\sum_{{u_{ij} \geq 0 (i
\notin I, 1 \leq j\leq n)}\atop u_{i1} + \cdots + u_{in} = 1}
\sum_{{v_{11}, v_{12}, \cdots , v_{n-1,n},v_{nn} \geq 0}\atop v_{11} +
v_{12} + \cdots + v_{nn} = k} {k! \over v_{11}!v_{12}! \cdots v_{nn}!}$$
$$\sum_{{s_1, \cdots , s_n}\atop 0 \leq s_j \leq 1} (-1)^{s_1 + \cdots +
s_n} \left [ \prod_{j=1}^n (s_j)^{p_j} \right ] \left [ \prod_{{i\notin
I}\atop 1 \leq j \leq n} (\vec{y}_i \cdot \vec{x}_j) ^{u_{ij}} \right ]
\left [ \prod_{j,j'=1}^n (\vec{x}_j \cdot \vec{x}_{j'}) ^{v_{jj'}}
\right ]$$

\noi where the exponent $p_j$ of $s_j$ is~:

$$p_j = \sum_{i\notin I} u_{ij} + \sum_{j'=1}^n (v_{j'j} + v_{jj'} ) \ .
\eqno({\rm A}.64)$$

According to the constraints on the $u_{ij}$'s and the $v_{jj'}$'s, one
has

$$\sum_{j=1}^n \ p_j = \sum_{{i\notin I}\atop 1 \leq j \leq n} u_{ij} +
2 \sum_{j,j'=1}^n v_{jj'} = n  \eqno({\rm A}.65)$$

\noi and, by the same arguments following eq. (A.49), one sees that only
the terms $p_1 = \cdots = p_n = 1$, $s_1 = \cdots = s_n = 1$ contributes
to (A.63). So, (A.63) reduces to~:

$$<\vec{y}_1 \otimes \cdots \otimes \vec{y}_n|\prod\nolimits_n|
\vec{x}_1 \otimes \cdots \otimes \vec{x}_n> =  \sum_{0\leq k \leq n/2}
2^k \ {(k!)^2 (n-2k)! \over (n!)^2} \ C_{n,k} \eqno({\rm A}.66)$$
$$\sum_{{I\subset \{1 \cdots n\}}\atop |I| = 2k} \left [ \sum_{{\cal J}
\in {\cal P}_2(I)} \prod_{\{i,i'\}\in {\cal J}} (\vec{y}_i \cdot
\vec{y}_{i'})\right ]$$ $$\sum_{{u_{ij}, v_{jj'} \geq 0 (i \notin I, 1
\leq j, j' \leq n)\atop \sum_{1\leq j\leq n} u_{ij} = 1 (i \notin
I)}\atop \sum_{i\notin I} u_{ij} + \sum_{1 \leq j'\leq n} (v_{jj'}
+v_{j'j}) = 1 (1 \leq j \leq n)} \left [ \prod_{{i\notin I}\atop 1 \leq
j \leq n} (\vec{y}_i \cdot \vec{x}_j) ^{u_{ij}} \right ] \left [
\prod_{j,j'=1}^n (\vec{x}_j \cdot \vec{x}_{j'}) ^{v_{jj'}} \right ] \ .
$$

\noi Consider the constraints on the $u_{ij}$'s~:

$$\sum_{j=1}^n u_{ij} = 1 \ \hbox{(for all $i\notin I$)}\ , \qquad
\sum_{i\not I} u_{ij} \leq 1 \ \hbox{(for all $1\leq j \leq n$)}\ .
\eqno({\rm A}.67)$$

\noi The solutions of (A.67) are in one to one correspondence with the
set of injective maps $\upsilon$~: ${\bf C}I \to \{1, \cdots ,n\}$ by~:

$$u_{ij} = 1 \ {\rm if}\ j = \upsilon (i) \ , \qquad u_{ij} = 0\ {\rm
if} \ j \not= \upsilon (i) \ . \eqno({\rm A}.68)$$

\noi Indeed, the first constraint (A.67) says that to each $i \notin I$
corrresponds one and only one $j = \upsilon (i)$ such that $u_{ij} = 1$,
and this defines a map $\upsilon : {\bf C}I \to \{ 1, \cdots , n\}$. The
second one says that, for each $j$, there is at most one $i \notin I$
such that $u_{ij} = 1$, and this means that the map $\upsilon$ is
injective. \par

We can therefore replace the $u_{ij}$'s, as summation variable, by the
injective maps $\upsilon : {\bf C}I \to \{1, \cdots , n\}$. It will be
furthermore convenient to replace $\upsilon$ by the pair $(J, \sigma )$
where $J = {\bf C} \upsilon ({\bf C}I)$ is the complementary subset of
the image of $\upsilon$, and $\sigma : {\bf C}I \to {\bf C}J$ is the
bijective map induced by $\upsilon$. Then formula (A.68) becomes~:

$$<\vec{y}_1 \otimes \cdots \otimes \vec{y}_n|\prod\nolimits_n|
\vec{x}_1 \otimes \cdots \otimes \vec{x}_n> = \sum_{0\leq k \leq n/2}
2^k {(k!)^2 (n-2k)! \over (n!)^2} C_{n,k} \eqno({\rm A}.69)$$
$$\sum_{{I,J\subset \{1 \cdots n\}}\atop |I| = |J| = 2k} \left [
\sum_{{\cal J} \in {\cal P}_2(I)} \prod_{\{i,i'\}\in {\cal J}}
(\vec{y}_i \cdot \vec{y}_{i'})\right ] \left [ \sum_{\sigma \in {\cal
B}(CI, CJ)} \prod_{i\notin I} (\vec{y}_i \cdot \vec{x}_{\sigma (i)})
\right ]$$ $$\sum_{{v_{jj'} \geq 0 (1 \leq j, j' \leq n)\atop
\sum_{1\leq j'\leq n} (v_{jj'} + v_{j'j}) = 0 (j \notin J)}\atop \sum_{1
\leq j'\leq n} (v_{jj'} +v_{j'j}) = 1 (j \in J)}  \left [
\prod_{j,j'=1}^n (\vec{x}_j \cdot \vec{x}_{j'}) ^{v_{jj'}} \right ] \ .
$$

This last summation on the $v_{jj'}$'s is exactly the same as the
summation on the $v_{ii'}$'s in formula (A.53), and is treated in the
same way. It is replaced by a summation on the partitions ${\cal J}' \in
{\cal P}_2(J)$ of $J$ by two-element subsets. To each ${\cal J}'$
corresponds $2^k$ solutions of the $v_{jj'}$'s constraints, which give
equal terms in (A.69). Finally we obtain

$$<\vec{y}_1 \otimes \cdots \otimes \vec{y}_n|\prod\nolimits_n|
\vec{x}_1 \otimes \cdots \otimes \vec{x}_n> = \sum_{0\leq k \leq n/2}
2^k {(k!)^2 (n-2k)! \over (n!)^2} C_{n,k} \sum_{{I,J\subset \{1 \cdots
n\}}\atop |I| = |J| = 2k}\eqno({\rm A}.70)$$ $$\left [ \sum_{{\cal J}
\in {\cal P}_2(I)} \prod_{\{i,i'\}\in {\cal J}} (\vec{y}_i \cdot
\vec{y}_{i'})\right ] \left [ \sum_{\sigma \in {\cal B}(CI, CJ)}
\prod_{i\notin I} (\vec{y}_i \cdot \vec{x}_{\sigma (i)}) \right ] \left
[ \sum_{{\cal J}' \in {\cal P}_2(I)} \prod_{\{j,j'\}\in {\cal J}'}
(\vec{x}_j \cdot \vec{x}_{j'})\right ] \ .$$

Formula (A.10) is just formula (A.70) with

$$(\vec{y}_1 , \cdots , \vec{y}_n ) = (\vec{e}_{i_1} , \cdots ,
\vec{e}_{i_n}) \ , \qquad (\vec{x}_1, \cdots , \vec{x}_n) =
(\vec{e}_{j_1}, \cdots , \vec{e}_{j_n})$$

\noi where ($\vec{e}_1, \vec{e}_2, \vec{e}_3)$ is the Cartesian basis of
${I\hskip - 1 truemm R}^3$. \par \vskip 0.5 truecm

\newpage
\noi {\bf Direct proof of the expression of the
projector} \vskip 5 truemm

The expression (A.10) for the projector $\prod_{i_1, \cdots , i_n;j_1,
\cdots , j_n}$ on the symmetric traceless tensors has been obtained by
rather lengthy and indirect arguments. However, {\it once} expression
(A.10) is known, it becomes possible to verify directly that it gives
the seeked projector. We do this now, obtaining a new proof of (A.10).
This new proof goes on, without added complications, for an arbitrary
dimension $D$ of space. \par

We thus consider now (A.10) as a {\it tentative formula}, with unknown
coefficients $f_{n,k}$. Let us enumerate the conditions for the
$\prod_{i_1, \cdots , i_n ; j_1, \cdots ,j_n}$ to be the components of
the projector $\prod_n$ on the subspace of the symmetric traceless
tensors (in the space of all $n$-rank tensors). \par

First, the image by $\prod_n$ of any tensor must be in the projection
subspace. For the components, this means that, for fixed $(j_1 , \cdots
, j_n)$, the $n$-rank tensor $\prod_{i_1, \cdots, i_n;j_1, \cdots ,
j_n}$ is symmetric traceless. Therefore we have the two conditions~:
\par

1) $\prod_{i_1,\cdots ,i_n;j_1, \cdots , j_n}$ is symmetric with respect
to the $i$'s.

2) $\sum\limits_{i_1,i_2 = 1}^D \delta_{i_1i_2} \prod_{i_1,\cdots
,i_n;j_1, \cdots , j_n} = 0$ \qquad (for all $i_3 , \cdots , i_n,j_1 ,
\cdots , j_n) \ . \hfill ({\rm A}.71)$ \par

\noi Next, $\prod_n$ must transform into itself any tensor in the
projection subspace. This gives the following third condition~:

3) $\sum\limits_{j_1,\cdots , j_n = 1}^D \prod_{i_1,\cdots ,i_n;j_1,
\cdots , j_n} \varepsilon_{j_1 , \cdots , j_n} = \varepsilon_{i_1 ,
\cdots , i_n}$ \hfill ({\rm A}.72) \par

\noi for any symmetric traceless tensor $\varepsilon_{i_1 , \cdots ,
i_n}$. \par

Conditions 1, 2, 3) says that $\prod_n$ is {\it some} projector on the
symmetric traceless tensors. To specify it completely, we must add that
$\prod_n$ is the {\it orthogonal} projector (i.e. it annihilates any
tensor orthogonal to all symmetric traceless tensors). It is equivalent
saying that $\prod_n$ is a symmetric (or hermitian) operator. This gives
a fourth and last condition~: \par

4) $\prod_{i_1,\cdots ,i_n;j_1, \cdots , j_n} = \prod_{j_1,\cdots
,j_n;i_1, \cdots , i_n}$ \hfill ({\rm A}.73) \par

Let us see now if we can fulfill these conditions with formula (A.10).
\par

The conditions 1) and 4) are easy. They are satisfied independently of
the coefficients $f_{n,k}$. The symmetry with respect to the $i$'s stems
from the fact that only the {\it set} $\{1 , \cdots , n\}$ of the
numbers which specify the $i$'s enters in (A.10). The symmetry with
respect to the exchange of the $i$'s and $j$'s is also clearly satisfied
by (A.10). \par

Condition 3) is not difficult to deal with. Notice that any term in
(A.10) containing a factor $\delta_{j_tj_{t'}}$ gives no contribution to
the left-hand side of (A.72). This is due to the fact that
$\varepsilon_{i_1 , \cdots , i_n}$ is symmetric traceless, so that

$$\sum_{j_t,j_{t'}=1}^D \delta_{j_tj_{t'}}\ \varepsilon_{j_1 , \cdots ,
j_n} = 0 \ . \eqno({\rm A}.74)$$

\noi The only terms in (A.10) without any factor $\delta_{j_tj_{t'}}$
are the ones with $J$ empty. But $J = \emptyset$ implies $k = 0$ and $I
= \emptyset$. So, with the trial formula (A.10), condition 3) takes the
form~:

$$f_{n,0} \sum_{j_1,\cdots ,j_n=1}^D \ \sum_{\sigma \in {\cal P}_n} \
\prod_{s=1}^n \delta_{i_sj_{\sigma (s)}} \ \varepsilon_{j_1 , \cdots ,
j_n} = \varepsilon_{i_1 , \cdots , i_n} \eqno({\rm A}.75)$$

\noi or, summing the $j_i$'s,

$$f_{n,0} \sum_{\sigma \in {\cal P}_n} \varepsilon_{i_{\sigma (1)} ,
\cdots ,  i_{\sigma (n)}} = \varepsilon_{i_1 , \cdots , i_n} \eqno({\rm
A}.76)$$

\noi where ${\cal P}_n$ is the set of permutations of the set $\{1,
\cdots , n\}$. Taking into account the symmetry of $\varepsilon_{i_1 ,
\cdots , i_n}$, and since $|{\cal P}_n| = n!$, this is the same as~:

$$n!\ f_{n,0}\ \varepsilon_{i_1 , \cdots , i_n} = \varepsilon_{i_1 ,
\cdots , i_n} \ . \eqno({\rm A}.77)$$

\noi Therefore, condition 3) just fixes $f_{n,k}$  for $k = 0$~:

$$f_{n,0} = {1 \over n!} \ . \eqno({\rm A}.78)$$

Condition 2) is the hard one. The first thing to be done is to rewrite
(A.10) in a form where the occurrences of the indices $i_1$ and $i_2$
are explicit. The result of this step is formula (A.83) below. To
alleviate the formulae, let us introduce the following notations~:

$$X(I) = \sum_{{\cal J} \in {\cal P}_2(I)} \ \prod_{\{r,r'\} \in {\cal
J}} \delta_{i_r,i_{r'}} \eqno({\rm A}.79)$$

$$Y(J) = \sum_{{\cal J}' \in {\cal P}_2(J)} \ \prod_{\{t,t'\} \in {\cal
J}'} \delta_{j_t,j_{t'}}$$

$$Z(I, J) = \sum_{\sigma \in {\cal B}(I,J)} \ \prod_{s \in I}
\delta_{i_s,j_{\sigma (s)}}$$

\noi for subsets $I$ and $J$ of $\{1, \cdots , n\}$. We will use the
following obvious relations~:

$$X(I) = \sum_{v \in I-\{u\}} \delta_{i_u,i_{v}} \ X(I - \{u , v\})
\qquad (u \in I\ {\rm fixed}) \eqno({\rm A}.80)$$

$$Y(J) = \sum_{v \in J-\{u\}} \delta_{j_u,j_{v}} \ Y(J - \{u , v\})
\qquad (u \in J\ {\rm fixed}) \eqno({\rm A}.81)$$

$$Z(I, J) = \sum_{v \in J} \delta_{i_u,j_{v}} \ Z(I - \{u \}, J - \{v\})
\qquad (u \in I\ {\rm fixed}) \ . \eqno({\rm A}.82)$$

\noi With these notations, (A.10) writes~:

$$\prod\nolimits_{i_1,\cdots ,i_n;j_1, \cdots , j_n} = \sum_{0 \leq k
\leq n/2} f_{n,k} \sum_{{J \subset \{1 \cdots n\}}\atop |J| = 2k} 
\sum_{{I \subset \{1 \cdots n\}}\atop |I| = 2k} X(I)\ Z({\bf C}I, {\bf
C}J)\ Y(J) \ . \eqno({\rm A}.83)$$

\noi The sum on the subsets $I \subset \{1, \cdots , n\}$ is decomposed
according to the four possible cases $(\emptyset , \{1\}, \{2\},
\{1,2\})$ of intersection of $I$ with the subset $\{1, 2\}$.

$$\prod\nolimits_{i_1,\cdots ,i_n;j_1, \cdots , j_n} = \sum_{0 \leq k
\leq n/2} f_{n,k} \sum_{{J \subset \{1 \cdots n\}}\atop |J| = 2k}
\eqno({\rm A}.84)$$ $$ \left \{  \sum_{{I \subset \{3 \cdots n\}}\atop
|I| = 2k} X(I)\ Z({\bf C}'I \cup \{1, 2\}, {\bf C}J) \right .$$ $$ +
\sum_{{I \subset \{3 \cdots n\}}\atop |I| = 2k-1} X(I \cup \{1\})\
Z({\bf C}'I \cup \{ 2\}, {\bf C}J)$$ $$ + \sum_{{I \subset \{3 \cdots
n\}}\atop |I| = 2k-1} X(I \cup \{2\})\ Z({\bf C}'I \cup \{ 1\}, {\bf
C}J)$$ $$\left .  + \sum_{{I \subset \{3 \cdots n\}}\atop |I| = 2k-2}
X(I \cup \{1,2\})\ Z({\bf C}'I, {\bf C}J)\right \} Y(J)\ . $$

\noi In this formula, the summation variable $I$ is obtained from the
one of (A.80) by possibly removing the elements 1 and 2. Furthermore
${\bf C}'I$ is the complementary set in $\{3, \cdots , n\}$ of the
subset $I$, while ${\bf C}J$ is as before the complementary set of $J$
in $\{1, \cdots, n\}$. The explicitation of the indices $i_1$ and $i_2$
is then effected by the following obvious formulae (written for subsets
$I \subset \{3, \cdots , n\}$ and $J \subset \{1, \cdots, n\}$), which
can be as well deduced from (A.80) and (A.82)~:

$$X(I \cup \{1\}) = \sum_{u \in I} \delta_{i_1i_u} \ X(I - \{ u \})
\eqno({\rm A}.85)$$ $$X(I \cup \{2\}) = \sum_{u \in I} \delta_{i_2i_u} \
X(I - \{ u \}) $$ $$X(I \cup \{1, 2\}) = \delta_{i_1i_2} \ X(I) +
\sum_{{u, v \in I}\atop u \not= v}  \delta_{i_1i_u} \ \delta_{i_2i_v}\
X(I - \{ u, v \})$$ $$Z(I \cup \{1\}, J ) = \sum_{v \in J}
\delta_{i_1j_v} \ Z(I,J - \{ v \})$$ $$Z(I \cup \{2\}, J ) = \sum_{v \in
J} \delta_{i_2j_v} \ Z(I,J - \{ v \})$$ $$Z(I \cup \{1, 2\},J) =
\sum_{{u, v \in J}\atop u \not= v}  \delta_{i_1j_u} \ \delta_{i_2j_v}\
Z(I,J - \{ u, v \}) \ . $$

\noi One obtains~:

$$\prod\nolimits_{i_1,\cdots ,i_n;j_1, \cdots , j_n} = \sum_{0 \leq k
\leq n/2} f_{n,k} \sum_{{J \subset \{1 \cdots n\}}\atop |J| = 2k}
\eqno({\rm A}.86)$$ $$ \left \{  \sum_{{I \subset \{3 \cdots n\}}\atop
|I| = 2k} \sum_{{u,v\notin J}\atop u\not= v} \delta_{i_1j_u}\
\delta_{i_2j_v}\ X(I)\ Z({\bf C}'I, {\bf C}J - \{u, v\}) \right .$$ $$ +
\sum_{{I \subset \{3 \cdots n\}}\atop |I| = 2k-1} \sum_{u\in I} \sum_{v
\notin J} \delta_{i_1i_u}\ \delta_{i_2j_v} \ X(I - \{u\})\ Z({\bf C}'I ,
{\bf C}J - \{ v\})$$ $$ + \sum_{{I \subset \{3 \cdots n\}}\atop |I| =
2k-1} \sum_{u\in I} \sum_{v \notin J} \delta_{i_2i_u}\ \delta_{i_1j_v} \
X(I - \{u\})\ Z({\bf C}'I , {\bf C}J -  \{ v\})$$ $$ + \sum_{{I \subset
\{3 \cdots n\}}\atop |I| = 2k-2} \delta_{i_1i_2}\ X(I)\ Z({\bf C}'I,
{\bf C}J)$$ $$\left .  + \sum_{{I \subset \{3 \cdots n\}}\atop |I| =
2k-2} \sum_{{u,v\in I}\atop u\not= v} \delta_{i_1i_u}\ \delta_{i_2i_v} \
X(I - \{u,v\})\ Z({\bf C}'I, {\bf C}J)\right \} Y(J)\ . $$ \vskip 8
truemm

\noi It is now straightforward to contract the indices $i_1$ and $i_2$~:

$$\sum_{i_1,i_2 = 1}^D \delta_{i_1i_2} \prod\nolimits_{i_1,\cdots
,i_n;j_1, \cdots , j_n} = \sum_{0 \leq k \leq n/2} f_{n,k}  \eqno({\rm
A}.87)$$ $$ \left \{ \sum_{{J \subset \{1 \cdots n\}}\atop |J| = 2k}
\sum_{{I \subset \{3 \cdots n\}}\atop |I| = 2k} \sum_{{u,v\notin J}\atop
u\not= v} \delta_{j_uj_v}\ \ X(I)\ Z({\bf C}'I, {\bf C}J - \{u, v\}) \
Y(J)\right .$$ $$ + 2 \sum_{{J \subset \{1 \cdots n\}}\atop |J| = 2k}
\sum_{{I \subset \{3 \cdots n\}}\atop |I| = 2k-1}\sum_{u\in I} \sum_{v
\notin J} \delta_{i_uj_v}\ X(I - \{u\})\ Z({\bf C}'I , {\bf C}J -  \{
v\})\ Y(J)$$ $$ + D \sum_{{J \subset \{1 \cdots n\}}\atop |J| = 2k}
\sum_{{I \subset \{3 \cdots n\}}\atop |I| = 2k-2} X(I)\ Z({\bf C}'I ,
{\bf C}J )\ Y(J)$$ $$\left .  + \sum_{{J \subset \{1 \cdots n\}}\atop
|J| = 2k} \sum_{{I \subset \{3 \cdots n\}}\atop |I| = 2k-2
}\sum_{{u,v\in I}\atop u\not= v} \delta_{i_ui_v}\ X(I - \{u,v\})\ Z({\bf
C}'I, {\bf C}J)\ Y(J)\right \} \ . $$

\noi Next we rewrite the first, second and last term in braces into the
same form as the third one. Let us transform a partial sum in the {\it
first} term as follows~:

$$\sum_{{J \subset \{1 \cdots n\}}\atop |J| = 2k} \sum_{{u,v\notin
J}\atop u\not= v} \delta_{j_uj_v}\ Z({\bf C}'I, {\bf C}J - \{u,v\})\
Y(J) \eqno({\rm A}.88)$$ $$= \sum_{{J \subset \{1 \cdots n\}}\atop |J| =
2k+2} \sum_{{u,v\in J}\atop u\not= v} \delta_{j_uj_v}\ Z({\bf C}'I, {\bf
C}J) Y(J - \{u,v\})\ . $$

\noi The first sum is, at fixed $u$ and $v$, over the subsets $J$ which
contain $u$ and $v$, and the second sum is over the subsets $J$ which do
not contain $u$ or $v$. To each subset of the first kind correspond the
subset of the second kind obtained by removing $u$ and $v$, and the
original first kind subset is recovered by including back $u$ and $v$.
This proves (A.88). The sum over $u$ and $v$ in the right-hand side of
(A.88) is then calculated. By summing (A.81) over $u \in J$, we have~:

$$\sum_{{u,v\in J}\atop u\not= v} \delta_{j_uj_v}\ Y(J - \{u,v\}) = |J|
\ Y(J) \ . \eqno({\rm A}.89)$$

\noi Combining (A.89) (with $|J| = 2k+2$) and (A.88), we have~:

$$\sum_{{J \subset \{1 \cdots n\}}\atop |J| = 2k} \sum_{{u,v\notin
J}\atop u\not= v} \delta_{j_uj_v}\ Z({\bf C}'I, {\bf C}J - \{u,v\})\
Y(J) \eqno({\rm A}.90)$$ $$= 2(k+1) \sum_{{J' \subset \{1 \cdots
n\}}\atop |J'| = 2k+2} Z({\bf C}'I, {\bf C}J') Y(J')\ . $$

\noi Let us transform a partial sum in the {\it second} term as
follows~:

$$\sum_{{I \subset \{3 \cdots n\}}\atop |I| = 2k-1} \ \sum_{u\in I}\
\sum_{v\notin J} \delta_{i_uj_v}\ X (I - \{u\}) \ Z({\bf C}'I, {\bf C}J
- \{v\}) \eqno({\rm A}.91)$$ $$= \sum_{{I \subset \{3 \cdots n\}}\atop
|I| = 2k-2} \ \sum_{u\in {\bf C}'I}\ \sum_{v\notin J} \delta_{i_uj_v}\
X(I) \ Z({\bf C}'I- \{u\}, {\bf C}J- \{v\})\ . $$

\noi The first sum is, at fixed $u \in \{3, \cdots ,n\}$, over the
subsets $I$ which contain $u$, and the second sum is over  the subsets
$I$ which do not contain $u$. Formula (A.91) results from the fact that
the two kinds of subsets are put in bijective correspondence by removing
or including $u$. The sum over $u$ and $v$ in the right-hand side of
(A.91) is then calculated. By summing (A.82) over $u \in I$, we have~:

$$\sum_{u \in I}\ \sum_{v\in J} \delta_{i_uj_v}\ Z(I - \{u\}, J - \{v\})
= |I| \ Z(I, J) \ . \eqno({\rm A}.92)$$

\noi Combining (A.92) (with $I \to {\bf C}'I$, $J \to {\bf C}J$, $|I|
\to n-2k$) and (A.91), we have~:

$$\sum_{{I \subset \{3 \cdots n\}}\atop |I| = 2k-1} \ \sum_{u\in I}\
\sum_{v\notin J} \delta_{i_uj_v}\ X (I - \{u\}) \ Z({\bf C}'I, {\bf C}J
- \{v\}) \eqno({\rm A}.93)$$ $$= (n-2k) \sum_{{I \subset \{3 \cdots
n\}}\atop |I| = 2k-2} X(I) \ Z({\bf C}'I, {\bf C}J)\ . $$

\noi For the {\it last} term, the sum over $u$ and $v$ is directly
calculated by summing (A.80) over $u \in I$, we have~:

$$\sum_{{u,v\in I}\atop u\not= v} \delta_{i_ui_v}\ X(I - \{u,v\}) = |I|
\ X(I) \ . \eqno({\rm A}.94)$$

\noi Using (A.90), (A.93), (A.94), the three last terms in (A.87)
combine, with a coefficient

$$2(n-2k) + D + 2(k-1) = 2(n-k+D/2 - 1) \eqno({\rm A}.95)$$

\noi and formula (A.87) becomes~:

$$\sum_{i_1,i_2 = 1}^D \delta_{i_1i_2} \prod\nolimits_{i_1,\cdots
,i_n;j_1, \cdots , j_n} = \eqno({\rm A}.96)$$ $$ \sum_{0 \leq k \leq
n/2-1} f_{n,k} \  2(k+1)  \sum_{{I \subset \{3 \cdots n\}}\atop |I| =
2k} \sum_{{J \subset \{1 \cdots n\}}\atop |J| = 2k+2} X(I)\ Z({\bf C}'I,
{\bf C}J)  \ Y(J)$$ $$ + \sum_{1 \leq k \leq n/2} f_{n,k} \ 2(n-k+D/2 -
1)  \sum_{{I \subset \{3 \cdots n\}}\atop |I| = 2k-2} \sum_{{J \subset
\{1 \cdots n\}}\atop |J| = 2k} X(I) \ Z({\bf C}'I , {\bf C}J\ Y(J) \ .
$$

\noi We have suppressed zero terms in the $k$ sums~: in the first sum,
the existence of $J \subset \{1, \cdots , n\}$ with $|J| = 2k+2$ needs
$2k \leq n-2$, and in the second sum, the existence of I with $|I| =
2k-2$ needs $2 k \geq 2$. This adjustment of the summation bounds is
important because now, after the change of variable $k \to k-1$, the
first sum combines exactly with the second one~:

$$\sum_{i_1,i_2 = 1}^D \delta_{i_1i_2} \prod\nolimits_{i_1,\cdots
,i_n;j_1, \cdots , j_n} = \eqno({\rm A}.97)$$ $$2 \sum_{1 \leq k \leq
n/2} \left [ k\ f_{n,k-1} + (n-k+D/2 - 1)\  f_{n,k} \right ]  \sum_{{I
\subset \{3 \cdots n\}}\atop |I| = 2k-2} \sum_{{J \subset \{1 \cdots
n\}}\atop |J| = 2k} X(I)\ Z({\bf C}'I, {\bf C}J)  \ Y(J) \ . $$

\noi We find thus that, with the trial formula (A.10), the tracelessness
condition 2) takes the form of the following recurrence relation for the
coefficients $f_{n,k}$~:

$$k\ f_{n,k-1} + (n-k+D/2 - 1)\  f_{n,k} = 0 \qquad (1 \leq k \leq n/2)
\eqno({\rm A}.98)$$

\noi and this recurrence relation, with the initial condition (A.78),
uniquely determines the $f_{n,k}$~:

$$f_{n,k} = (-1)^k {k!\Gamma (n-k+D/2 - 1) \over n!\Gamma (n+D/2-1)} =
(-1)^k \ {1 \over n!} {n+D/2-1\choose k}^{-1}\eqno({\rm A}.99)$$

\noi where the second expression is defined for all required values $D
\geq 1$, $n \geq 0$, $0 \leq k \leq n/2$, while the more explicit first
one is ambiguous ($\infty / \infty)$ for $D = 2$, $n = 0$, $k = 0$. \par

This result can be expressed with factorials, directly in case of $D$
even and by use of the duplication formula of the $\Gamma$ function in
case of $D$ odd~:

$$f_{n,k} = (-1)^k \ {k!(n-k+D/2 - 2)! \over n!(n+D/2-2)!} \qquad (D\
{\rm even}) \eqno({\rm A}.100)$$

$$f_{n,k} = (-1)^k \ 2^{-2k} \ {k!(n+D/2 - 3/2)! (2n-2k+D-3)! \over
n!(n-k+D/2-3/2)!(2n+D-3)!} \qquad (D\ {\rm odd}) \ . \eqno({\rm
A}.101)$$

\noi For the spatial dimension $D=3$, the expression (A.11) of $f_{n,k}$
is recovered. \par \vskip 5 truemm

 \noi {\bf Calculation of the particular matrix elements from
the expression of the projector} \vskip 5 truemm

By using the explicit expression (A.10) of the projector, it is of
course possible to compute any matrix element needed. As a simple
example, we present here the calculation of the particular matrix
elements (A.3). \par

We may do all the calculations for a particle at rest ($v = (1,
\vec{0}))$, and write the final result covariantly in term of the
4-vectors $v$, $v_i$, $v_f$. Then we have~:

$$v_f^{\mu_1} \cdots v_f^{\mu_n} \prod\nolimits_{\mu_1 \cdots
\mu_n;\nu_1 \cdots \nu_n} \ v_i^{\nu_1} \cdots v_i^{\nu_n} \eqno({\rm
A}.102)$$ $$= \sum_{i_1,\cdots i_n,j_1,\cdots j_n} v_f^{i_1 } \cdots
v_f^{i_n} \prod\nolimits_{i_1, \cdots , i_n;j_1, \cdots, j_n} v_i^{j_1}
\cdots v_i^{j_n} \ . $$

\noi The calculation of this from (A.10) simply amounts to the
substitutions~:

$$\delta_{i_ri_{r'}} \to (\vec{v}_f)^2 \ , \quad \delta_{i_sj_{\sigma
(x)}} \to (\vec{v}_f\cdot \vec{v}_i) \ , \delta_{j_tj_{t'}} \to
(\vec{v}_i)^2 \eqno({\rm A}.103)$$

\noi and we have~:

$$\sum_{i_1,\cdots i_n,j_1,\cdots j_n} v_f^{i_1 } \cdots v_f^{i_n}
\prod\nolimits_{i_1, \cdots , i_n;j_1, \cdots, j_n} v_i^{j_1} \cdots
v_i^{j_n} \eqno({\rm A}.104)$$ $$= \sum_{0 \leq k \leq n/2} f_{n,k}
\sum_{{I,J \subset \{1 \cdots n\}}\atop |I| = |J| = 2k} \ \sum_{{\cal J}
\in {\cal P}_2(I)} \ \sum_{\sigma \in {\cal B}({\bf C}I,{\bf C}J)} \
\sum_{{\cal J}' \in {\cal P}_2(J)} (\vec{v}_f)^{2k} \ (\vec{v}_f\cdot
\vec{v}_i)^{n-2k} \ (\vec{v}_i)^{2k}\ . $$

\noi The summand is independent of the summations variables $I$, $J$,
${\cal J}$, $\sigma$, ${\cal J}'$, so that we have just to count the
number of values they take. As well known, the number of subsets of
cardinality $q$ in a set of cardinality $n$ is given by the binomial
coefficient ${n \choose q} = {n! \over q! (n-q)!}$, and the number
$|{\cal B}(X,Y)|$ of bijections of a set $X$ on a set $Y$ is $|X|!$ if
$|X| = |Y|$ (and 0 else). The number $|{\cal P}_2(X)|$ of partitions of
a set $X$ by two-element subsets is ${(2k)! \over 2^k k!}$ when $|X| =
2k$. This last number is easily obtained by first considering the set of
decompositions of $X$ into ordered $k$-uples of ordered pairs. The
number of such $k$-uples is the same $(2k)!$ as the number of
permutations of $X$, and when the orderings (of the $k$-uples and of the
pairs) are disregarded, there are $2^kk!$ $k$-uples giving each
partition by two-element subsets. Therefore the numbers of values taken
by the summation variables in (A.104) are~:

$\begin{array}{ll} \displaystyle{{n! \over (2k)!(n-2k)!}} &\qquad
\hbox{for $I$ and for $J$} \\ &\\ \displaystyle{{(2k)! \over 2^kk!}}
&\qquad \hbox{for ${\cal J}$ and for ${\cal J}'$} \\ &\\ (n-2k)! &\qquad
\hbox{for $\sigma$}\ . \end{array}$ \par \vskip 5 truemm

\noi and from (A.104) we have~:

$$\sum_{i_1,\cdots i_n,j_1,\cdots j_n} v_f^{i_1 } \cdots v_f^{i_n}
\prod\nolimits_{i_1, \cdots , i_n;j_1, \cdots, j_n} v_i^{j_1} \cdots
v_i^{j_n} \eqno({\rm A}.105)$$ $$= \sum_{0 \leq k \leq n/2} f_{n,k}
\left ( {n! \over (2k)!(n-2k)!}\right )^2 \left ( {(2k)! \over
2^kk!}\right )^2 (n-2k)! \ (\vec{v}_f)^{2k} \ (\vec{v}_f\cdot
\vec{v}_i)^{n-2k} \ (\vec{v}_i)^{2k}$$ $$= \sum_{0 \leq k \leq n/2}
2^{-2k} {(n!)^2 \over (k!)^2(n-2k)!}f_{n,k} \ (\vec{v}_f)^{2k} \
(\vec{v}_f\cdot \vec{v}_i)^{n-2k} \ (\vec{v}_i)^{2k}\ . $$

\noi With $(\vec{v}_f)^{2}$, $(\vec{v}_f\cdot \vec{v}_i)$,
$(\vec{v}_i)^{2}$ written in covariant form using (A.41), we recover
(A.5) with $C_{n,k}$ given by

$$C_{n,k} = (-1)^k\ 2^{-2k} {n! \over (k!)^2(n-2k)!} {n+D/2-2\choose
k}^{-1}\eqno({\rm A}.106)$$

\noi for an arbitrary spatial dimension $D$. A compact expression like
(A.3) is obtained from (A.5) and (A.106) by introducing the Gegenbauer
polynomials $C_n^{\lambda}$, which can be defined by the generating
functions~:

$$\sum\limits_{n \geq 0} C_n^{\lambda}(x)\ t^n = {1 \over (1 -
2xt+t^2)^{\lambda}} \eqno({\rm A}.107)$$

\noi and have the following expressions~:

$$C_n^{\lambda}(x) = \sum\nolimits_{k} (-1)^k {n-k\choose k} {n-k +
\lambda -1 \choose n-k} (2x)^{n-2k} \eqno({\rm A}.108)$$

$$C_n^{\lambda}(x) = \sum\nolimits_{k} (-1)^k {k+ \lambda - 1\choose k}
{n + 2\lambda -1 \choose n-2k} x^{n-2k}\ (1 - x^2)^k \ .\eqno({\rm
A}.109)$$

\noi Indeed, (A.106) writes~:

$$C_{n,k} = (-1)^k\ 2^{-2k} {n+ D/2 -2 \choose n}^{-1} {n-k \choose k}
{n-k+D/2-2\choose n-k} \eqno({\rm A}.106')$$

\noi and using (A.108) we have

$$\sum_{0\leq k\leq n/2} C_{n,k} \ x^{n-2k} = {1 \over 2^n} {n+ D/2 -2
\choose n}^{-1} C_n^{D/2-1}(x) \eqno({\rm A}.110)$$

\noi obtaining~:

$$\sum_{i_1,\cdots i_n,j_1,\cdots j_n} v_f^{i_1 } \cdots v_f^{i_n}
\prod\nolimits_{i_1, \cdots , i_n;j_1, \cdots, j_n} v_i^{j_1} \cdots
v_i^{j_n} \eqno({\rm A}.111)$$ $$= {1 \over 2^n} {n+ D/2 -2 \choose
n}^{-1} \ |\vec{v}_f|^{n} \ |\vec{v}_i|^{n} \ C_n^{D/2-1} \left (
{\vec{v}_f\cdot \vec{v}_i\over |\vec{v}_f|\ |\vec{v}_i|} \right ) \ . $$

\noi When $D=3$, expression (A.3) is recovered from (A.111) using
$C_n^{1/2}(x) = P_n(x)$ and ${n-1/2 \choose n} = 2^{-2n} {(2n)! \over
(n!)^2}$. Using in (A.111) the expression (A.109) of $C_n^{\lambda}(x)$,
one obtains expression (A.6) with $C'_{n,k}$ given by~:

$$C'_{n,k} = (-1)^k\ {1 \over 2^n} {n+ D/2 -2 \choose n}^{-1} {k+D/2-2
\choose k} {n+D-3\choose n-2k} \ . \eqno({\rm A}.112)$$

Notice that the expression (A.106$'$) is ambiguous for $D=2$, $n \geq 1$
(while (A.106) is defined for all required values $D \geq 1$, $n \geq
0$, $0 \leq k \leq n/2$), and the ambiguity propagates to (A.111) and
(A.112). If we want the case $D=2$ to be included in our formulae, we
may use, instead the Gegenbauer polynomials, the Jacobi polynomials
simply related to them by~:

$$C_n^{\lambda}(x) = {(2\lambda )_n \over (\lambda + 1/2)_n} \
P_n^{(\lambda - 1/2, \lambda -1/2)}(x)\eqno({\rm A}.113)$$

\noi which have the following expressions~:

$$P_n^{(\alpha , \alpha )}(x) = 2^{-n} \sum\nolimits_{k} (-1)^k {n+
\alpha \choose k} {2n-2k + 2\alpha \choose n-2k} x^{n-2k}\eqno({\rm
A}.114)$$

$$P_n^{(\alpha , \alpha )}(x) = \sum\nolimits_{k} (-1)^k \ 2^{-2k} {n-k
\choose k} {n+ \alpha \choose n-k} x^{n-2k} \ (1 - x^2)^k \ .\eqno({\rm
A}.115)$$

\noi Rewriting (A.106) as~:

$$C_{n,k} = (-1)^k\ {2n+D-3 \choose n}^{-1} {n+D/2-3/2\choose k}
{2n-2k+D-3 \choose n-2k} \eqno({\rm A}.106'')$$

\noi from (A.114) we see that~:

$$\sum_{i_1,\cdots i_n,j_1,\cdots j_n} v_f^{i_1 } \cdots v_f^{i_n}
\prod\nolimits_{i_1, \cdots , i_n;j_1, \cdots, j_n} v_i^{j_1} \cdots
v_i^{j_n} \eqno({\rm A}.111')$$ $$= 2^n {2n+ D -3 \choose n}^{-1} \
|\vec{v}_f|^{n} \ |\vec{v}_i|^{n} \ P_n^{(D-3)/2,(D-3)/2)} \left (
{\vec{v}_f\cdot \vec{v}_i\over |\vec{v}_f|\ |\vec{v}_i|} \right ) \ . $$

\noi and using (A.115) we obtain~:

$$C'_{n,k} = (-1)^k\ 2^{n-2k} {2n+D-3 \choose n}^{-1} {n-k\choose k}
{n+D/2-3/2 \choose n-k} \ . \eqno({\rm A}.112')$$

\noi The expression (A.106$''$) is ambiguous only for $D = 1$, $n=1$,
and so are (A.111$'$) and (A.112$'$). \par

Finally, let us consider the case $D=1$. It is in fact trivial. All
$n$-rank tensor spaces ($n \geq 0$) are of dimension 1. All $n$-rank
tensors are symmetric. The subspace of $n$-rank symmetric traceless
tensors is the whole space when $n = 0$ or 1, and is the {\it zero
subspace} when $n \geq 2$. Therefore, the projector on this subspace is
the identity operator when $n = 0$ or 1 (as for any dimension $D$), and
is, when $n \geq 2$, the {\it zero operator}.\par

To see how this particular case is obtained with our results, we may
apply eq. (A.111), which is well defined for $D=1$. First, due to
$\vec{v}_f\cdot \vec{v}_i = |\vec{v}_f|\ |\vec{v}_i|{\rm sign}\
(\vec{v}_f\cdot \vec{v}_i)$, (A.111) writes

$$\sum_{i_1,\cdots i_n,j_1,\cdots , j_n} v_f^{i_1 } \cdots v_f^{i_n}
\prod\nolimits_{i_1, \cdots , i_n;j_1, \cdots, j_n} v_i^{j_1} \cdots
v_i^{j_n} = {1 \over 2^n} {n-3/2 \choose n}^{-1} \ C_n^{-1/2}(1) \
(\vec{v}_f\cdot \vec{v}_i)^n  \eqno({\rm A}.116)$$

\noi (where the sum has only one term). Next $C_n^{-1/2}(1)$ is given by
eq. (A.109)~:

$$C_n^{-1/2}(1) = {n - 2 \choose n} = (-1)^n {1 \choose n} =
\delta_{n,0} - \delta_{n,1} \ . \eqno({\rm A}.117)$$

\noi Then we have

$$\left . {1 \over 2^n} {n - 3/2 \choose n}^{-1} \right |_{n=0} = 1
\quad , \quad \left . {1 \over 2^n} {n - 3/2 \choose n}^{-1} \right
|_{n=1} = - 1 \eqno({\rm A}.118)$$

\noi so that (A.116) becomes~:

$$\sum_{i_1,\cdots i_n,j_1,\cdots , j_n} v_f^{i_1 } \cdots v_f^{i_n}
\prod\nolimits_{i_1, \cdots , i_n;j_1, \cdots, j_n} v_i^{j_1} \cdots
v_i^{j_n} = \delta_{n,0} + (\vec{v}_f\cdot \vec{v}_i) \delta_{n,1}
\eqno({\rm A}.119)$$

\noi and, as expected, this vanishes for $n \geq 2$.

\vskip 2 truecm

\noi {\large \bf Appendix B.} \\

In this Appendix we give a manifestly covariant derivation of Bjorken
and Uraltsev SR using the states and currents considered by Uraltsev
\cite{5r}. He considers the forth component of the vector current, and
initial and final $B^*$ states, allowing for spin flip transitions,
i.e., with our notation, he takes $\Gamma_1 = \Gamma_2 = \gamma^0$, the
initial and final states ${\cal B}_i = B^{*(\lambda_i)}(1,{\bf 0})$,
${\cal B}_f = B^{*(\lambda_f)}(v_f^0,{\bf v}_f)$, and performing an
expansion for small velocities. In the covariant language adopted here,
the case he considers is $$\Gamma_1 = \Gamma_2 = {/\hskip - 2 truemm
v}_i $$ $${\cal B}_i = P_{i+} {/ \hskip - 2 truemm \varepsilon}_i \qquad
\qquad {\cal B}_f = P_{f+} {/ \hskip - 2 truemm \varepsilon}_f \
.\eqno({\rm B}.1)$$

\noi We realize that this case does not present the symmetry of the 
simple choice of Sections 3 and 4, since both currents, projected in 
the $v_i$ direction, appear
in a non-symmetric way relative to the initial and final states, that 
have four-velocities $v_i$ and $v_f$. This aspect, plus the $B^*$ 
polarization, complicates
the calculation in a considerable way, since then all states 
$2^+_{3/2}$, $1^+_{3/2}$, $0^+_{1/2}$ and $1^+_{1/2}$ contribute. We 
give now the covariant version of
Uraltsev calculation. \par

After a good deal of algebra, the r.h.s. of the general SR 
(\ref{38e}) writes, for the choice (B.1)~:
$$R(w_i, w_f, w_{if}) = \xi (w_{if}) \Big \{ (\varepsilon_i 
\cdot \varepsilon_f)(w_i + w_f) - (\varepsilon_i\cdot v_f) 
(\varepsilon_f \cdot v') -
(\varepsilon_f \cdot v_i) (\varepsilon_i \cdot v')$$
$$ - (2w_i + 1) \left [ (\varepsilon_i \cdot \varepsilon_f)(w_{if} + 
1) - (\varepsilon_i\cdot v_f) (\varepsilon_f\cdot v_i)\right
] \Big \} \eqno({\rm B}.2)$$

\noi while the contribution of the different intermediate states is 
given by~:\\

\noi $L(0^-_{1/2}) = 0 \hfill ({\rm B}.3)$\\

\noi $L(1^-_{1/2}) = \Big \{ - (w_i + 1) \left [ (\varepsilon_i \cdot 
\varepsilon_f) (w_{if} + w_i) - (\varepsilon_f \cdot v_i) 
(\varepsilon_i \cdot v_f) +
(\varepsilon_f \cdot v_i) (\varepsilon_i \cdot v') \right ]$ \par 
\vskip 2 truemm
\noi $+ (\varepsilon_i \cdot v') \left [ ( \varepsilon_f \cdot v_i) 
(2w_i-w_f+1) + (\varepsilon_f \cdot v')(w_{if}-1) \right ] \Big \} 
\displaystyle{\sum_n}
\xi^{(n)}(w_i) \xi^{(n)}(w_f) \hfill({\rm B}.4)$\\

\noi $L(2^+_{3/2} ) = \Big \{ \displaystyle{{1 \over 2}}  (w_{if} - 
w_f w_i) (w_i + 1) \Big [ (\varepsilon_i \cdot \varepsilon_f) w_{if} 
- (\varepsilon_f \cdot
v_i) (\varepsilon_i \cdot v_f)$\par \vskip 2 truemm
\noi $+ (\varepsilon_i \cdot \varepsilon_f) w_i + (\varepsilon_f 
\cdot v_i) (\varepsilon_i \cdot v') \Big ] - \displaystyle{{1 \over 
2}} (w_{if} - w_f w_i)
(\varepsilon_i \cdot v') \Big [ (\varepsilon_f \cdot v_i) w_i + 
(\varepsilon_f \cdot v_i) w_i - (\varepsilon_f \cdot v') \Big ]$\par 
\vskip 2 truemm
\noi $ + \displaystyle{{1 \over 6}} (-2 -2w_i - 2w_f -3w_{if} + 
4w_iw_f) (\varepsilon_i \cdot v') \Big [ (\varepsilon_f \cdot v_i) (1 
- w_f) + (\varepsilon_f \cdot
v')w_{if} \Big ]$\par \vskip 2 truemm
\noi $- \displaystyle{{1 \over 2}} \Big [ (\varepsilon_i \cdot 
v_f)(w_i+1) - (\varepsilon_i \cdot v') w_{if} \Big ]  (\varepsilon_f 
\cdot v')$\par \vskip 2 truemm
\noi $- \displaystyle{{1 \over 2}} w_i \Big [ (\varepsilon_i \cdot 
v_f)(w_i+1) - (\varepsilon_i \cdot v') w_{if} \Big ] \Big [ 
(\varepsilon_f \cdot v_i)(1 - w_f) +
(\varepsilon_f \cdot v') w_{if} \Big ]$\par \vskip 2 truemm
\noi $+ \displaystyle{{1 \over 2}} w_f (\varepsilon_i \cdot 
v')(\varepsilon_f \cdot v') + w_{i} \Big [ (\varepsilon_i \cdot 
v_f)(w_i +1) - (\varepsilon_i \cdot
v') (v_i \cdot v_f) \Big ] (\varepsilon_f \cdot v_i) $\par \vskip 2 truemm
\noi $ - w_i w_f (\varepsilon_i \cdot v') (\varepsilon_f \cdot v_i) 
\Big \} 3 \displaystyle{\sum_n} \tau_{3/2}^{(n)}(w_i) 
\tau_{3/2}^{(n)}(w_f) \hfill({\rm
B}.5)$\\

\noi $L(1^+_{3/2}) = \Big \{ - \displaystyle{{1 \over 6}}  (1 + w_i) 
(1 + w_f) \displaystyle{{1 \over 4}} Tr \Big [ {/\hskip - 2 truemm 
v}_i {/\hskip
- 2 truemm \varepsilon}_i \gamma^{\sigma} {/\hskip - 2 truemm 
v'}\gamma_5 \Big ] \displaystyle{{1 \over 4}} Tr \Big [ {/\hskip - 2 
truemm v}_i {/\hskip
- 2 truemm \varepsilon}_f \gamma_{\sigma} {/\hskip - 2 truemm 
v}_f\gamma_5 \Big ]$\par \vskip 2 truemm
\noi $+ \displaystyle{{1 \over 6}}  (1 + w_i) (1 + w_f) 
\displaystyle{{1 \over 4}} Tr \Big [ {/\hskip - 2 truemm v}_i {/\hskip
- 2 truemm \varepsilon}_i {/\hskip - 2 truemm v'} \gamma^{\sigma} 
\gamma_5 \Big ] \displaystyle{{1 \over 4}} Tr \Big [ {/\hskip - 2 
truemm v}_i {/\hskip
- 2 truemm \varepsilon}_f {/\hskip - 2 truemm v'}\gamma_{\sigma} 
\gamma_5 \Big ]$\par \vskip 2 truemm
\noi $- \displaystyle{{1 \over 2}}  (1 + w_i) \displaystyle{{1 \over 
4}} Tr \Big [ {/\hskip - 2 truemm v}_i {/\hskip
- 2 truemm \varepsilon}_i {/\hskip - 2 truemm v'} {/\hskip - 2 truemm 
v}_f \gamma_5 \Big ] \displaystyle{{1 \over 4}} Tr \Big [ {/\hskip - 
2 truemm v}_i {/\hskip
- 2 truemm \varepsilon}_f {/\hskip - 2 truemm v'}{/\hskip - 2 truemm 
v}_f \gamma_5 \Big ]\Big \} 3 \displaystyle{\sum_n} 
\tau_{3/2}^{(n)}(w_i)
\tau_{3/2}^{(n)}(w_f) \hfill({\rm B}.6)$\\

\noi $L(0^+_{1/2} ) = (\varepsilon_i \cdot v')\Big [ (\varepsilon_f 
\cdot v_i) (1 - w_f) + (\varepsilon_f \cdot v') (v_i \cdot v_f) \Big 
] 4\displaystyle{\sum_n}
\tau_{1/2}^{(n)}(w_i) \tau_{1/2}^{(n)}(w_f) \hfill({\rm B}.7)$\\

\noi $L(1^+_{1/2} ) = \Big \{ - \displaystyle{{1 \over 4}}Tr \Big [ 
{/\hskip - 2 truemm v}_i {/\hskip
- 2 truemm \varepsilon}_i   {/\hskip - 2 truemm v'} 
\gamma^{\sigma}\gamma_5 \Big ] \displaystyle{{1 \over 4}} Tr \Big [ 
{/\hskip - 2 truemm v}_i {/\hskip
- 2 truemm \varepsilon}_f  {/\hskip - 2 truemm v}_f\gamma_{\sigma} 
\gamma_5 \Big ]$\par \vskip 2 truemm
\noi $+ \displaystyle{{1 \over 4}}  Tr \Big [ {/\hskip - 2 truemm v}_i {/\hskip
- 2 truemm \varepsilon}_i  {/\hskip - 2 truemm v'} \gamma^{\sigma} 
\gamma_5 \Big ] \displaystyle{{1 \over 4}} Tr \Big [ {/\hskip - 2 
truemm v}_i {/\hskip
- 2 truemm \varepsilon}_f  {/\hskip - 2 truemm v'}\gamma_{\sigma} 
\gamma_5 \Big ]\Big \} 4 \displaystyle{\sum_n} \tau_{1/2}^{(n)}(w_i)
\tau_{1/2}^{(n)}(w_f) \hfill({\rm B}.8)$\\

\noi $L(2^-_{3/2} ) =  \displaystyle{{1 \over 2}} \Big \{ (w_{if} - 
w_iw_f ) (w_i + 1) \displaystyle{{1 \over 4}} Tr \Big [ {/\hskip - 2 
truemm v}_i {/\hskip
- 2 truemm \varepsilon}_i {/\hskip - 2 truemm v}'\gamma^{\sigma} 
\gamma_5 \Big ] \displaystyle{{1 \over 4}} Tr \Big [ {/\hskip - 2 
truemm v}_i {/\hskip
- 2 truemm \varepsilon}_f {/\hskip - 2 truemm v}_f\gamma_{\sigma} 
\gamma_5 \Big ]$\par \vskip 2 truemm
\noi $- \ (w_{if} -w_iw_f ) \displaystyle{{1 \over 4}} Tr \Big [ 
{/\hskip - 2 truemm v}_i {/\hskip
- 2 truemm \varepsilon}_i {/\hskip - 2 truemm v}'\gamma^{\sigma} 
\gamma_5 \Big ] \displaystyle{{1 \over 4}} Tr \Big [ {/\hskip - 2 
truemm v}_i {/\hskip
- 2 truemm \varepsilon}_f {/\hskip - 2 truemm v}'\gamma_{\sigma} 
\gamma_5 \Big ] \Big\}$\par \vskip 2 truemm
\noi $+ \ w_i \displaystyle{{1 \over 4}} Tr \Big [ {/\hskip - 2 
truemm v}_i {/\hskip
- 2 truemm \varepsilon}_i {/\hskip - 2 truemm v}' {/\hskip - 2 truemm 
v}_f\gamma_5 \Big ] \displaystyle{{1 \over 4}} Tr \Big [ {/\hskip - 2 
truemm
v}_i {/\hskip - 2 truemm \varepsilon}_f {/\hskip - 2 truemm v}' 
{/\hskip - 2 truemm v}_f \gamma_5 \Big ]\Big \} 3 
\displaystyle{\sum_n}
\sigma_{3/2}^{(n)}(w_i) \sigma_{3/2}^{(n)}(w_f)$\\

\noi $L(1^-_{3/2} ) = \Big \{ - \displaystyle{{1 \over 6}}  (w_i - 1) 
(w_f - 1) (w_i + 1)\Big [ (\varepsilon_i \cdot \varepsilon_f) (w_{if} 
+ w_i) +
(\varepsilon_f \cdot v_i) (\varepsilon_i \cdot v')- (\varepsilon_f 
\cdot v_i) (\varepsilon_i \cdot v_f)\Big ]$\par \vskip 2 truemm
\noi $+ \displaystyle{{1 \over 6}} (w_i - 1) (w_f - 1) (\varepsilon_i 
\cdot v') \Big [2 (\varepsilon_f \cdot v_i) w_i - (\varepsilon_f 
\cdot v')\Big ]$\par \vskip 2
truemm \noi $ + \displaystyle{{1 \over 6}} (1 - 9w_{if}+ 4w_iw_f + 
2w_i + 2w_f) (\varepsilon_i \cdot v') \Big [ (\varepsilon_f \cdot 
v_i) (1 - w_f) + (\varepsilon_f
\cdot v') w_{if} \Big ]$\par \vskip 2 truemm
\noi $ + \displaystyle{{1 \over 2}} (w_f - 1) (\varepsilon_i \cdot 
v') \Big [ 2w_i (\varepsilon_f \cdot v_i) - (\varepsilon_f \cdot v') 
\Big ]$\par \vskip 2 truemm
\noi $+ \displaystyle{{1 \over 2}} (w_i - 1) \Big [ (\varepsilon_i 
\cdot v_f)(w_i+1) - (\varepsilon_i \cdot v') w_{if} \Big ] \Big [ 
(\varepsilon_f \cdot v')
w_{if} + (\varepsilon_f \cdot v_i) (1 - w_f)\Big ] \Big \}$\par 
\vskip 2 truemm
\noi  $3 \displaystyle{\sum_n} \sigma_{3/2}^{(n)}(w_i) 
\sigma_{3/2}^{(n)}(w_f)$\\

\noi From
$${1 \over 4} Tr \left [ {/\hskip - 2 truemm a} \ {/\hskip - 2 truemm 
b} \ {/\hskip - 2 truemm c} \ {/\hskip - 2 truemm d} \ \gamma_5 
\right ] = - i
\varepsilon_{\mu\nu\rho\sigma} \ a^{\mu}b^{\nu}c^{\rho}d^{\sigma} 
\eqno({\rm B}.9)$$

\noi one can express the contributions $L(1^+_{3/2} )$, $L(1^+_{1/2} 
)$ and $L(2^-_{3/2} )$ in terms of scalar products. Indeeed, the 
product of two tensors
$\varepsilon_{\mu\nu\rho\sigma}$ even {\it non contracted} can be 
expressed in terms of the tensor $g_{\mu\nu}$~:
$$\varepsilon_{\mu\nu\rho\sigma} \ \varepsilon_{\mu '\nu '\rho 
'\sigma '} = - \det (g_{\alpha \alpha '}) \qquad (\alpha = \mu , \nu 
, \rho , \sigma \ ; \
\alpha ' = \mu ', \nu ', \rho ', \sigma ')\eqno({\rm B}.10)$$
$$g^{\mu\mu '} \varepsilon_{\mu\nu\rho\sigma} \ \varepsilon_{\mu '\nu 
'\rho '\sigma '} = - \det (g_{\alpha \alpha '}) \qquad (\alpha = \nu 
, \rho , \sigma \ ; \
\alpha ' = \nu ', \rho ', \sigma ') \ .\eqno({\rm B}.11)$$
\vskip 3 truemm

\noi From these relations one obtains, for the traces involved in 
$L(1^+_{3/2} )$, $L(1^+_{1/2} )$ and $L(2^-_{3/2} )$~:  \\

\noi $\displaystyle{{1 \over 4}}  Tr \Big [ {/\hskip - 2 truemm v}_i {/\hskip
- 2 truemm \varepsilon}_i  {/\hskip - 2 truemm v}' \gamma^{\sigma} 
\gamma_5 \Big ] \displaystyle{{1 \over 4}} Tr \Big [ {/\hskip - 2 
truemm v}_i {/\hskip
- 2 truemm \varepsilon}_f {/\hskip - 2 truemm v}_f \gamma_{\sigma} 
\gamma_5 \Big ]$\par
\noi $= - (\varepsilon_i \cdot v_f) (\varepsilon_f \cdot v') + w_i 
(\varepsilon_i \cdot v_f) (\varepsilon_f \cdot v_i) - (w_{if} w_i - 
w_f) (\varepsilon_i \cdot
\varepsilon_f)\hfill({\rm B}.12)$\\

\noi $\displaystyle{{1 \over 4}}  Tr \Big [ {/\hskip - 2 truemm v}_i {/\hskip
- 2 truemm \varepsilon}_i {/\hskip - 2 truemm v}' \gamma^{\sigma} 
\gamma_5 \Big ] \displaystyle{{1 \over 4}} Tr \Big [ {/\hskip - 2 
truemm v}_i {/\hskip
- 2 truemm \varepsilon}_f {/\hskip - 2 truemm v}' \gamma_{\sigma} 
\gamma_5 \Big ]$\par
\noi $= - (\varepsilon_i \cdot v') (\varepsilon_f \cdot v') + w_i 
(\varepsilon_i \cdot v') (\varepsilon_f \cdot v_i) - (w_i^2 - 1) 
(\varepsilon_i \cdot
\varepsilon_f)\hfill({\rm B}.13)$\\

\noi $\displaystyle{{1 \over 4}}  Tr \Big [ {/\hskip - 2 truemm v}_i {/\hskip
- 2 truemm \varepsilon}_i {/\hskip - 2 truemm v}' {/\hskip - 2 truemm 
v}_f \gamma_5 \Big ] \displaystyle{{1 \over 4}} Tr \Big [ {/\hskip - 
2 truemm
v}_i {/\hskip - 2 truemm \varepsilon}_f {/\hskip - 2 truemm v}' 
{/\hskip - 2 truemm v}_f \gamma_5 \Big ]$\par
\noi $= (w_{if}^2 - 1)  (\varepsilon_i \cdot v') (\varepsilon_f \cdot 
v') -(w_{if} w_f -w_i)  (\varepsilon_i \cdot v') (\varepsilon_f \cdot 
v_i)$\par
\noi $- (w_{if}w_i -w_f) (\varepsilon_i \cdot v_f) (\varepsilon_f 
\cdot v') + (w_{if} - w_i w_f)  (\varepsilon_i \cdot v_f) 
(\varepsilon_f \cdot v_i)$\par
\noi$+ (2w_{if} w_iw_f -w_{if}^2 - w_i^2 -w_f^2 + 1) (\varepsilon_i 
\cdot \varepsilon_f)\ . \hfill({\rm B}.14)$\\

\noi From the latter expressions (B.12)-(B.14) and from (B.2)-(B.8) 
one gets finally for the equation (\ref{38e})~: \\

\noi $ \Big \{ - (w_i + 1) (w_{if} + w_i) (\varepsilon_i \cdot 
\varepsilon_f) + (w_i + 1) (\varepsilon_i \cdot v_f) (\varepsilon_f 
\cdot v_i) + (w_i - w_f)
(\varepsilon_i \cdot v') (\varepsilon_f \cdot v_i)$\par
\noi $+ (w_{if} - 1) (\varepsilon_i \cdot v') (\varepsilon_f \cdot 
v') \Big \} \displaystyle{\sum_n} \xi^{(n)} (w_i) \xi^{(n)}(w_f)$\par
\noi $ + \Big \{ - (w_i + 1) (4w_fw_iw_{if} + 2w_fw_i^2 -w_i^2 -w_f^2 
- 2w_{if} w_i - 3w_{if}^2 + 1) (\varepsilon_i \cdot 
\varepsilon_f)$\par
\noi $ + (w_i+1) (4w_iw_f - 3w_{if} + w_i) (\varepsilon_i \cdot v_f) 
(\varepsilon_f\cdot v_i) - (w_i +1) (w_f+1) (\varepsilon_i  \cdot 
v_f) (\varepsilon_f \cdot
v')$\par
\noi $ + \Big [ (3w_{if} - 2w_iw_f + w_f)(w_f - w_i) - (w_i + 1)^2 
\Big ] (\varepsilon_i \cdot v') (\varepsilon_f \cdot v_i)$\par
\noi $+ (2w_iw_f-3w_{if}-w_i-w_f-1)(w_{if} - 1) (\varepsilon_i \cdot 
v') (\varepsilon_f \cdot v')\Big \} \displaystyle{\sum_n} 
\tau_{3/2}^{(n)}(w_i)
\tau_{3/2}^{(n)}(w_f)$\par
\noi $+ \ 4 \Big \{ (w_{if}w_i-w_i^2 -w_f + 1) (\varepsilon_i \cdot 
\varepsilon_f)-w_i(\varepsilon_i \cdot v_f) (\varepsilon_f \cdot v_i) 
+ (\varepsilon_i \cdot
v_f) (\varepsilon_f \cdot v')$\par
\noi $+ (1 + w_i - w_f)(\varepsilon_i \cdot v') (\varepsilon_f \cdot 
v_i)+ (w_{if} - 1) (\varepsilon_i \cdot v') (\varepsilon_f \cdot v') 
\Big\}
\displaystyle{\sum_n} \tau_{1/2}^{(n)}(w_i) \tau_{1/2}^{(n)}(w_f)$\par
\noi $\Big \{ \left  (w_i + 2w_i w_f - 3w_f^2 w_i - w_i^3 - 2w_fw_i^3 
- 2w_{if} + 2w_f w_{if} + 2w_i^2 w_{if} \right .$  \par
\noi $\left . +\ 4w_f w_i^2 w_{if} - 3w_i w_{if}^2 \right ) 
(\varepsilon_i \cdot \varepsilon_f)$  \par
\noi $+\ \left ( - 1 + w_f + w_i^2 - 4w_f w_i^2 + 3w_i w_{if} \right 
) (\varepsilon_f \cdot v_i) (\varepsilon_i \cdot v_f)$  \par
\noi $+ \ 3 \left ( w_iw_f - w_{if}\right ) (\varepsilon_i \cdot v_f) 
(\varepsilon_f \cdot v')$  \par
\noi $+\ \left ( w_f - w_f^2 - w_i + 3w_i w_f - 2w_f^2 w_i + w_i^2 + 
2w_f w_i^2 - 3w_{if} \right .$  \par
\noi $\left . +\ 3w_f w_{if} - 3w_i w_{if} \right ) (\varepsilon_f 
\cdot v_i) (\varepsilon_i \cdot v')$  \par
\noi $+\left ( w_{if} - 1) (-1 + w_i + w_f + 2w_iw_f - 3w_{if}\right 
) (\varepsilon_i \cdot v')(\varepsilon_f \cdot v') \Big \} 
\displaystyle{\sum_n}
\sigma_{3/2}^{(n)} (w_i) \sigma_{3/2}^{(n)}(w_f)$  \par
\noi $= \xi (w_{if}) \Big \{ \Big [ (w_i + w_f) - (2w_i + 1) 
(w_{if} + 1) \Big ] (\varepsilon_i \cdot \varepsilon_f) + (2w_i + 1) 
(\varepsilon_i \cdot
v_f) (\varepsilon_f \cdot v_i) + \cdots $\par
\noi $- (\varepsilon_i \cdot v_f)  (\varepsilon_f \cdot v') - 
(\varepsilon_i \cdot v')(\varepsilon_f \cdot v_i) \Big \} \ . 
\hfill({\rm B}.15)$\\

\noi This expression is considerably more complicated than equation 
(\ref{48e}), that readily gives Uraltsev SR. We can choose the 
particular polarizations~:

$$\begin{array}{ll} \qquad \varepsilon_i^{(1)} = \displaystyle{{v_f - 
w_{if}v_i \over \sqrt{w_{if}^2 - 1}}} &\qquad \varepsilon_f^{(1)} = 
\displaystyle{{v_i -
w_{if}v_f \over \sqrt{w_{if}^2 - 1}}} \\
&\\
\qquad \varepsilon_i^{(2)} = \displaystyle{{v' - w_{i}v_i \over 
\sqrt{w_{i}^2 - 1}}} &\qquad \varepsilon_f^{(2)} = \displaystyle{{v' 
- w_{f}v_f \over
\sqrt{w_{f}^2 - 1}}}\hskip 5.7 truecm ({\rm B}.16)\end{array}$$

\noi that satisfy $\varepsilon_i^2 = - 1$, $\varepsilon_i \cdot v_i = 
0$, $\varepsilon_f^2 = -1$, $\varepsilon_f \cdot v_f = 0$. \par

We can consider the following different cases~:

$${\rm (1)}\qquad  \varepsilon_i = \varepsilon_i^{(1)}\quad , \quad 
\varepsilon_f = \varepsilon_f^{(1)}\ ;$$
$${\rm (2)}\qquad  \varepsilon_i = \varepsilon_i^{(2)}\quad , \quad 
\varepsilon_f = \varepsilon_f^{(1)}\ ; $$
$${\rm (3)}\qquad  \varepsilon_i = \varepsilon_i^{(1)} \quad , \quad 
\varepsilon_f = \varepsilon_f^{(2)}\ ;$$
$${\rm (4)}\qquad  \varepsilon_i = \varepsilon_i^{(2)} \quad , \quad 
\varepsilon_f = \varepsilon_f^{(2)} \ . \eqno({\rm B}.17)$$

These four different cases exhaust the number of independent SR in 
the case under consideration, characterized by (B.1). \par

That there are only four independent SR can be seen by the following 
argument. If, in the general SR (B.15), we make the replacements (the 
sum over $\alpha$
denotes the sum over the different polarizations)~:
$$\varepsilon_i^{\mu} \to \sum_{\alpha} \varepsilon_i^{(\alpha )\mu} 
= g^{\rho\mu} - v_i^{\rho} v_i^{\mu}$$
$$\varepsilon_f^{\nu} \to \sum_{\alpha} \varepsilon_f^{(\alpha )\nu} 
= g^{\sigma\nu} - v_f^{\sigma} v_i^{\nu} \eqno({\rm B}.18)$$

\noi one obtains a set of tensorial identities, that depend only on 
$v_i$, $v_f$ and $v'$~:
$$ X^{\rho\sigma}(v_i, v_f, v') = 0 \ . \eqno({\rm B}.19)$$

\noi From these 16 identities one obtains 9 scalar identities 
saturating with all the pairs $v_{i\rho} v_{i\sigma}$, $v_{f\rho} 
v_{f\sigma}$, $v'_{\rho}
v'_{\sigma}$, $v_{i\rho} v_{f\sigma}$, ...

\par \vskip 5 truemm

$$v_{i\rho} v_{i\sigma}\ X^{\rho\sigma} (v_iv_f, v') = 0$$
\hskip 5.5 truecm $\cdots \hfill  ({\rm B}.20) $ \par

\noi plus 3 other scalar identities, identically vanishing,
$$\varepsilon_{\mu \nu \rho \sigma} v_i^{\mu} v_f^{\nu}\ 
X^{\rho\sigma} (v_iv_f, v') \equiv 0$$
\hskip 5 truecm $\cdots \hfill  ({\rm B}.21)$ \par

\noi However, among these equations, only 4 are independent, 
corresponding to the two non-vanishing products
$$v_{f\rho} \left ( g^{\rho \mu} - v_i^{\rho} v_i^{\mu} \right ) = 
v_f^{\mu} - w_{if} v_i^{\mu}$$
$$v'_{\rho} \left ( g^{\rho \mu} - v_i^{\rho} v_i^{\mu} \right ) = 
v'^{\mu} - w_{i} v_i^{\mu} \eqno({\rm B}.22)$$

\noi that must be combined with the other two four-vectors~:
$$v_{i\rho} \left ( g^{\rho \mu} - v_f^{\rho} v_f^{\mu} \right ) = 
v_i^{\mu} - w_{if} v_f^{\mu}$$
$$v'_{\rho} \left ( g^{\rho \mu} - v_f^{\rho} v_f^{\mu} \right ) = 
v'^{\mu} - w_{f} v_f^{\mu} \ .\eqno({\rm B}.23)$$

  Let us now write this SR (B.15) for the different cases. We need the 
following scalar products~:

$$\begin{array}{ll} \varepsilon_i^{(1)}\cdot v_f  = - 
\displaystyle{{w_{if}^2 - 1 \over \sqrt{w_{if}^2 - 1}}} 
&\varepsilon_i^{(1)} \cdot v' = -
\displaystyle{{w_{if}w_i- w_f \over \sqrt{w_{if}^2 - 1}}} \\ &\\
\varepsilon_i^{(2)} \cdot v_f  = - \displaystyle{{w_{if}w_i -w_f\over 
\sqrt{w_{i}^2 - 1}}} &\varepsilon_i^{(2)} \cdot v'  = - 
\displaystyle{{w_{i}^2-1\over
\sqrt{w_{i}^2 - 1}}}\\
& \\
\varepsilon_f^{(1)}\cdot v_i  = - \displaystyle{{w_{if}^2 - 1 \over 
\sqrt{w_{if}^2 - 1}}} &\varepsilon_f^{(1)} \cdot v' = -
\displaystyle{{w_{if}w_f -w_i \over \sqrt{w_{if}^2 - 1}}} \\ &\\
\varepsilon_f^{(2)} \cdot v_i  = - \displaystyle{{w_{if}w_f -w_i\over 
\sqrt{w_{f}^2 - 1}}} &\varepsilon_f^{(2)} \cdot v'  = - 
\displaystyle{{w_{f}^2-1\over
\sqrt{w_{f}^2 - 1}}}\\
&\\
\varepsilon_i^{(1)}\cdot \varepsilon_f^{(1)} = 
\displaystyle{{w_{if}(w_{if}^2 - 1) \over \sqrt{w_{if}^2 - 
1}\sqrt{w_{if}^2 - 1}}} &\quad \varepsilon_i^{(2)} \cdot
\varepsilon_f^{(1)}  = \displaystyle{{w_{if}(w_{if} w_i- w_f) \over 
\sqrt{w_{i}^2 - 1}\sqrt{w_{if}^2 - 1}}} \\ \end{array}$$
$$\begin{array}{ll}
\varepsilon_i^{(1)} \cdot  \varepsilon_f^{(2)}=
\displaystyle{{w_{if}(w_{if} w_f -w_i)\over \sqrt{w_{f}^2 - 
1}\sqrt{w_{if}^2 - 1}}} &\quad \varepsilon_i^{(2)} \cdot 
\varepsilon_f^{(2)}   =
\displaystyle{{w_{if}w_iw_f - w_i^2 - w_f^2 + 1\over \sqrt{w_{if}^2 - 
1}\sqrt{w_{if}^2 - 1}}}\qquad \quad ({\rm B}.24)\end{array}$$
\vskip 3 truemm

\noi Since equation (B.15) is linear in $\varepsilon_i$ and in 
$\varepsilon_f$, in deducing the equation for the different cases we 
can multiply (B.15) by the
denominators defining the polarizations in (B.16). We thus obtain 
from (B.15) four different equations for the different cases 
(B.17).\par

If, in particular, we make $w_i = w_f = w$, we obtain the following 
equations, for the different cases considered~:\\

(1) $\varepsilon_i = \varepsilon_i^{(1)}$, $\varepsilon_f = 
\varepsilon_f^{(1)}$~: \\

\noi $- (w_{if} - 1)(1 + w - w^2 - w_{if} + 2w w_{if} +3w^2 w_{if} +w 
w_{if}^2) \displaystyle{\sum_n} \left [ \xi^{(n)} (w)\right ]^2$\par
\noi $ + (w_{if} - 1) (w - 2w^2 - 4w^3 + 2w^4 + 2w_{if} + w w_{if} - 
4w^2 w_{if} - 6w^4 w_{if} + 2w_{if}^2 + 3 ww_{if}^2 + 6w^2 
w_{if}^2$\par
\noi $- 4w^3 w_{if}^2 + 3ww_{if}^3) \displaystyle{\sum_n} \left 
[\tau_{3/2}^{(n)} (w)\right ]^2$\par
\noi $-4(w_{if} - 1) (w-w^2 -w_{if} + 3w^2 w_{if} -w_{if}^2 - w 
w_{if}^2 ) \displaystyle{\sum_n} \left [\tau_{1/2}^{(n)} (w)\right 
]^2$\par
  \noi $+ (w_{if} - 1) \left ( 1 - w - 2w^2 + 2w^4 + w_{if} + 3w 
w_{if} - 4w^3 w_{if} - 6 w^4 w_{if} - 3 w_{if}^2 \right .$ \par
\noi $\left . + \ w(w_{if})^2 + 10 w^2 w_{if}^2 + 4 w^3 w_{if}^2 - 3 
w_{if}^3 - 3 ww_{if}^3 \right )  \displaystyle{\sum_n} \left [ 
\sigma_{3/2}^{(n)} (w)\right
]^2 + \cdots $ \par \noi $=- (w_{if}-1)(w_{if}+1) (1+w_{if}+2w 
w_{if}) \xi (w_{if})\hfill ({\rm B}.25)$\\

(2) $\varepsilon_i = \varepsilon_i^{(2)}$, $\varepsilon_f = 
\varepsilon_f^{(1)}$~:  \\

\noi $- w(w+1) (w_{if}-1) (w+w_{if}) \displaystyle{\sum_n} \left [ 
\xi^{(n)} (w)\right ]^2$\par
\noi $$ + (w + 1) (w_{if}-1) (1 - 2w^2-2w^4 + w_{if} +w w_{if} +2w^2 
w_{if} - 4w^3 w_{if} + 3w w_{if}^2) \displaystyle{\sum_n} \left [ 
\tau_{3/2}^{(n)}
(w)\right ]^2$$\par
\noi $+4(w-1) (w+1) (w_{if} -1) (1 - w +w_{if})\displaystyle{\sum_n} 
\left [ \tau_{1/2}^{(n)} (w)\right ]^2$\par
\noi $+(w-1) (w+1) (w_{if}-1) \left ( 2w^2 - 2w^3 - 3w_{if} + 
2ww_{if} + 4w^2w_{if} - 3w_{if}^2\right ) \displaystyle{\sum_n} \left 
[
\sigma_{3/2}^{(n)} (w)\right ]^2$\par
\noi $ + \cdots =- (w+1) (w_{if}+1) (w_{if}- 1) (2w-1) 
\xi (w_{if})\hfill ({\rm B}.26)$\\

(3) $\varepsilon_i = \varepsilon_i^{(1)}$, $\varepsilon_f = 
\varepsilon_f^{(2)}$~: \\

\noi $- w(w+1) (w_{if}-1) (w+w_{if}) \displaystyle{\sum_n} \left [ 
\xi^{(n)} (w)\right ]^2$\par
\noi $$ + (w + 1) (w_{if}-1) (1 - 2w^2-2w^4 + w_{if} +w w_{if} +2w^2 
w_{if} - 4w^3 w_{if} + 3w w_{if}^2) \displaystyle{\sum_n} \left [ 
\tau_{3/2}^{(n)}
(w)\right ]^2$$\par
\noi $+4(w-1) (w+1) (w_{if} -1) (1 - w +w_{if})\displaystyle{\sum_n} 
\left [ \tau_{1/2}^{(n)} (w)\right ]^2$\par
\noi $+ (w-1) (w_{if}-1) \left ( -w^2 + 3w^3 -2w^4 - 3w_{if} - 
ww_{if} + 9w^2w_{if} + w^3w_{if} - 3w_{if}^2  - 3ww_{if}^2\right 
)$\par
\noi $ \displaystyle{\sum_n} \left [
\sigma_{3/2}^{(n)} (w)\right ]^2 + \cdots = - (w+1) (w_{if}+1) 
(w_{if}- 1) (2w-1) \xi (w_{if})\hfill ({\rm B}.27)$\\

(4) $\varepsilon_i = \varepsilon_i^{(2)}$, $\varepsilon_f = 
\varepsilon_f^{(2)}$~: \\

\noi $(w+1)^2 (-1+w+w^2-ww_{if}) \displaystyle{\sum_n} \left [ 
\xi^{(n)} (w)\right ]^2$\par
\noi $+(w+1)^2(-w-2w^2 + 4w^3 + 2w^4 + 2w_{if} -2ww_{if}-2w^2w_{if} + 
4w^3 w_{if} + 3ww_{if}^2 )$\par
\noi $\displaystyle{\sum_n} \left [ \tau_{3/2}^{(n)}(w)\right ]^2 + 
4(w-1)^2 (w+1) (w+w_{if})\displaystyle{\sum_n}\left [ 
\tau_{1/2}^{(n)}(w)\right ]^2$\par
\noi $+ (w-1)^2 (w+1) \left ( 1 - 2w^2 + 2w^3 - 2ww_{if} + 4w^2 
w_{if} - 3w_{if}^2\right )\displaystyle{\sum_n} \left [ 
\sigma_{3/2}^{(n)}
(w)\right ]^2 + \cdots$ \par
\noi $  =(w+1)(-1 + 3w -w_{if}-3ww_{if} + 2w^2 w_{if}) 
\xi (w_{if})\ . \hfill ({\rm
B}.28)$\\

\noi Let us now obtain the SR that can be obtained without deriving 
the function $\xi(w_{if})$~:\\

(1) $\varepsilon_i = \varepsilon_i^{(1)}$, $\varepsilon_f = 
\varepsilon_f^{(1)}$~: \\

\noi Dividing by $(w_{if}-1)$ and taking the limit $w_{if} \to 1$, one gets~:\\

\noi $- \ 2(w+1)^2 \ \displaystyle{\sum_n} \ \left [ \xi^{(n)} 
(w)\right ]^2 - 4(w-1) (w+1)^3 \ \displaystyle{\sum_n} \ \left [ 
\tau_{3/2}^{(n)}(w)\right ]^2$\par
\noi $ -\ 8(w-1) (w+1) \ \displaystyle{\sum_n} \ \left 
[\tau_{1/2}^{(n)} (w)\right ]^2 - 4(w^2-1)^2 \displaystyle{\sum_n} 
\left [ \sigma_{3/2}^{(n)}(w)\right
]^2 + \cdots $\par
  \noi $=- \ 4(w+1)\hfill ({\rm B}.29)$\\

(2) $\varepsilon_i = \varepsilon_i^{(2)}$, $\varepsilon_f = 
\varepsilon_f^{(1)}$ and (3) $\varepsilon_i = \varepsilon_i^{(1)}$, 
$\varepsilon_f =
\varepsilon_f^{(2)}$~: \\

\noi Dividing by $(w_{if}-1)$ and taking the limit $w_{if} \to 1$, 
one obtains~:\\

\noi $- \ w(w+1)^2 \ \displaystyle{\sum_n} \ \left [\xi^{(n)} 
(w)\right ]^2 - 2(w-1) (w+1)^4 \ \displaystyle{\sum_n} \ \left 
[\tau_{3/2}^{(n)} (w)\right ]^2$\par
\noi $ -\ 4(w-1) (w+1) (w-2) \ \displaystyle{\sum_n} \ \left 
[\tau_{1/2}^{(n)} (w)\right ]^2 + 2 (w^2-1)^2 (3-w) 
\displaystyle{\sum_n}  \left [
\sigma_{3/2}^{(n)}(w)\right ]^2 + \cdots$\par
\noi $=- \ 2(w+1)(2w-1)\hfill ({\rm B}.30)$\\

  (4) $\varepsilon_i = \varepsilon_i^{(2)}$, $\varepsilon_f = 
\varepsilon_f^{(2)}$~: \\

\noi Taking the limit $w_{if} \to 1$, one gets~: \\

\noi $(w-1)(w+1)^3 \ \displaystyle{\sum_n} \ \left [ \xi^{(n)} 
(w)\right ]^2 + 2(w+1)^4 (w-1)^2 \ \displaystyle{\sum_n} \ \left 
[\tau_{3/2}^{(n)} (w)\right
]^2$\par  \noi $ +\ 4(w-1)^2 (w+1)^2 \ \displaystyle{\sum_n} \ \left 
[\tau_{1/2}^{(n)} (w)\right ]^2 + 2 (w^2-1)^3 \displaystyle{\sum_n} 
\left [
\sigma_{3/2}^{(n)}(w)\right ]^2 + \cdots $\par
\noi $= 2(w+1)^2(w-1)\hfill ({\rm B}.31)$\\

\noi From (B.26)-(B.27) one gets two sum rules \\

\noi $\displaystyle{{w+1 \over 2}} \ \displaystyle{\sum_n}\  \left 
[\xi^{(n)}(w)\right ]^2 + (w-1) \left \{ 2 \displaystyle{\sum_n}\ 
\left
[\tau_{1/2}^{(n)}(w)\right ]^2 + (w+1)^2 \ \displaystyle{\sum_n}\ 
\left [\tau_{3/2}^{(n)}(w)\right ]^2 \right \}$\par
\noi $+ \ (w+1)(w-1)^2 \displaystyle{\sum_n}  \left [ 
\sigma_{3/2}^{(n)}(w)\right
]^2 + \cdots = 1$ \hfill ({\rm B}.32)\\

\noi $w\ \displaystyle{{w+1} \over 2} \ \displaystyle{\sum_n}\  \left 
[\xi^{(n)}(w)\right ]^2 + (w-1) \Big \{(w+1)^3 \ 
\displaystyle{\sum_n}\
\left [\tau_{3/2}^{(n)}(w)\right ]^2 $\par \noi $+ 2(w-2) \ 
\displaystyle{\sum_n}\  |\tau_{1/2}^{(n)}(w)|^2 \Big \} - (w+1) 
(w-1)^2 (3-w) \displaystyle{\sum_n}  \left [ 
\sigma_{3/2}^{(n)}(w)\right
]^2 + \cdots = 2w-1  \hfill ({\rm B}.33)$\\

\noi The first SR is Bjorken SR (\ref{56e}). \par

Eliminating ${w+1 \over 2} \sum\limits_n |\xi^{(n)}(w)|^2$ between 
(B.28) and (B.29) one obtains~: \\
$$(w+1)^2 \sum_n |\tau_{3/2}^{(n)}(w)|^2 - 4 \sum_n 
|\tau_{1/2}^{(n)}(w)|^2 - 3(w^2-1) \sum_n \left [ 
\sigma_{3/2}^{(n)}(w)\right
]^2 + \cdots = 1 \eqno({\rm B}.34)$$

\vskip 3 truemm

\noi Equation (B.34) is another generalization of Uraltsev SR for $w 
\not = 1$~; it reduces indeed to (\ref{60e}) for $w = 1$. Notice that 
the states ${3 \over
2}^-$ contribute at order $(w - 1)$ to (B.34), while they do not 
contribute at all to the generalization of Uraltsev SR for $w \not= 
1$ (\ref{75e}). There is no
contradiction~: these are two different generalizations, and the 
difference can be traced back to the fact that the former is obtained 
from asymmetric currents $\{
{/\hskip - 2 truemm v}_i  , {/\hskip - 2 truemm v}_i \}$ while the 
latter is obtained from symmetric currents $\{ {/\hskip - 2 truemm 
v}_i \gamma_5 , {/\hskip - 2
truemm v}_f \gamma_5 \}$ relative to the initial and final four-velocities. \par \vskip 5 truemm

\noi {\large \bf Acknowledgements} \par

We are indebted to Zoltan Ligeti for encouragements and for pointing out
to us the interest of publishing the derivation of Uraltsev sum rule
given in Section 3, and to Nikolai Uraltsev for long standing
discussions on QCD in the heavy quark limit. We acknowledge support from
the EC contract HPRN-CT-2002-00311 (EURIDICE).

\end{document}